\documentclass[journal]{IEEEtran}
\ifCLASSINFOpdf
\else
\fi

\usepackage{times}
\usepackage{epsfig}
\usepackage{graphicx}
\usepackage{amsmath}
\usepackage{amssymb}
\usepackage{enumitem}
\usepackage{subcaption}
\usepackage{txfonts}
\usepackage{multirow}
\usepackage{array}
\usepackage{adjustbox}
\usepackage{xcolor}
\usepackage{cite}
\usepackage{url}

\definecolor{RevisionColor}{rgb}{1, 0.549, 0}

\begin{document}
\title{Exploit Camera Raw Data for Video Super-Resolution via Hidden Markov Model Inference}

\author{Xiaohong~Liu,~\IEEEmembership{Student~Member,~IEEE},
Kangdi Shi,~
Zhe Wang,~
and~Jun~Chen,~\IEEEmembership{Senior Member,~IEEE}% <-this % stops a space
\thanks{X. Liu, K. Shi, and J. Chen (corresponding author) are with the Department of Electrical and Computer Engineering, McMaster University, Hamilton, ON L8S 4K1, Canada (e-mail: liux173@mcmaster.ca; shik9@mcmaster.ca; chenjun@mcmaster.ca). 
	
Z. Wang is with the David R. Cheriton School of Computer Science, University of Waterloo, Waterloo, ON N2L 3G1, Canada (e-mail: zhe.wang@uwaterloo.ca).% <-this % stops a space

%This paper has supplementary downloadable material available at http://ieeexplore.ieee.org., provided by the author.
}}

\maketitle

\begin{abstract}
To the best of our knowledge, the existing deep-learning-based Video Super-Resolution (VSR) methods  exclusively make use of videos produced by the Image Signal Processor (ISP) of the camera system as inputs. Such methods are 1) inherently suboptimal due to information loss incurred by non-invertible operations in ISP, and 2) inconsistent with the real imaging pipeline where VSR  in fact serves as a pre-processing unit of ISP. To address this issue,
we propose a new VSR method that can directly exploit camera sensor data, accompanied by a carefully built Raw Video Dataset (RawVD) for training, validation, and testing. This method consists of a Successive Deep Inference  (SDI) module and a reconstruction module, among others.
The SDI module is designed according to the architectural principle suggested by a canonical decomposition result for Hidden Markov Model (HMM) inference; it estimates the target high-resolution frame by repeatedly performing pairwise feature fusion using deformable convolutions. 
The reconstruction module, built with elaborately designed Attention-based Residual Dense Blocks (ARDBs), serves the purpose of 1) refining the fused feature and 2) learning the color information needed to generate a spatial-specific  transformation for accurate color correction. Extensive experiments demonstrate that owing to the informativeness of the camera raw data,  the effectiveness of the  network architecture, and the separation of super-resolution and color correction processes, the proposed method achieves superior VSR results compared to the state-of-the-art and can be adapted to any specific camera-ISP. Code and dataset are available at \url{https://github.com/proteus1991/RawVSR}.

\end{abstract}

\begin{IEEEkeywords}
Video super-resolution, camera raw data, hidden Markov model inference
\end{IEEEkeywords}

%%%%%%%%% BODY TEXT

\section{Introduction}
Super-resolution (SR) is a promising technique that can restore high-resolution (HR) pictorial data from their low-resolution (LR) counterpart without requiring hardware upgrades. Over the past few decades, it has found applications in a wide range of areas such as medical imaging~\cite{MedicalImaging1,MedicalImaging2}, satellite imaging~\cite{satellite1,satellite2} and surveillance~\cite{surveillance2,surveillance3}. SR is also useful for improving the quality of data dedicated to
high-level vision tasks~\cite{hao2018deep,recoginition1,na2018object}. There are two major categories of SR: Single Image Super-Resolution (SISR) and Video Super-Resolution (VSR). Although 
early studies~\cite{liao2015video,kappeler2016video,shahar2011space} largely treat VSR as a simple extension of SISR by focusing on intra-frame spatial correlation, more attention has been paid in recent works~\cite{sajjadi2018frame,jo2018deep,haris2019recurrent,wang2019edvr} to strategically exploiting inter-frame temporal correlation (typically in the form of motion estimation and compensation) to further improve the VSR results. The availability of an extra dimension in VSR creates both opportunities and challenges. Indeed, there is considerable freedom in the ways that multiple LR frames can be leveraged to reconstruct one target HR frame, especially  with the advent of deep learning techniques. Even though many effective heuristics have been proposed, a theoretical guideline for the fusion process is still lacking.

\begin{figure}[t]
	\centering
	\begin{minipage}[h]{0.49\linewidth}
		\centering
		\includegraphics[width=\linewidth]{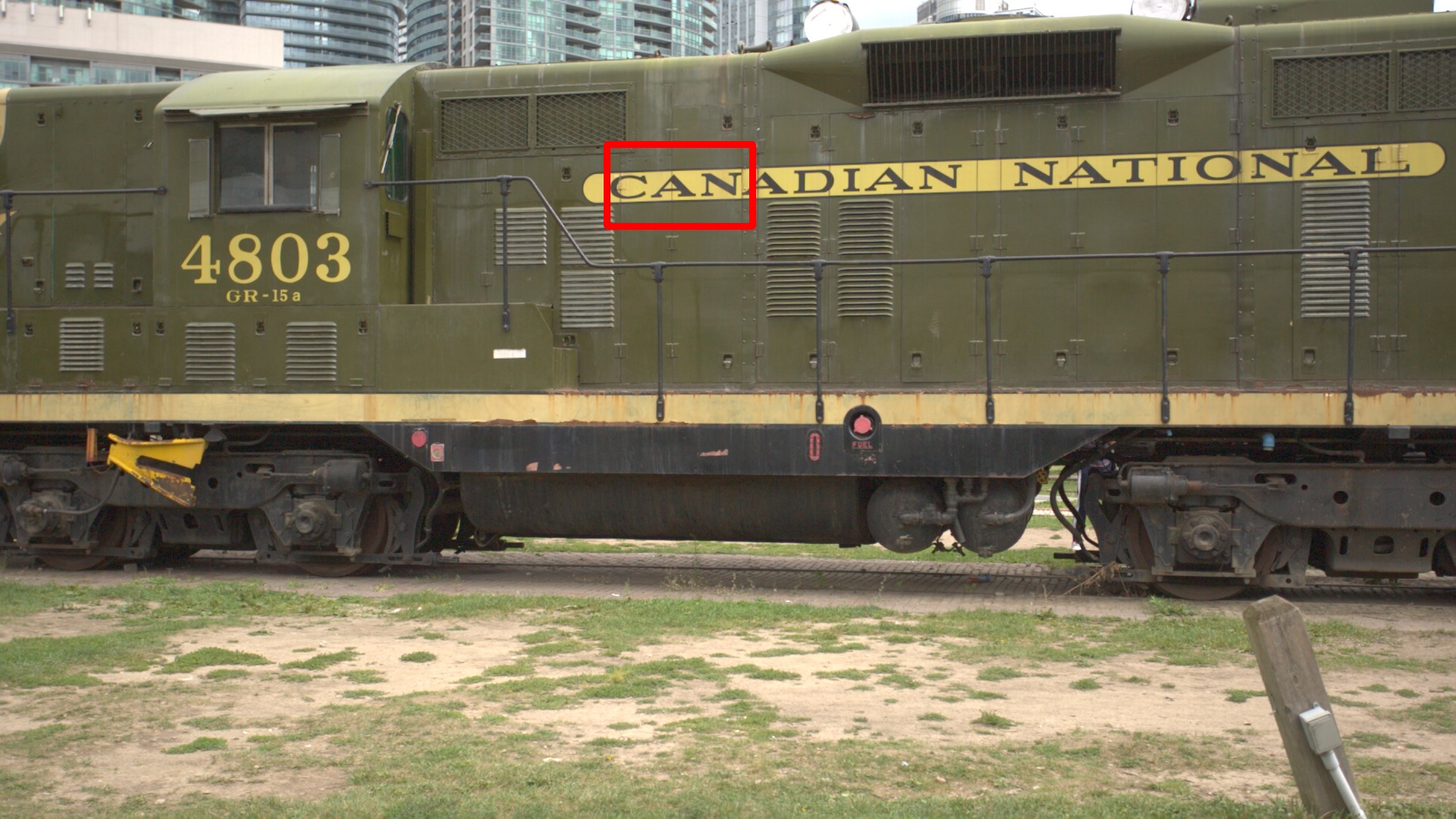}
		\scriptsize{(a) Train}
	\end{minipage}
	\begin{minipage}[h]{0.49\linewidth}
		\centering
		\includegraphics[width=\linewidth]{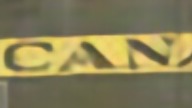}
		\scriptsize{(b) w/ camera processed data}
	\end{minipage}
	\begin{minipage}[h]{0.49\linewidth}
		\centering
		\includegraphics[width=\linewidth]{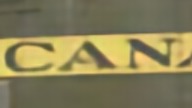}
		\scriptsize{(c) w/ raw data}
	\end{minipage}
	\begin{minipage}[h]{0.49\linewidth}
		\centering
		\includegraphics[width=\linewidth]{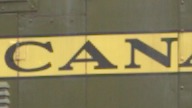}
		\scriptsize{(d) ground truth }
	\end{minipage}
	\caption{Illustrations of the proposed RawVSR based on camera processed data and raw data, respectively, for $4\times$ VSR using the  testing video ``Train'' in RawVD.}
	\label{fig:firstpage}
\end{figure}

The data-driven approach has become increasingly popular in VSR research. To the best of our knowledge, the existing data-driven VSR methods~\cite{sajjadi2018frame,jo2018deep,haris2019recurrent,wang2019edvr} exclusively  make use of camera processed data as inputs, despite the fact that modern cameras are capable of providing raw data which are potentially more informative.
Note that camera processed data are produced by the Image Signal Processor (ISP) from raw data through several operations, including image demosaicing, denoising, sharpening, color converting, tone adjustment,  compression, among others~\cite{jiang2017learning,schwartz2018deepisp,xu2019towards}. Overall, the operations in ISP are non-invertible and tend to degrade the information content of the original data. For example, the bit-depth is reduced from $12$-$14$ bits to $8$ bits~\cite{bychkovsky2011learning} due to quantization, and various artifacts are introduced by  compression (JPEG is the default image format for many cameras). Therefore, using camera processed data as inputs is inherently suboptimal. Moreover, such methods are inconsistent with the real imaging pipeline where VSR in fact serves as a pre-processing unit rather than a post-processing unit of ISP. 

The main contributions of this paper are summarized as follows: 
\begin{itemize}
	\item[1)] To avoid information loss and better fit the imaging pipeline, we propose a new Raw VSR method, named RawVSR, that can directly exploit camera sensor data. Furthermore, we carefully build the first Raw Video Dataset (RawVD) for training, validation, and testing.
	\item[2)] Assuming a Hidden Markov 
	Model (HMM), we show that the minimal sufficient statistic of LR frames with respect to the target HR frame can be computed via an iterative pairwise fusion process; 
	this result provides the architectural principle for the design of  RawVSR, especially its Successive Deep Inference  (SDI) module. 
	\item[3)] We design an Attention-based Residual Dense Block (ARDB) and leverage it to build the reconstruction module of RawVSR; this module is capable of simultaneously refining the fused feature and learning the color information needed to generate a spatial-specific  transformation for accurate color correction.  
\end{itemize}

Our extensive experimental results demonstrate that owing to the informativeness of the camera raw data, the  effectiveness of the neural network architecture, and the separation of super-resolution and color correction processes,  the proposed RawVSR achieves superior performance compared to the state-of-the-art and can be adapted to any specific camera-ISP.

Fig.~\ref{fig:firstpage} illustrates  the proposed RawVSR based on camera processed data and raw data, respectively, for $4\times$ VSR using the  testing video ``Train'' in RawVD. It can be seen that the latter yields clearer texture and sharper edges.

\section{Related Work}
To place our work in a proper context, we give a review of the existing VSR methods and discuss some relevant issues.

%We provide a short review of existing VSR methods and introduce some relevant works about raw image processing, in which we indicate some drawbacks and put this paper in proper context. 

%Recent years have seen the increasing popularity of deep-learning-based methods for  low-level vision tasks  \cite{chen2018learning,liu2019griddehazenet,shao2020domain,jiang2018super,xu2019quadratic}. Dong  \textit{et al.} \cite{dong2014learning} pioneered this approach to SR by building a convolutional neural network (CNN) to extract LR features for HR reconstruction via non-linear mapping. This has led to a paradigm shift in VSR research in particular, enabling  advances beyond the reach of traditional prior-based methods~\cite{liao2015video,xingjian2015convolutional,kappeler2016video,caballero2017real,tao2017detail,sajjadi2018frame,jo2018deep,haris2019recurrent,wang2019edvr, liu2020end}. Moving from SISR to VSR, it is essential to exploit temporal correlations existent across video frames. Existing works~\cite{liao2015video,xingjian2015convolutional,kappeler2016video,caballero2017real,tao2017detail,sajjadi2018frame,jo2018deep,haris2019recurrent,wang2019edvr, liu2020end} have demonstrated the superiority of deep-learning-based methods as compared to traditional prior-based ones.

%\noindent \textbf{Video Super-Resolution}
 VSR can be viewed as an underdetermined problem. Many traditional methods tackle this problem via regularized optimization, aided by certain image priors~\cite{farsiu2004fast,liu2014bayesian,ma2015handling,kohler2016robust,liux2017robust,liu2018robust}. The advent of deep learning has led to a paradigm shift in VSR and more generally low-level vision research  \cite{chen2018learning,liu2019griddehazenet,shao2020domain,xu2019quadratic}. Indeed, following the pioneering work by Dong  \textit{et al.} \cite{dong2014learning} on SISR, the recently proposed VSR methods are predominantly deep-learning-based
 	~\cite{liao2015video,xingjian2015convolutional,kappeler2016video,caballero2017real,tao2017detail,sajjadi2018frame,jo2018deep,haris2019recurrent,wang2019edvr, liu2020end}, achieving performances out of reach of the traditional prior-based approach. 	
 	Here we just describe a few representative ones. 	Kappeler \textit{et al.}~\cite{kappeler2016video} resort to the hand-crafted optical flow~\cite{drulea2011total} for motion compensation across input frames, and adopt a pre-trained CNN for SR operation. Caballero \textit{et al.}~\cite{caballero2017real} improve the quality of estimated optical flow using a spatial transformer module to facilitate the subsequent spatial-temporal network~\cite{shi2016real} for VSR. Tao \textit{et al.}~\cite{tao2017detail} design a Sub-Pixel Motion Compensation (SPMC) layer  to perform motion compensation and resolution upsampling simultaneously. A novel ConvLSTM is put forward in~\cite{xingjian2015convolutional}, which can be integrated into the autoencoder structure to effectively exploit temporal correlations among input frames. The 3D convolution finds its first application to motion estimation and compensation in 	
 	\cite{jo2018deep}, where the spatial and temporal correlations are explored coherently via 3D convolving operations. Inspired by back-projection~\cite{haris2018deep}, Haris \textit{et al.}~\cite{haris2019recurrent} develop a fusion strategy that treats each input frame as a stand-alone source carrying distinct information and progressively embeds 
 	 extracted correlation from each source-target pair into the back-projection network for motion compensation. In consideration of the fact that inter-frame displacements could spread over a wide range, Tian \textit{et al.}~\cite{tian2018tdan} substitute the conventional convolution with a deformable one that is capable of exploring spatial correlation  via learnable sampling points tailored to each pixel, not constrained by any prescribed kernel shape. Wang \textit{et al.}~\cite{wang2019edvr} propose a refined version of deformable convolution and construct a Pyramid, Cascading and Deformable (PCD) module for motion alignment; the resulting method is the winner of the NTIRE 2019 challenge on video deblurring and super-resolution~\cite{nah2019ntire}, and achieves the best VSR performance to date.

  Below we identify several issues with the existing VSR methods that will be addressed in the present work.
  
  \noindent \textbf{Raw Data Processing}
  The advantage of untouched camera raw data over processed data has been recognized in several areas of low-level vision.
  For instance, Chen \textit{et al.}~\cite{chen2018learning} make use of raw data to perform fast imaging in low-light and achieved favorable results in terms of texture detail. This success can be attributed to the primitive radiance information retained by raw data. Indeed, using camera processed data as a substitute of raw data fails to deliver comparable results due to  the information loss caused by quantization in ISP, which obscures subtle differences in pixel values that are essential for exhibiting delicate textures.
  %Ignatov \textit{et al.}~\cite{ignatov2020replacing} demonstrate that raw data captured by mobile devices can be directly converted to color images with visual quality approaching those photographed by a professional camera.
  Ignatov \textit{et al.}~\cite{ignatov2020replacing} demonstrate that color images suitably converted from smartphone's raw data are visually comparable to those produced by a professional camera.
  A comprehensive study of raw-data-based SISR is conducted by Xu \textit{et al.}~\cite{xu2019towards}, showing that the benefits of raw data are intrinsic, \textit{i.e.}, not specific to a particular SISR method~\cite{zhang2018residual}. However, to the best of our knowledge,the existing VSR methods are still exclusively based on camera processed data, and there is a lack of understanding of the potential gain offered by raw data (beyond that seen in the SISR setting).
    
    %in that context  (especially, )

    %and the effective way to exploit raw data
      
      %However, none of the above methods exploits raw data for VSR in consideration of both spatial and temporal correlations among video frames.
      
      %However, to the best of our knowledge, none of them takes camera raw data into consideration for VSR. The existing methods simply ignore the potential of favorable informativeness recorded by raw data that could be additionally beneficial to restoration process. In this paper, we aim to clearly illustrate the benefits of employing raw data for VSR rather than the current paradigm that utilizes camera-processed data.

  %Schwartz \textit{et al.}~\cite{schwartz2018deepisp} leveraged a deep CNN that is capable of simulating the camera ISP to process raw data. Ignatov \textit{et al.}~\cite{ignatov2020replacing} were dedicated to convert raw data captured by mobile device directly into the appealing color image that visually approaches the one photographed by a professional camera.

%studied  raw-data-based SISR, and 

%validated this kind of data can facilitate other SISR methods~\cite{zhang2018residual} by simply substituting the camera processed data to raw one.

\noindent \textbf{Motion Estimation} Due to the temporal dependencies among video frames,
VSR can  benefit substantially from precise motion estimation, which can be performed either explicitly or implicitly. Explicit motion estimation usually relies on optical flows~\cite{kappeler2016video,caballero2017real,tao2017detail,sajjadi2018frame}. However, for video frames with large pixel displacement and object occlusion, it could be extremely challenging and arguably impossible to obtain precise optical flows. Moreover, as indicated by~\cite{xue17toflow}, even completely accurate optical flows might not be adequate for motion estimation since they do not fully capture all possible motion effects.  Therefore, many recent VSR methods such as  DUF~\cite{jo2018deep}, TDAN~\cite{tian2018tdan}, and EDVR~\cite{wang2019edvr} choose to estimate motion implicitly, resulting in better final reconstructions compared to the flow-based ones. In a certain sense, implicit motion estimation provides a new mechanism for multi-frame fusion. Nevertheless, many aspects of the fusion process remain unspecified, and a theoretical guideline is much needed. %Our work makes some progress in this respect.

\noindent \textbf{Color Correction}
Different from camera processed images, raw images only record the primitive radiance information, which has not been projected to any color space. 
In \cite{chen2018learning}, an end-to-end mapping is learned that transforms raw images to colored images directly. However, this mapping does not have the desired flexibilities due to its camera-specific nature. 
\cite{schwartz2018deepisp} addresses this issue by generating a global color transformation that
can cope with a variety of cameras. But it has been observed that such a transformation may cause significant local color distortions. This problem is solved by \cite{xu2019towards}
through the introduction of a spatial-specific color transformation. Note that \cite{xu2019towards} makes use of
a stand-alone deep convolutional network  for this purpose, which increases the overall model size and computational load. In contrast, we show that spatial-specific color transformation can actually be accomplished together with HR image reconstruction by a single network. Our new design leads to more efficient implementation and facilitates joint learning for both tasks.

\section{Raw Video Dataset}

\begin{figure}[t]
	\begin{minipage}[t]{0.188\linewidth}
		\centering
		\includegraphics[width=\linewidth]{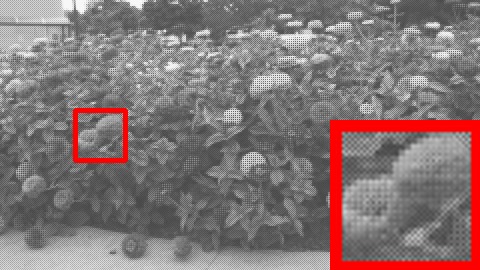}
		%\caption{fig2}
	\end{minipage}
	\begin{minipage}[t]{0.188\linewidth}
		\centering
		\includegraphics[width=\linewidth]{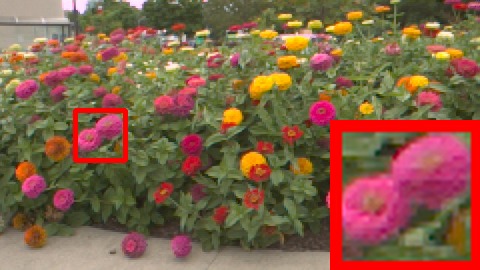}
		%\caption{fig1}
	\end{minipage}
	\begin{minipage}[t]{0.188\linewidth}
		\centering
		\includegraphics[width=\linewidth]{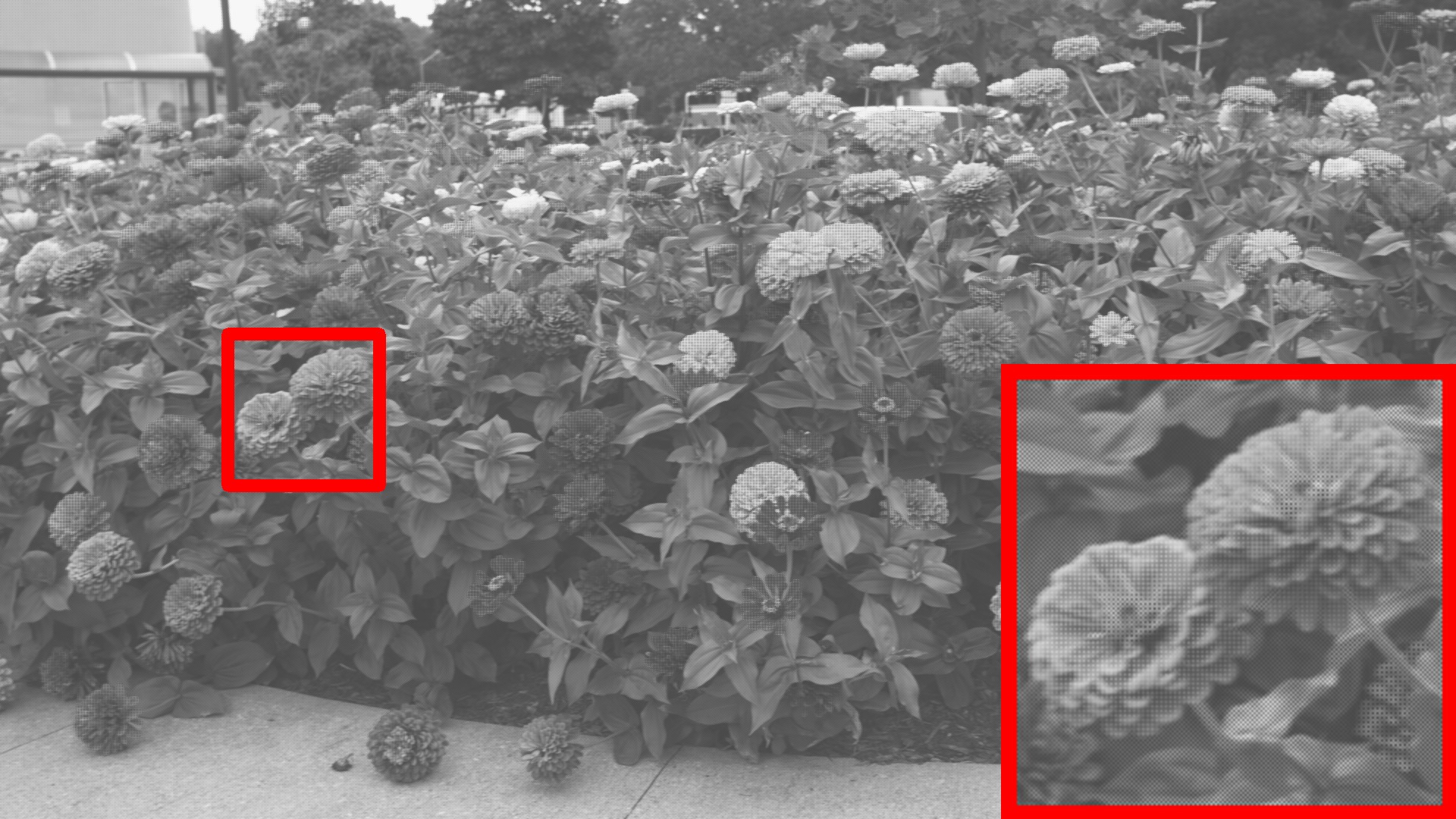}
		%\caption{fig2}
	\end{minipage}
	\begin{minipage}[t]{0.188\linewidth}
		\centering
		\includegraphics[width=\linewidth]{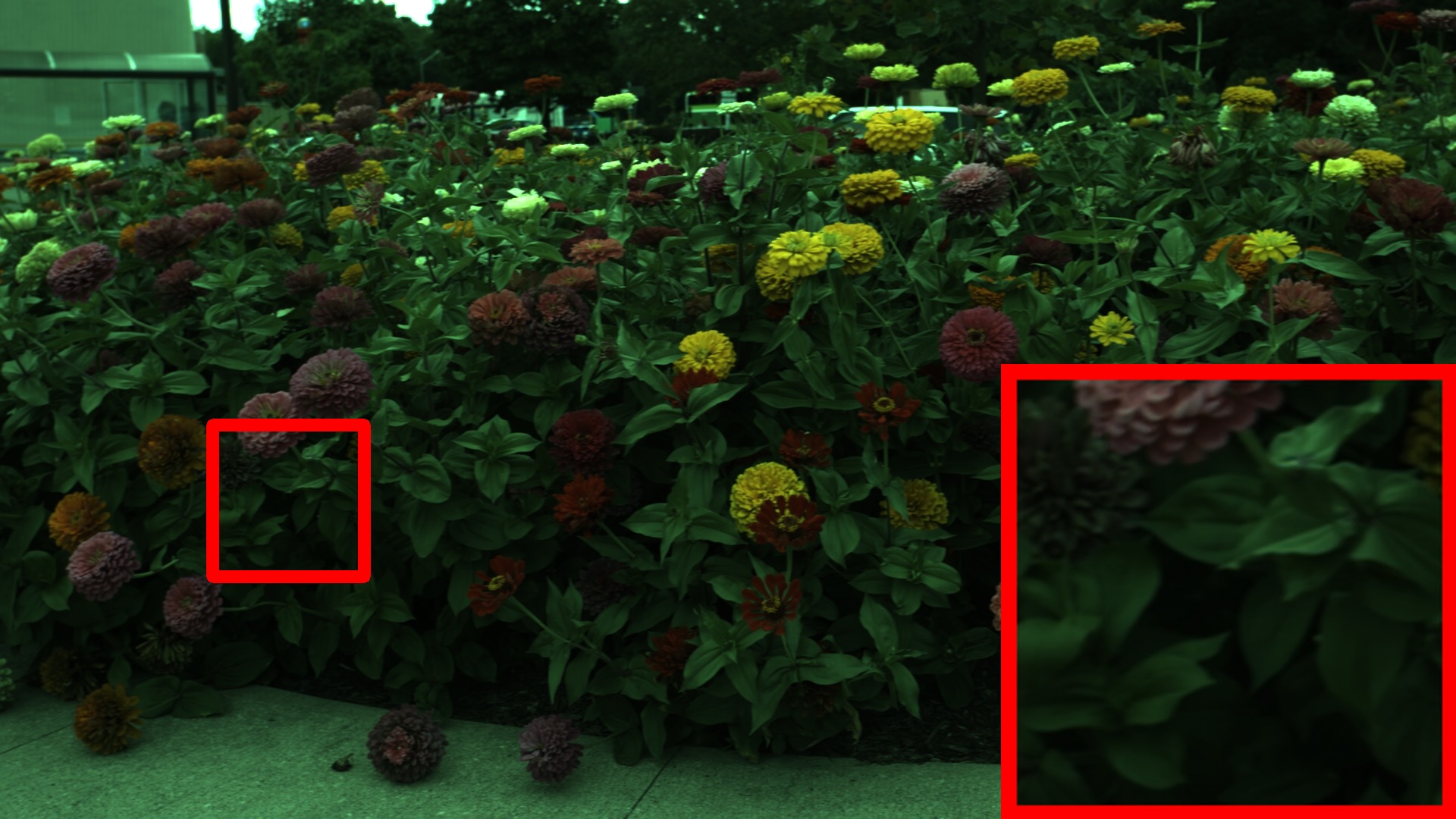}
		%\caption{fig2}
	\end{minipage}
	\begin{minipage}[t]{0.188\linewidth}
		\centering
		\includegraphics[width=\linewidth]{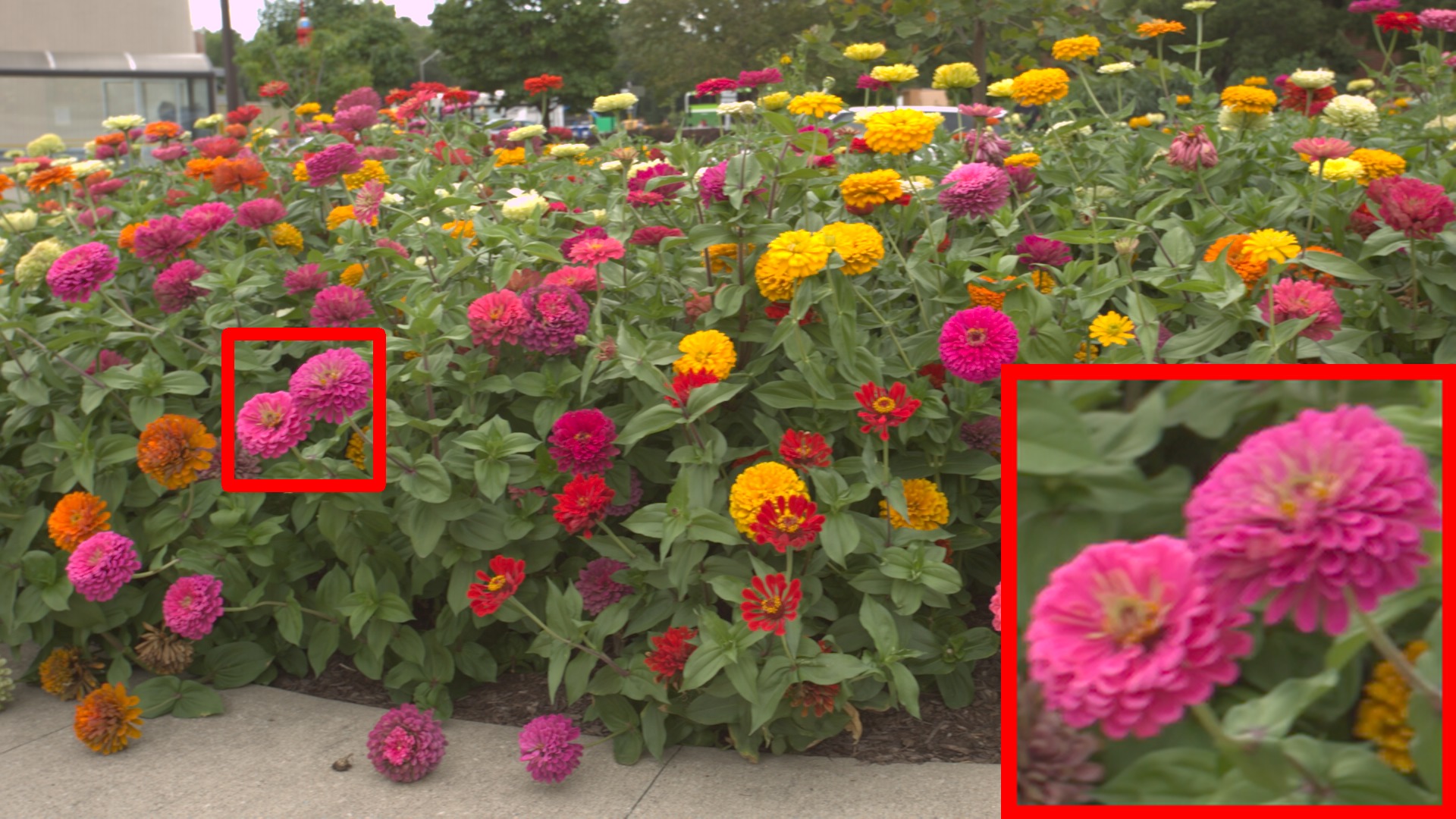}
		%\caption{fig2}
	\end{minipage}
	\begin{minipage}[t]{0.188\linewidth}
		\centering
		\includegraphics[width=\linewidth]{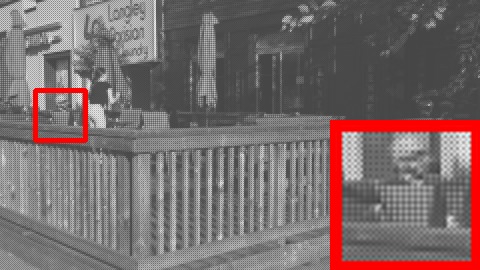}
		%\caption{fig2}
	\end{minipage}
	\begin{minipage}[t]{0.188\linewidth}
		\centering
		\includegraphics[width=\linewidth]{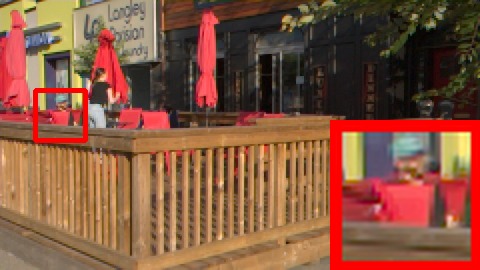}
		%\caption{fig1}
	\end{minipage}
	\begin{minipage}[t]{0.188\linewidth}
		\centering
		\includegraphics[width=\linewidth]{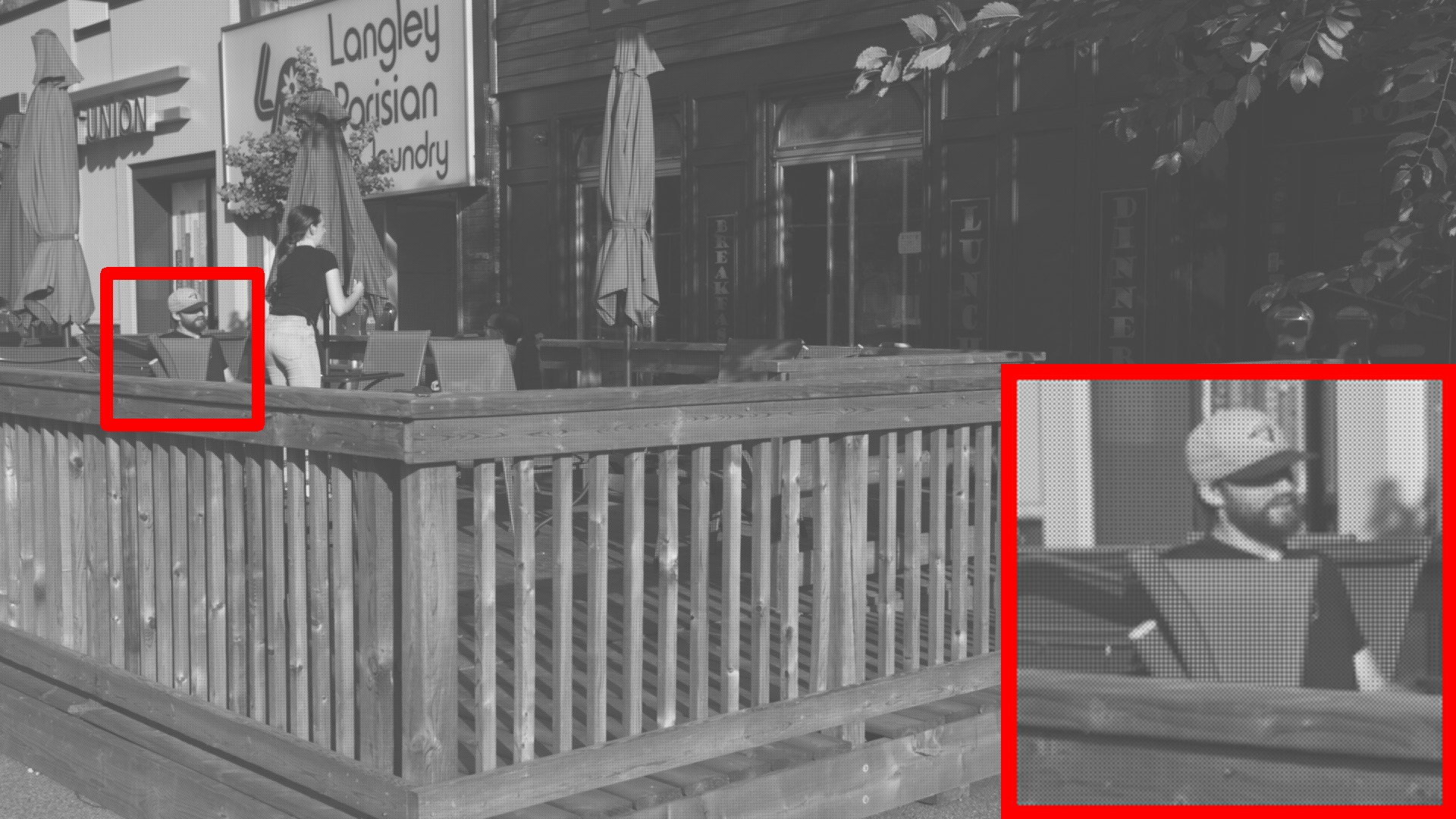}
		%\caption{fig2}
	\end{minipage}
	\begin{minipage}[t]{0.188\linewidth}
		\centering
		\includegraphics[width=\linewidth]{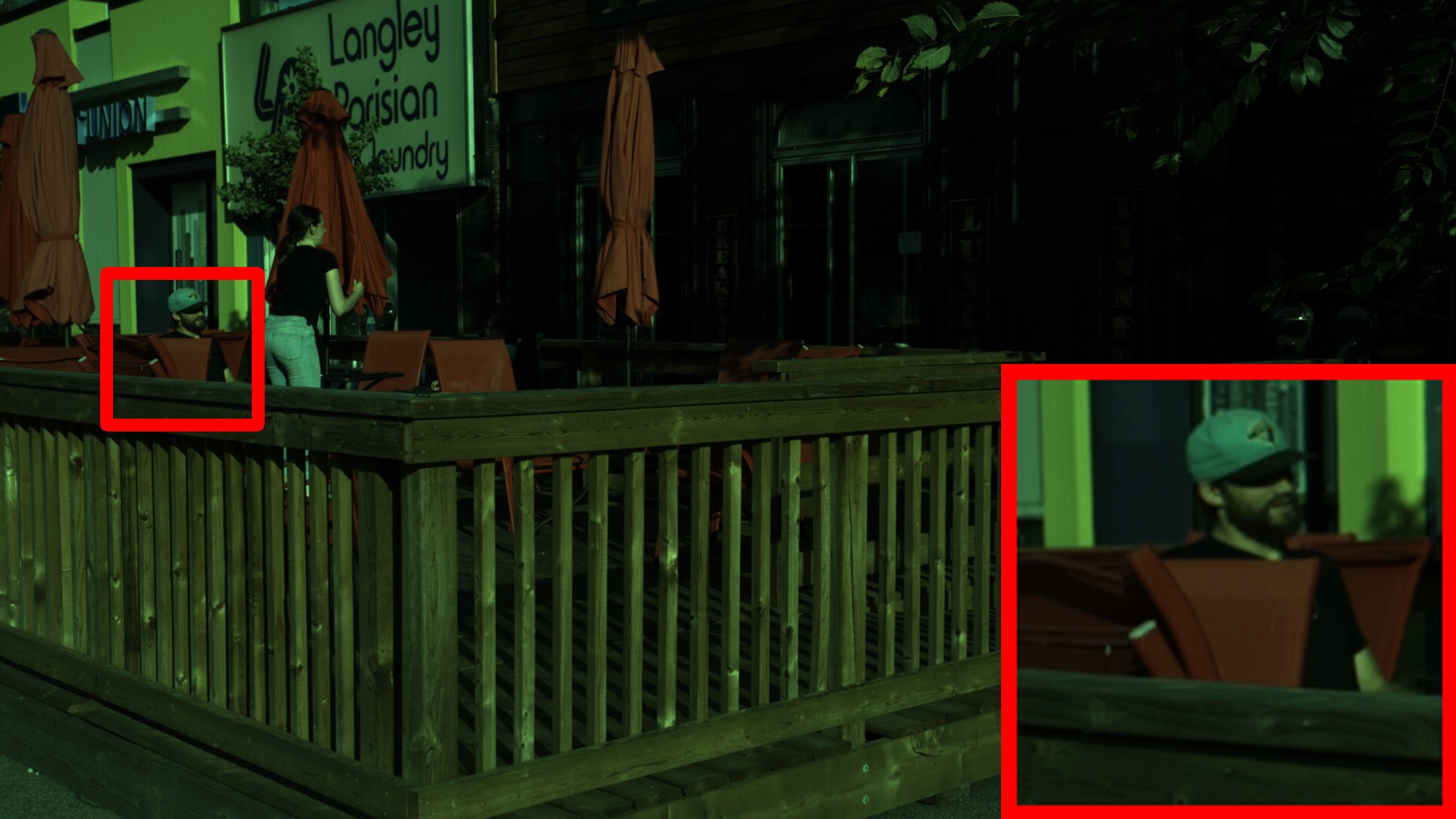}
		%\caption{fig2}
	\end{minipage}
	\begin{minipage}[t]{0.188\linewidth}
		\centering
		\includegraphics[width=\linewidth]{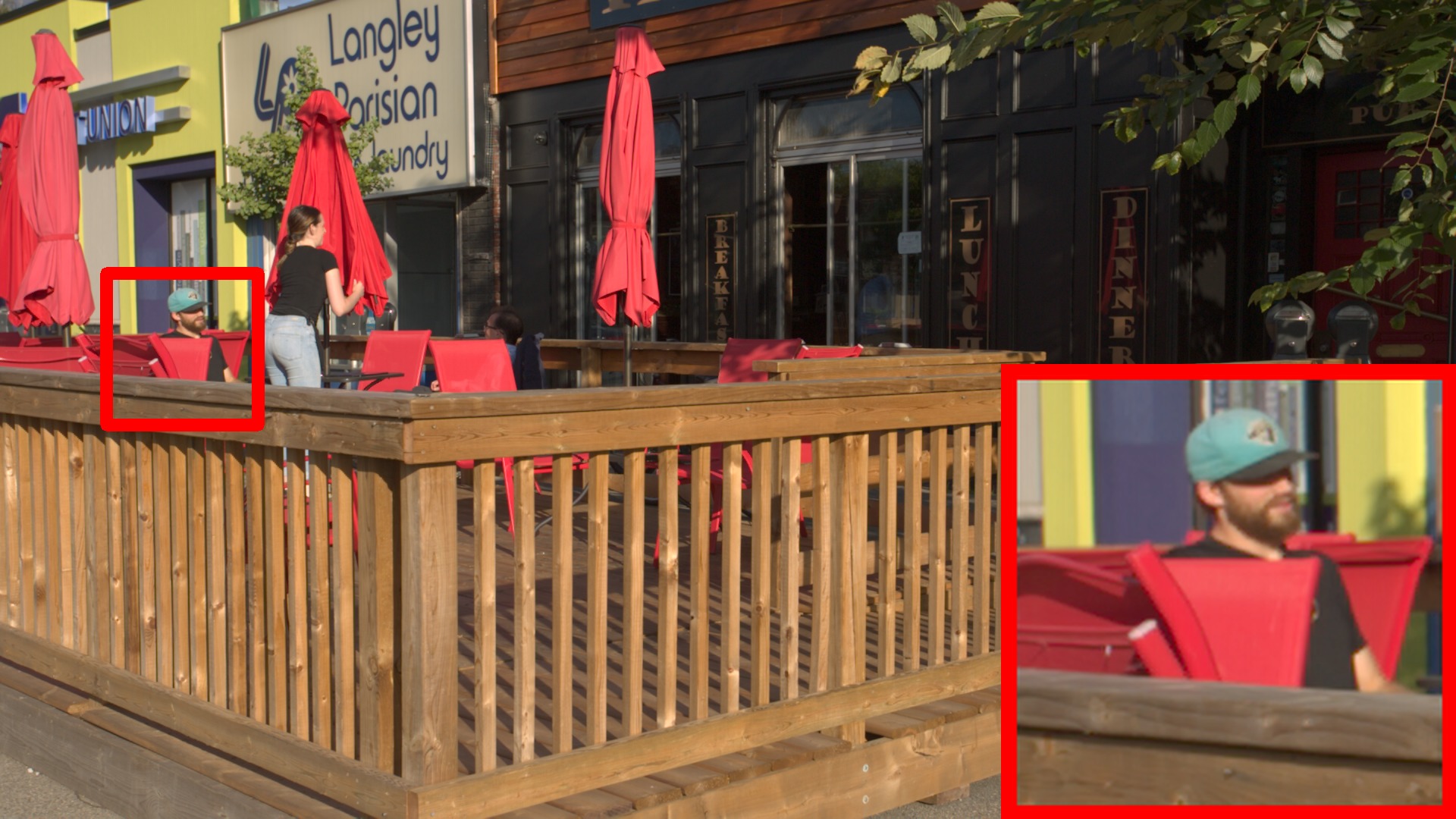}
		%\caption{fig2}
	\end{minipage}
	\begin{minipage}[t]{0.188\linewidth}
		\centering
		\includegraphics[width=\linewidth]{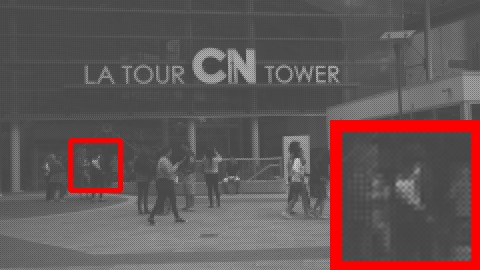}
		%\caption{fig2}
	\end{minipage}
	\begin{minipage}[t]{0.188\linewidth}
		\centering
		\includegraphics[width=\linewidth]{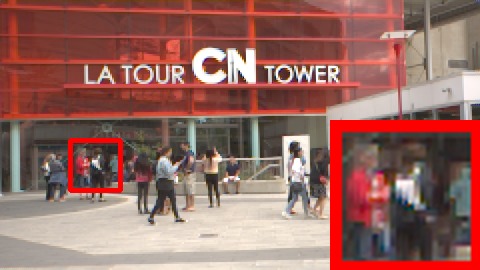}
		%\caption{fig1}
	\end{minipage}
	\begin{minipage}[t]{0.188\linewidth}
		\centering
		\includegraphics[width=\linewidth]{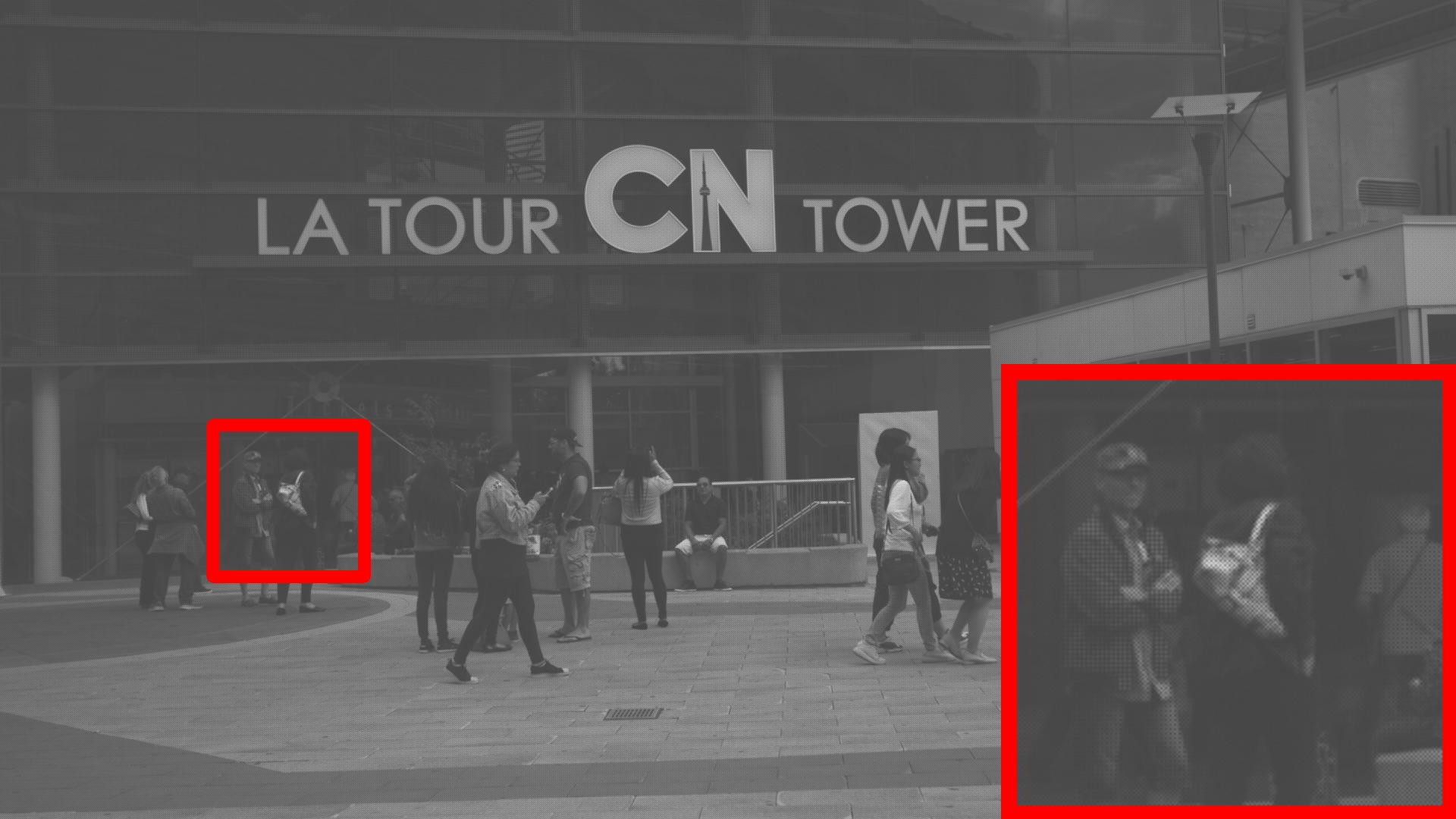}
		%\caption{fig2}
	\end{minipage}
	\begin{minipage}[t]{0.188\linewidth}
		\centering
		\includegraphics[width=\linewidth]{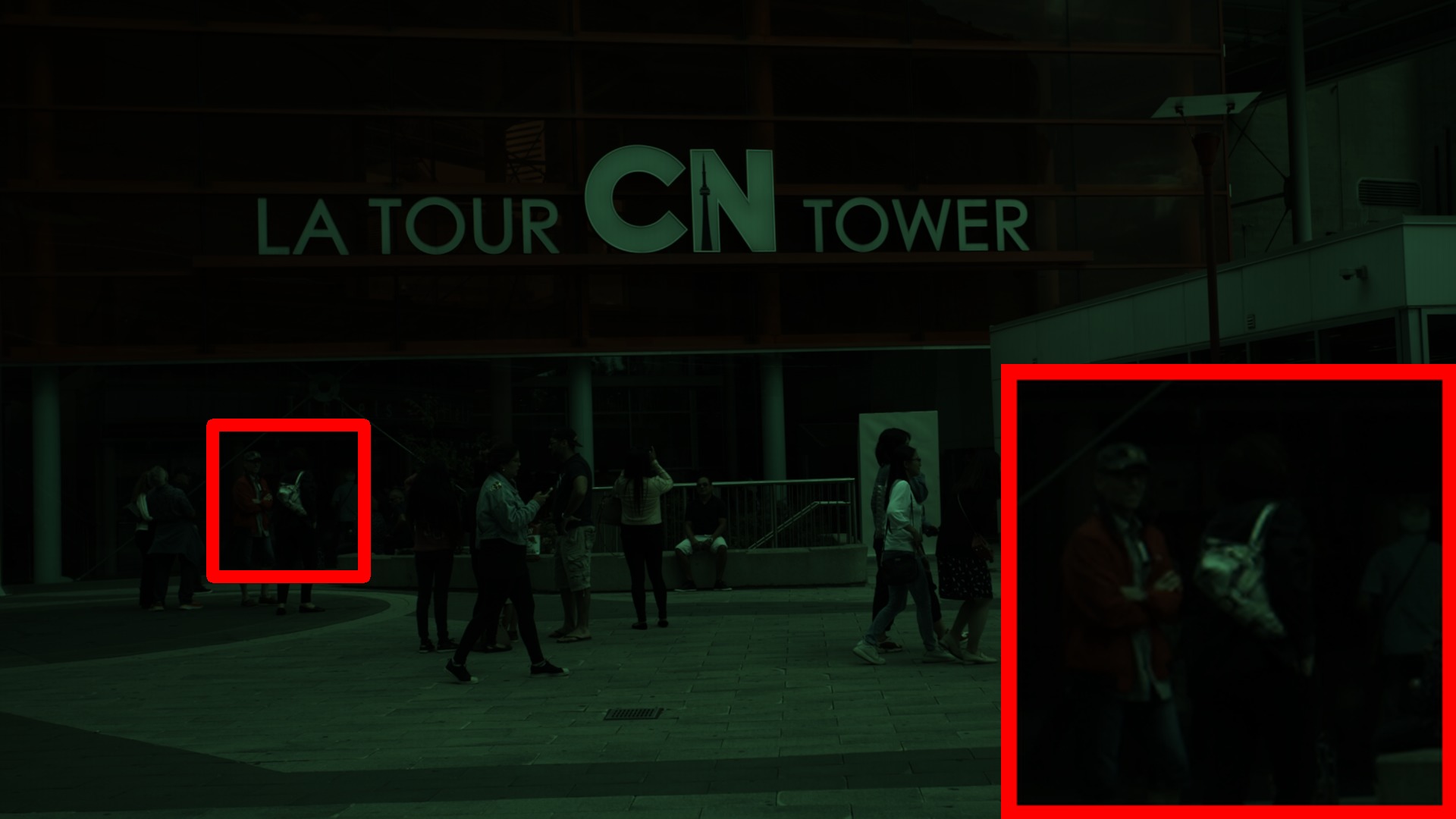}
		%\caption{fig2}
	\end{minipage}
	\begin{minipage}[t]{0.188\linewidth}
		\centering
		\includegraphics[width=\linewidth]{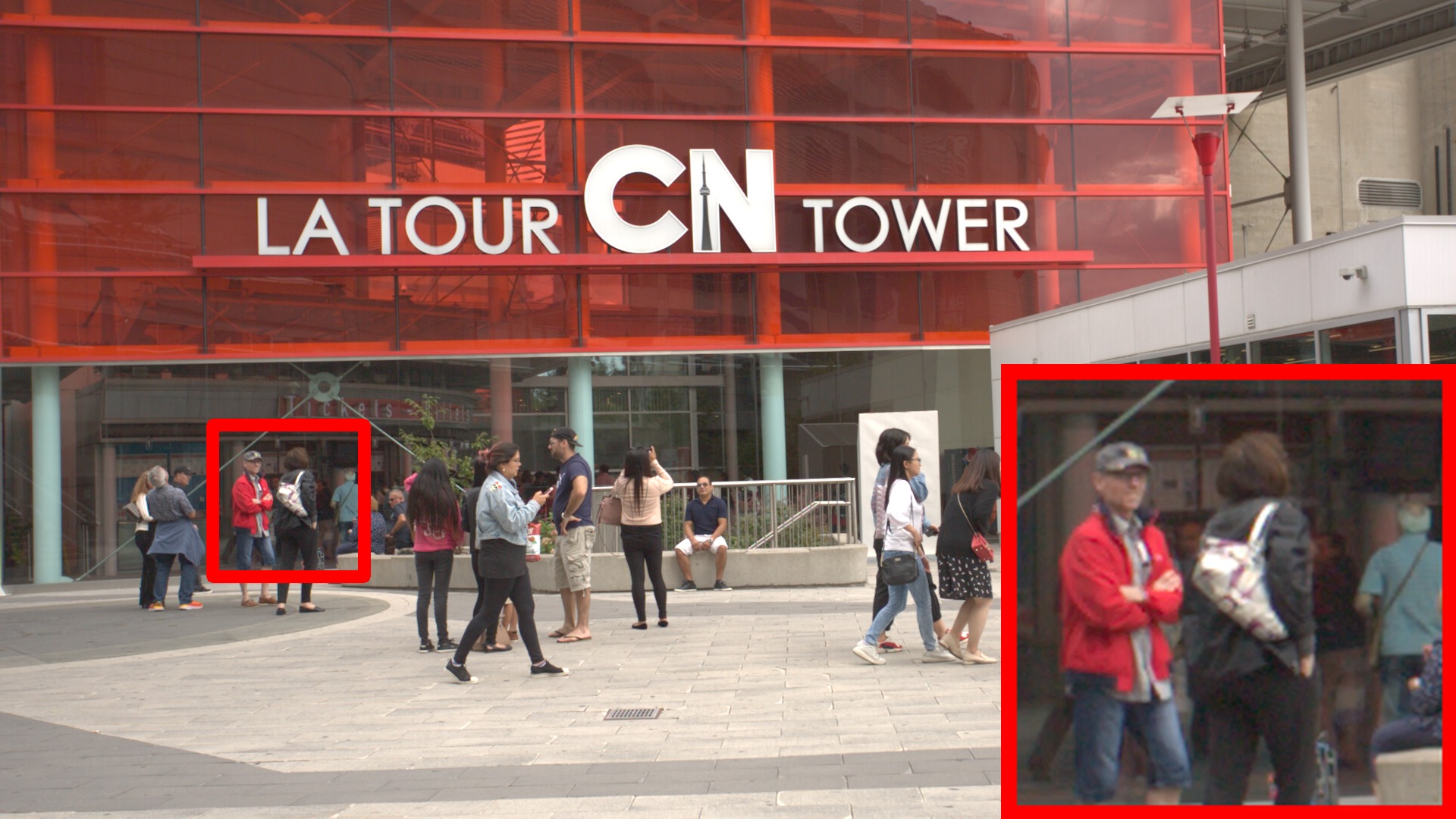}
		%\caption{fig2}
	\end{minipage}
	\begin{minipage}[t]{0.188\linewidth}
		\centering
		\includegraphics[width=\linewidth]{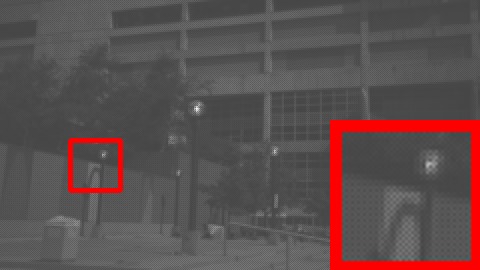}
		\scriptsize{(a) $Y_{raw}$}
		%\caption{fig2}
	\end{minipage}
	\begin{minipage}[t]{0.188\linewidth}
		\centering
		\includegraphics[width=\linewidth]{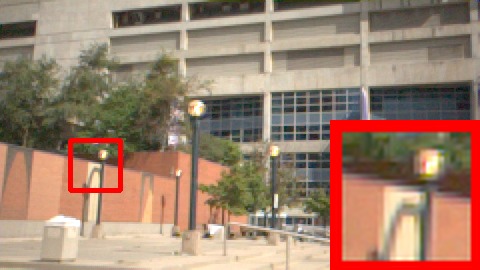}
		\scriptsize{(b) $Y_{rgb}$}
		%\caption{fig1}
	\end{minipage}
	\begin{minipage}[t]{0.188\linewidth}
		\centering
		\includegraphics[width=\linewidth]{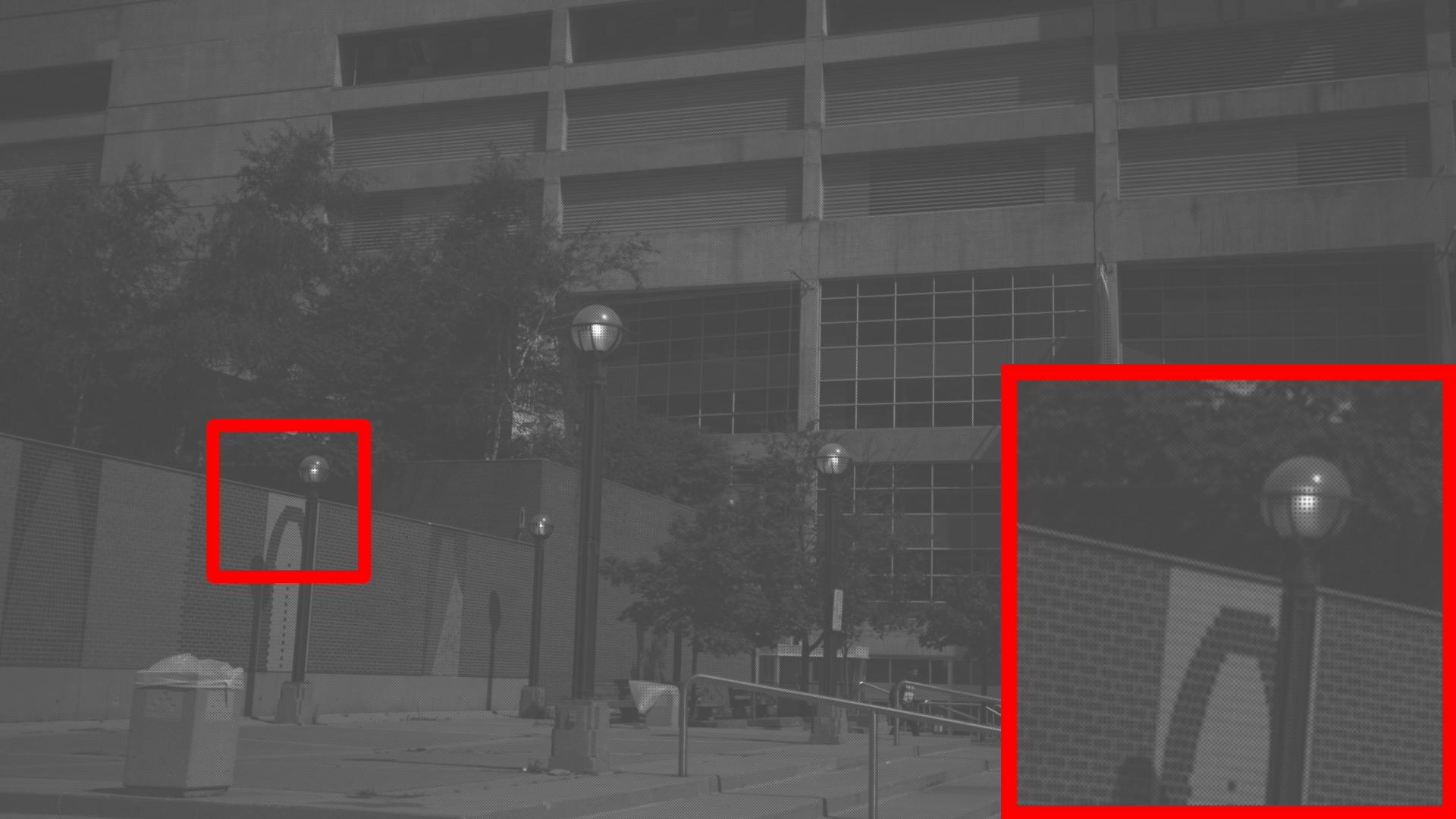}
		\scriptsize{(c) $X_{raw}$}
		%\caption{fig2}
	\end{minipage}
	\begin{minipage}[t]{0.188\linewidth}
		\centering
		\includegraphics[width=\linewidth]{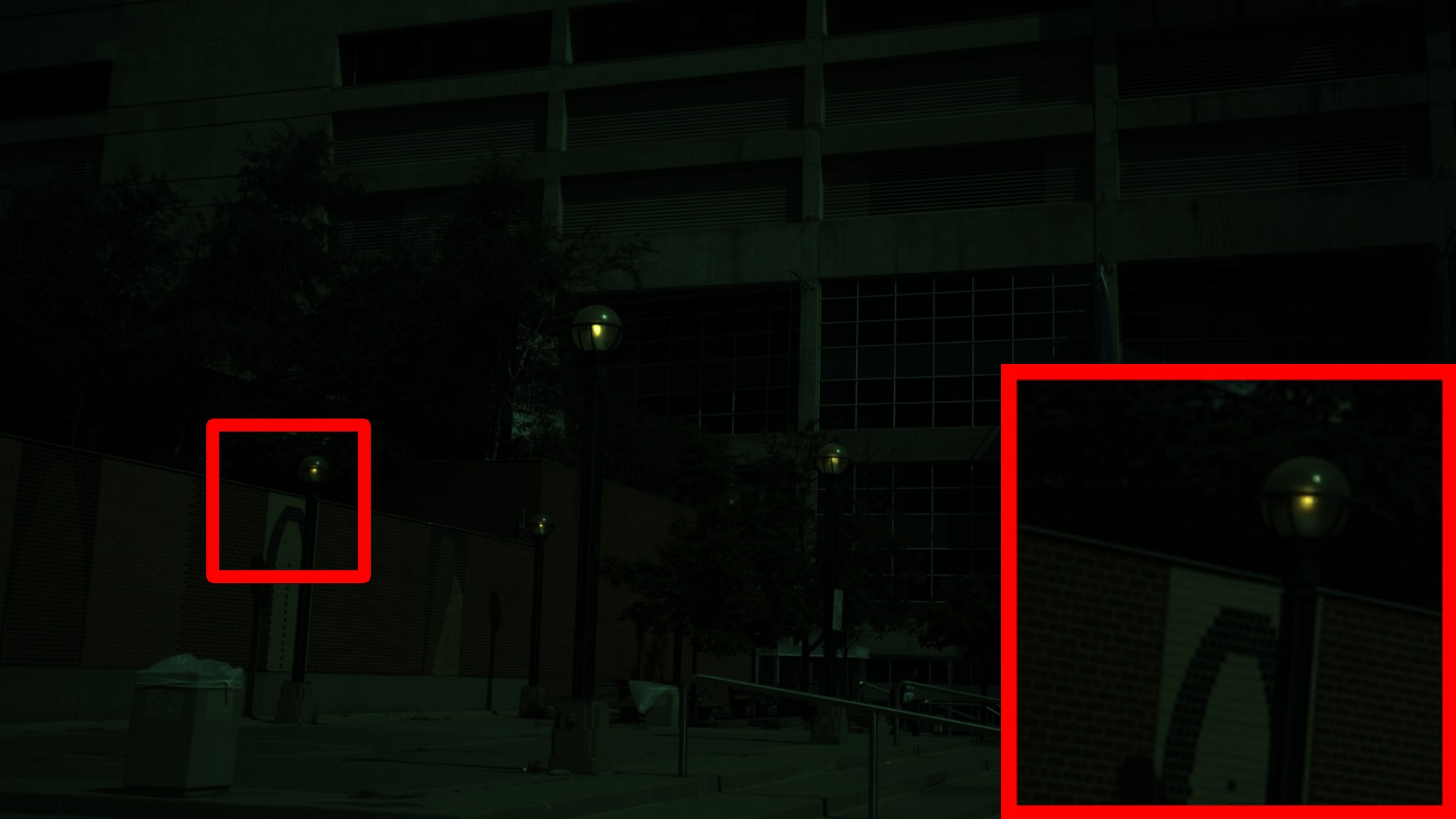}
		\scriptsize{(d) $X_{lin}$}
		%\caption{fig2}
	\end{minipage}
	\begin{minipage}[t]{0.188\linewidth}
		\centering
		\includegraphics[width=\linewidth]{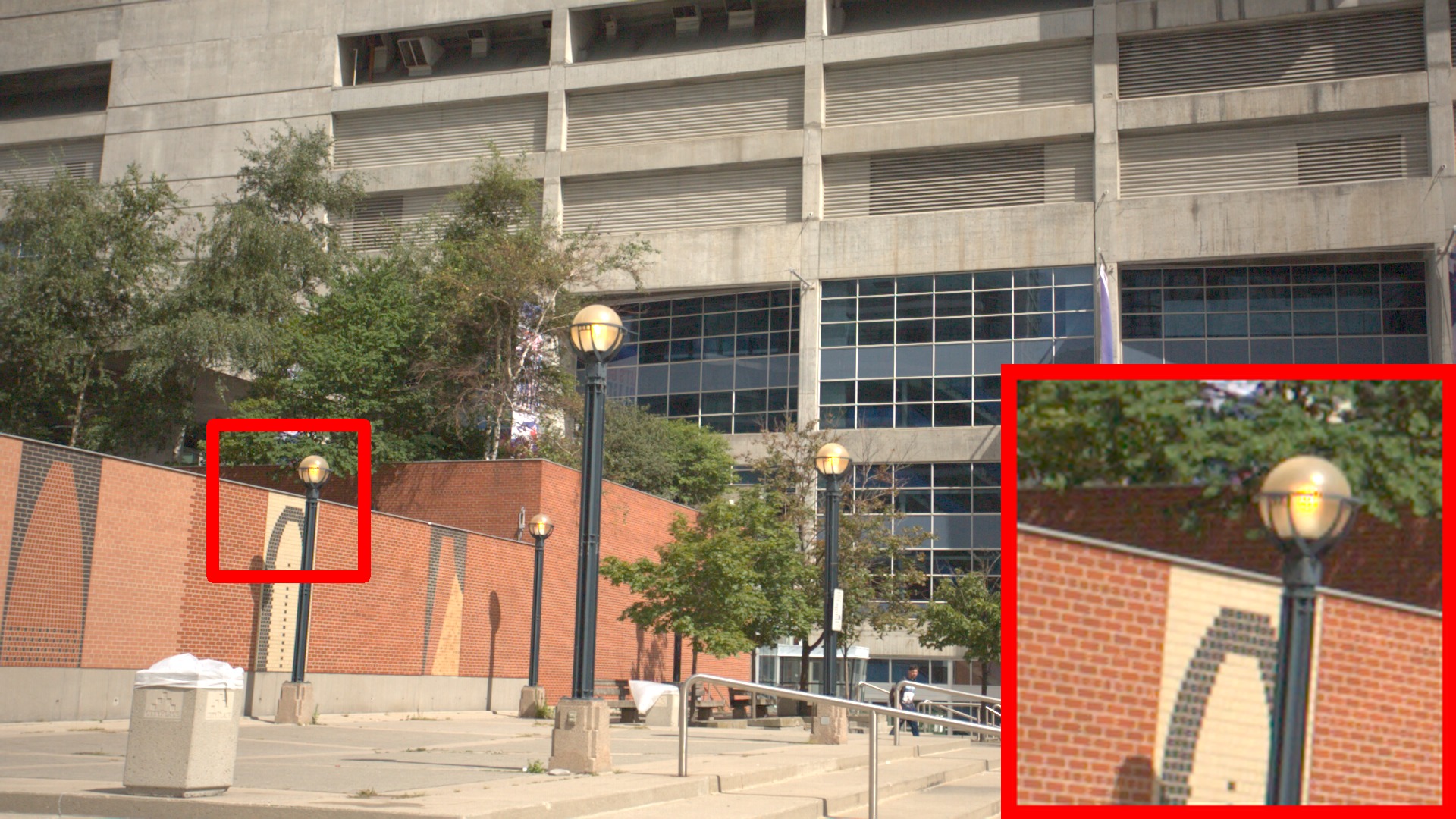}
		\scriptsize{(e) $X_{rgb}$}
		%\caption{fig2}
	\end{minipage}
	\centering
	\caption{Examples of videos in RawVD with the brightness and contrast of raw frames adjusted for better visualization.}
	\label{fig:RawVD}
\end{figure}

The widely-used VSR training datasets such as Vimeo-90K~\cite{xue17toflow} and REDS~\cite{nah2019ntire} are constructed with post-ISP videos (\textit{e.g.}, H.264 videos) and, as a  consequence, are inherently unsuitable for raw-data-based VSR.
We are not aware of the existence of publicly available raw video datasets for VSR. An important contribution of this work is a new raw video dataset (RawVD) for training and benchmarking. In fact, it will be seen that RawVD consists of both HR/LR raw video pairs and their processed counterparts, thus is suitable  for essentially all VSR methods, regardless whether they are raw-data-based or not. To build this dataset, we use a Canon 5D3 camera with the third-party upgrade\footnote {https://magiclantern.fm/} to film videos in Magic Lantern Video (MLV) format (which is a raw video format), and utilize the MLV App\footnote{https://mlv.app/} to obtain corresponding raw frames in DNG format (which is a raw image format). Note that different cameras might have different Bayer patterns. For Canon 5D3, the Bayer pattern is RGGB, which means $50\%$ sensors for green color, $25\%$ for red color, and $25\%$ for blue color. 
A degraded LR raw frame $Y_{raw}\in\mathbb{R}^{H/S\times W/S\times 1}$ is generated for each filmed HR raw frame $X_{raw}\in\mathbb{R}^{H\times W\times 1}$, where  $H$ and $W$ stand respectively for frame height and width, and $S$ represents the VSR scale ratio.  Specifically, inspired by~\cite{chen2018learning,xu2019towards}, we first use AHD~\cite{hirakawa2005adaptive} to produce a  demosaiced frame $X_{lin}\in\mathbb{R}^{H\times W\times 3}$ based on $X_{raw}$ without any post-processing such as white balance and gamma correction (where the subscript of $X_{lin}$ reflects the fact that each of its pixel value is a linear measurement of some radiance information contained in $X_{raw}$), then generate $Y_{raw}$ from $X_{lin}$ through a sequence of degradation operations involving, among others, blurring, downsampling, and noising. More precisely, we have
\begin{equation}
Y_{raw} = f_{Bayer}(f_{Down}(X_{lin}\ast K_{Blur})) + n.
\end{equation}
Here $K_{Blur}$ is a uniformly distributed defocus blur kernel, which simulates the out-of-focus effect in camera; $\ast$ denotes the convolutional operation;  $n$ is a heteroscedastic Gaussian noise~\cite{plotz2017benchmarking,xu2019towards} with its variance specified by two parameters $\sigma^2_1$ and $\sigma^2_2$; 
$f_{Down}$ and $f_{Bayer}$ stand respectively for downsampling and mosaicing operations (in particular, $f_{Bayer}$ converts a three-channel frame back to a one-channel frame that obeys the RGGB pattern). Finally, given $X_{raw}$ and $Y_{raw}$, we use Rawpy, a Python version of LibRaw (with parameters adjusted to closely approximate Canon-ISP), to generate their corresponding color frames $X_{rgb}\in\mathbb{R}^{H\times W\times 3}$ and $Y_{rgb}\in\mathbb{R}^{H/S\times W/S\times 3}$. An illustration of $Y_{raw}$, $Y_{rgb}$, $X_{raw}$, $X_{lin}$, and $X_{rgb}$ can be found in Fig.~\ref{fig:RawVD}; note that  the color of $X_{lin}$ is biased towards green
because the dominant measurement recorded in raw data is from green sensors. In total,  $110$ videos (with a resolution of  $1920 \times 1080$ and at least $100$ frames each) are filmed. These videos capture a wide variety of scenes.
We divide these $110$ videos into three groups: $100$ videos for training, $5$ for validation, and the rest $5$ for testing. The $5$ test videos are named ``Store'', ``Painting'', ``Train'', ``City'', and ``Walk'' according to their respective content.

As mentioned above, each video has four different versions (in addition to the linear measurement version ($X_{lin}$)): the original HR raw version ($X_{raw}$), the LR raw version ($Y_{raw}$), the HR color version ($X_{rgb}$), and the LR color version ($Y_{rgb}$). The LR raw version and the LR color version serve as the inputs of raw-data-based VSR methods and processed-data-based VSR methods, respectively. We adopt the HR color version as the ground truth for both types of methods to facilitate direct comparisons of their outputs. The linear measure version is not directly used for supervised training in the present work; nevertheless, we choose to include it in our RawVD as it visually reveals the intimate connections among the other four versions and potentially provides valuable insights that can be fruitfully exploited in future research.

%\textcolor{magenta}{In our paper, even though the $X_{lin}$ is not adopted as a intermediate supervision in training, this version is still encompassed in our RawVD as it visually provides some favorable insights that disclose the intrinsic connections among the other four versions, which can be potentially used by follow-up works.}

\begin{figure*}[t]
	\centering
	\includegraphics[width=\linewidth]{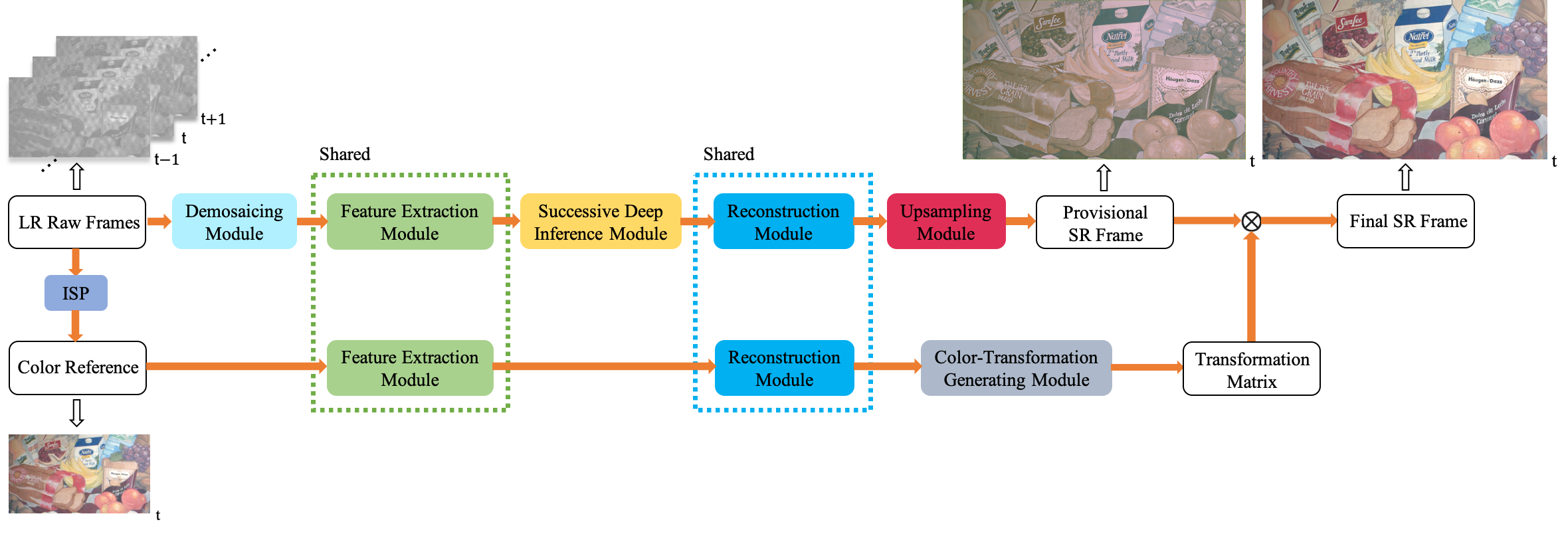}
	\caption{The architecture of the proposed RawVSR.}
	\label{fig:RawVSR}
\end{figure*}

Remark: We have also considered collecting real-world raw LR-HR pairs through optical zooming \cite{cai2019toward}. However, this approach has several limitations. First, each LR-HR pair should have exactly the same motion (otherwise the pixel-level alignment is difficult to accomplish), which severely limits the varieties of suitable videos that can be collected in practice. 
Second, since raw data only record primitive radiance information, some back-and-forth conversions between raw data and processed data  (\textit{e.g.}, sRGB) are typically needed for raw LR-HR pair alignment, causing information loss.
As such, this approach is tailored to SISR (without inter-frame motion) with sRGB data. In contrast, our approach does not suffer from these problems and is better suited to raw VSR. It is also worth noting that even though AHD is not the state-of-the-art for demosaicing, it performs very stable and is widely adopted in practice \cite{xu2019towards}. Indeed, upon close scrutiny,  no ground-truth video in the constructed RawVD suffers evident demosaicing artifacts.  

\section{Method}
Now we are in a position to present the proposed raw-data-based VSR method, \textit{i.e.}, RawVSR. A general overview is provided in Section \ref{overview}. Sections \ref{SDI} and \ref{RM} are devoted to describing two key components of RawVSR: the SDI module and the reconstruction module. We address some potential practical concerns regarding  raw-data-based VSR in Section \ref{usefuleness}. The implementation details can be found in the supplementary material.

\subsection{Overview} \label{overview}

To effectively exploit LR raw videos, one might be inclined to build a deep neural network to learn an end-to-end mapping that can generate HR color videos directly. However, in reality,  raw videos only record the radiance information read from camera sensors and the associated color videos are in fact ISP-dependent. That is to say, different ISPs may produce different color videos based on the same raw video. As such, an end-to-end mapping can only be tailored to a specific type of camera ISP and lacks the flexibilities needed for accommodating diverse VSR requirements. To tackle this problem, we use one certain color reference generated by an  ISP-adaptive operation to guide the  transformation from the raw space to the color space.

The proposed RawVSR consists of several components: the demosaicing module, the feature extraction module, the SDI module, the reconstruction module, the upsampling module, and the color-transformation generating module. It takes $7$ consecutive LR raw frames $Y^{t-3}_{raw},\cdots, Y^{t+3}_{raw}$  and outputs a SR color frame $X^t_{sr}\in\mathbb{R}^{H\times W\times 3}$. The goal is to make $X^t_{sr}$ as close to the HR color frame $X^t_{rgb}$ as possible. We use $Y^t_{rgb}$ as the color reference frame to ensure that the provided color information is consistent with that of $X^t_{rgb}$. 
%Note that the color reference frame should not be viewed as an additional input since it can be derived from the given LR raw frames (actually just the middle frame in the current case) through an  ISP-adaptive operation (which is Rawpy in the current case) and provides no new information regarding the target HR color frame beyond what is already available in the given LR raw frames except some ISP-specific information needed for color transformation and correction.
Note that the color reference frame is derived from the middle frame in the raw video sequence through an ISP-adaptive operation, and thus should not be viewed as an additional input. Indeed, except some ISP-specific parameters needed for color transformation and correction, this middle frame already contains all necessary information to produce the color reference frame.

As shown in Fig.~\ref{fig:RawVSR}, RawVSR has two branches: the texture restoration branch (upper branch) produces a provisional SR frame $\hat{X}^t_{sr}\in\mathbb{R}^{H\times W\times 3}$ while the color correction branch (lower branch) generates a spatial-specific color transformation $T$. The final SR frame $X^t_{sr}$ is obtained from $\hat{X}^t_{sr}$ through pixel-wise color correction specified by $T$. The orange arrow shows the image/feature flow direction, and the black arrow points to the illustration of input/color reference/provisional SR/final SR frame respectively using the ``Painting'' video in RawVD.

In the texture restoration branch, the input LR raw frames are first processed by the demosaicing module, where each LR raw frame is converted to a $64$-channel frame (with the spatial size unchanged) via packing, convolution, and pixel shuffling.
We utilize the U-net~\cite{ronneberger2015u} for  frame-by-frame multi-scale feature extraction.
The outputs of the feature extraction module induced by the given seven LR raw frames
are  jointly fed into the SDI module to perform feature alignment and fusion. The fused feature is then refined by the reconstruction module. The upsampling module consists of three convolutional layers and one pixel-shuffle operator. It 
enlarges the spatial size of the refined feature to generate a provisional SR frame. It can be seen that although the provisional SR frame suffers severe color distortion, the texture  details are faithfully restored.

\begin{figure*}[htbp]
	\begin{minipage}[t]{0.55\linewidth}
		\centering
		\includegraphics[width=\linewidth]{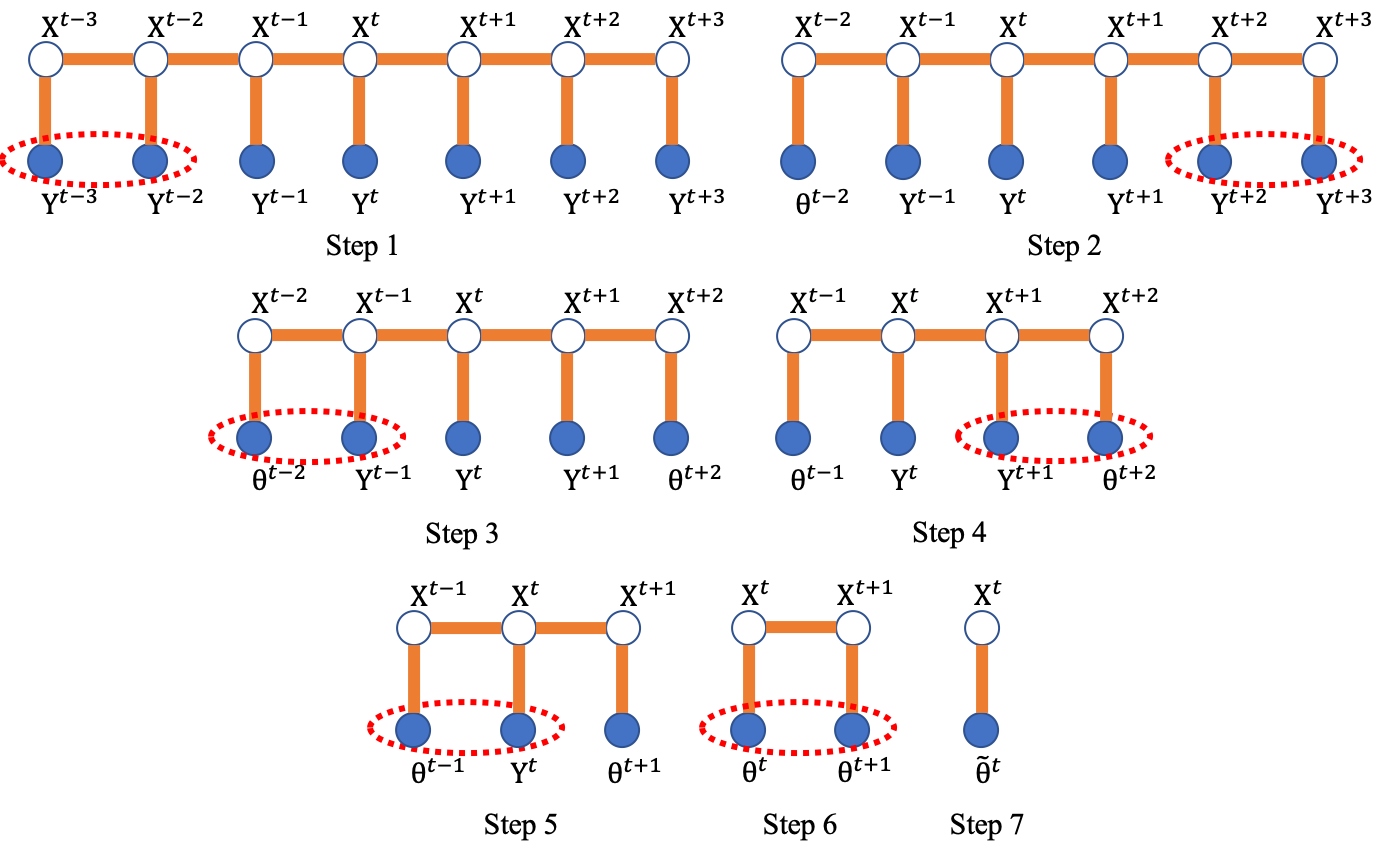}
		\caption{A graphical illustration of HMM inference via iterative pairwise fusion.}
		\label{fig:SDITheory}
	\end{minipage}
	\begin{minipage}[t]{0.44\linewidth}
		\centering
		\includegraphics[width=\linewidth]{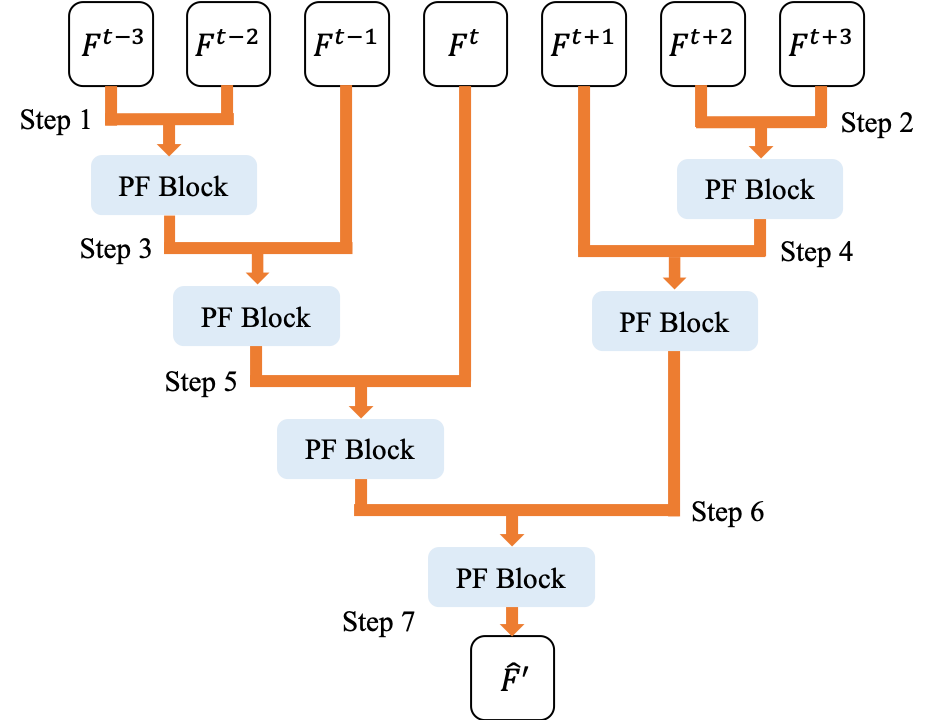}
		\caption{The architecture of the SDI module.}
		\label{fig:SDI}
	\end{minipage}
	\centering
	
\end{figure*}

In the color correction branch, both the feature extraction module and the reconstruction module are shared from the texture restoration branch. To ensure dimension compatibility, the color reference frame is pre-processed by a single convolutional layer before fed into the feature extraction module. Different from its main role in the texture restoration branch (which is feature refinement), the reconstruction module is used in the color correction branch to learn the ISP information needed for generating a spatial-specific color transformation. The structure of the color-transformation generating module is the same as that of the upsampling module except for the last convolutional layer, which outputs a $9$-channel feature map  rather than a $3$-channel one. The output feature map is reshaped from $H\times W\times 9$ to $H\times W\times 3 \times 3$, and the reshaped map is denoted by $T$. Note that $T$ can be viewed as a collection of $3\times 3$ matrices $\{T(i,j)\}$ with each $(i,j)$ pair specifying a pixel position. These matrices are leveraged to perform a  spatial-specific color transformation on the provisional SR frame $\hat{X}^t_{sr}$ to produce the final SR frame $X^t_{sr}$ as follows: 
\begin{equation} \label{color_matrix}
X^t_{sr}(i,j) = T(i,j) \otimes \hat{X}^t_{sr}(i,j),
\end{equation}
where  $X^t_{sr}(i,j)\in\mathbb{R}^{3\times 1}$ ($\hat{X}^t_{sr}(i,j)\in\mathbb{R}^{3\times 1}$) is the RGB vector of the $(i,j)$ pixel of $X^t_{sr}$ ($\hat{X}^t_{sr}$), and
$\otimes$ denotes matrix multiplication. 

It is worth noting that due to the sharing of the feature extraction module and the reconstruction module as well as the lightweight design of the color-transformation generating module, the number of independent parameters that come from the color correction branch is negligible (accounting for about $5\%$ of the model size of RawVSR). On the other hand, from the perspective of building an interpretable device-independent network, it is preferable to have a dual-branch architecture that disentangles the super-resolution process from the color correction process, and one may argue that weight sharing does not suit this purpose. However, since the generated transformation matrix needs to be properly aligned with the provisional SR frame to perform pixel-wise color correction, it is essential to have a mechanism to facilitate the coordination of two processes. In this regard,  weight sharing contributes to this coordination by helping maintain the latent consistency of intermediate features in these two branches. Moreover, it will be seen from the experimental results in Section \ref{sec:experiment} that weight sharing actually does not jeopardize the interpretability and device independence of the resulting design.

\subsection{Successive Deep Inference Module} \label{SDI}
Although many existing VSR methods exploit temporal redundancy, the underlying alignment and fusion rules are often heuristic in nature. In contrast, we shall propose a systematic approach based on a canonical decomposition result for HMM inference.

First note that the VSR problem can be mathematically formulated as
\begin{align}
\tilde{X}^t\triangleq\min\limits_{f}\mathbb{E}[d(X^t-f(Y^{t-3},\cdots,Y^{t+3}))],
\end{align}
where $(Y^{t-3},\cdots,Y^{t+3})\triangleq(Y^{t-3}_{raw},\cdots,Y^{t+3}_{raw})$, $X^t\triangleq X^t_{rgb}$, $\tilde{X}^t\triangleq X^t_{sr}$, and
$d(\cdot,\cdot)$ is the loss function. However, $\tilde{X}^t$ in general depends on the choice of  $d(\cdot,\cdot)$. To gain insight into the fundamental architectural principle that is applicable under any loss function, it is instructive to consider the following alternative formulation with hard reconstruction $\tilde{X}^t$ replaced by soft reconstruction
\begin{align}
\tilde{\Theta}^t\triangleq p(X^t|Y^{t-3},\cdots,Y^{t+3}),\label{eq:alterform}
\end{align}
where $p(X^t|Y^{t-3},\cdots,Y^{t+3})$ denotes the conditional distribution of $X^t$ given $(Y^{t-3},\cdots,Y^{t+3})$. 
The SDI module is designed to solve (\ref{eq:alterform}), \textit{i.e.}, to compute $\tilde{\Theta}^t$ based on $(Y^{t-3},\cdots,Y^{t+3})$. 
Note that $\tilde{\Theta}^t$ is a minimal sufficient statistic of $(Y^{t-3},\cdots,Y^{t+3})$ with respect to $X^t$ in the sense that $(Y^{t-3},\cdots,Y^{t+3})$ and $X^t$ are conditionally independent given $\tilde{\Theta}^t$ and $\tilde{\Theta}^t$ is a function of any statistic of $(Y^{t-3},\cdots,Y^{t+3})$ with this conditional independence property.
As a consequence, it is possible to deduce $\tilde{X}^t$ from $\tilde{\Theta}^t$ once the loss function is specified; for example, $\tilde{X}^t$ is simply the mean of $\tilde{\Theta}^t$ (\textit{i.e.}, $\tilde{X}^t=\mathbb{E}[X^t|Y^{t-3},\cdots,Y^{t+3}]$) if the loss function is chosen to be Mean Squared Error (MSE).
This deduction is basically accomplished by the reconstruction module to be described in Section~\ref{RM}. 
It is also worth noting that  (\ref{eq:alterform}) is invariant
under  isomorphic representations of $(Y^{t-3},\cdots,Y^{t+3})$.
This is the reason why 
the features extracted from the LR raw frames can be used in lieu of the LR frames themselves as the input of the SDI module.  
Similarly, the output of the SDI module can be any equivalent representation of $\tilde{\Theta}^t$.

%soft vs. hard

%Under this alternative formulation, the representation of $(Y^{t-\ell},\cdots,Y^{t+\ell})$ and $\tilde{P}^t$

%under equivalent representations of $(Y^{t-\ell},\cdots,Y^{t+\ell})$ and $\tilde{P}^t$

However, solving (\ref{eq:alterform}) through a data-driven approach is very difficult. For one thing, it requires learning joint patterns existent in the (features of) LR raw frames. Presumably the number of such patterns is already very large even for 2 frames and the number is likely to increase by several orders of magnitude with every additional frame. Therefore, it is unrealistic to expect that one can learn enough joint patterns needed to solve (\ref{eq:alterform}) reliably based on limited training data. Fortunately, it turns out that the learning complexity can be greatly reduced under a proper fusion strategy. For ease of exposition, we assume that the HR raw frames $X^{t-3}\leftrightarrow\cdots\leftrightarrow X^{t+3}$ form a Markov chain in this order. Since each LR raw frame (or its feature) is generated from the corresponding HR raw frame through an independent degradation process, $(X^{t-3},\cdots, X^{t+3})$ and $(Y^{t-3},\cdots, Y^{t+3})$ jointly constitute an HMM. This HMM enables us to decompose (\ref{eq:alterform}) into a sequence of simpler problems. Note that $\Theta^{t-2}\triangleq p(X^{t-2}|Y^{t-3},Y^{t-2})$ is a minimal sufficient statistic of $(Y^{t-3},Y^{t-2})$ with respect to $X^{t-2}$. Therefore, 
\begin{align}
\tilde{\Theta}^{t}=p(X^t|\Theta^{t-2},Y^{t-1},\cdots,Y^{t+3}).
\end{align}
Moreover, removing $X^{t-3}$ and replacing $(Y^{t-3},Y^{t-2})$ with $\Theta^{t-2}$ resulting in a new HMM. One can iterate the above argument to show that
\begin{align}
\tilde{\Theta}^{t}=p(X^t|\Theta^t,\Theta^{t+1}),\label{PF}
\end{align}
where $\Theta^{t+\ell}\triangleq p(X^{t+\ell}|\Theta^{t+\ell-1},Y^{t+\ell})$ for $\ell=-2,-1, 0$, and $\Theta^{t+\ell}\triangleq p(X^{t+\ell}|Y^{t+\ell},\Theta^{t+\ell+1})$ for $\ell=1, 2$, with $\Theta^{t-3}\triangleq Y^{t-3}$ and $\Theta^{t+3}\triangleq Y^{t+3}$. Fig.~\ref{fig:SDITheory} provides an intuitive illustration of this iterative argument using probabilistic graphical models. It is worth mentioning that this is similar to the reasoning underlying the well-known forward-backward algorithm for HMM inference. The main difference is that here we are mostly concerned with the fundamental architectural principle rather  than the computational complexity aspect. 
%It is worth mentioning that for the special case where $(X^{t-3},\cdots,X^{t+3})$ and $(Y^{t-3},\cdots,Y^{t+3})$ are jointly Gaussian, one can replace the conditional distributions in the above derivation by the conditional expectations and recover 

\begin{figure}[t]
	\centering
	\includegraphics[width=\linewidth]{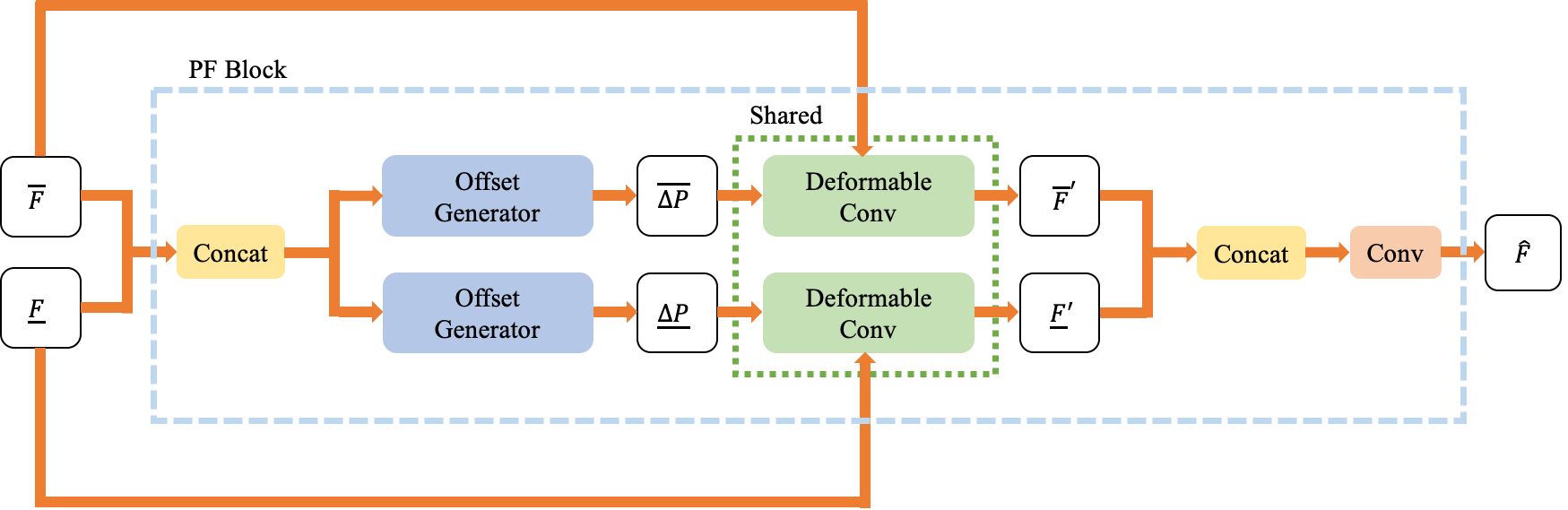}
	\caption{The structure of the PF block.}
	\label{fig:SDI_block}
\end{figure}

Our  fusion strategy has several advantages over the existing ones \cite{bao2019memc,wang2019edvr}. First, an important implication of (\ref{PF}) is that (\ref{eq:alterform}) can be solved by repeatedly performing pairwise fusion, which is much less demanding in terms of the amount of training data needed for reliable learning. Second, 
%The above strategy also ensures that in each pairwise fusion step, 
the two inputs are of comparable importance with respect to the target object and consequently are more likely to be thoroughly exploited. In contrast, for other fusion strategies where the inputs under consideration have a clear difference in importance (for example, $Y^t$ is clearly more relevant than $Y^{t-3}$ for the purpose of estimating $X^t$), the subtle details contained in the less important ones may easily get overlooked.

%\textcolor{magenta}{It is worth noting that the asymmetric structure of this module is a little bit surprising nevertheless inevitable consequence of our analysis; in fact, it can be shown that symmetric pairwise fusion (PF) rules are strictly suboptimal even in the simple Gaussian case.}

As shown in Fig.~\ref{fig:SDI}, the SDI module is designed according to the architectural principle suggested by (\ref{PF})  (note that the slight asymmetry in (\ref{PF}) also manifests in the structure of this module).
It takes the features $F^{t-3},\cdots,F^{t+3}$ extracted respectively from the LR raw frames $Y^{t-3}_{raw},\cdots,Y^{t+3}_{raw}$ and outputs a fused feature $\hat{F}^{t}$.  Each pairwise fusion (PF) step is realized by a PF block (see Fig.~\ref{fig:SDI_block} for its structure). Specifically, the PF block first concatenates its two input features $\overline{F}$ and $\underline{F}$; it then uses two structurally-identical offset generators (without  weight-sharing) to learn positional offsets $\overline{\Delta P}\triangleq(\overline{\Delta P}_1,\cdots,\overline{\Delta P}_K)$ and $\underline{\Delta P}\triangleq(\underline{\Delta P}_1,\cdots,\underline{\Delta P}_K)$, which are leveraged to align $\overline{F}$ and $\underline{F}$ using the modulated deformable convolution~\cite{zhu2019deformable}; finally, the aligned features $\overline{F}'$ and $\underline{F}'$ are concatenated and passed through two convolutional layers to produce the fusion result $\hat{F}$. Note that the alignment procedure can be interpreted as implicit motion estimation and compensation in the feature space, and the relationship between pre- and post-alignment features at position $P$ can be expressed as
\begin{align} \label{equ:n2m}
&\overline{F}'(P) = \sum_{i=1}^{K}\overline{w}_i\cdot \overline{F}(P+\overline{\Delta P}_i)\cdot \Delta \overline{S}_i,\\
\label{equ:m2n}
&\underline{F}'(P) = \sum_{i=1}^{K}\underline{w}_i\cdot \underline{F}(P+\underline{\Delta P}_i)\cdot \Delta \underline{S}_i,
\end{align}
where $\overline{w}_i$ ($\underline{w}_i$) and $\overline{\Delta S}_i$ ($\underline{\Delta S}_i$) denote the weight and the modulation scalar of the deformable convolution for  $\overline{F}(P)$ ($\underline{F}(P)$) with respect to offset $\overline{\Delta P}_i$ ($\underline{\Delta P}_i$), $i=1,\cdots,K$ (we set $K=9$ in this work).  It is worth noting that different from the flow-based method, the proposed SDI module does  not require any pre-training and can be embedded into another network for end-to-end training. We would also like to point out that the first-order Markov assumption for HR raw frames is not required for the derivation of the architecture in Fig.~\ref{fig:SDI}. Indeed, a higher-order Markov chain (which is more suitable to model scenarios with occlusion and disocclusion) can be reduced to a first-order one by grouping several consecutive states in a sliding-window fashion, after which our previous argument can be invoked with no essential change. In fact, the only constraining factor for the SDI module is the dimension of the output of each PF block in the sense that higher dimensional outputs are likely needed to accommodate more sophisticated Markov structures.

%This assumption is quite reasonable in view of the fact that $X^{t-3},\cdots, X^{t+3}$ are consecutive frames captured by camera sensors without undergoing any lossy processing. 

%this assumption is just to simplify the analysis,  actually not essential for the architecture. with higher-order Markov chain, one will get essentially the same architecture. The only constraining factor is the dimension of the output of each PF block. one needs to increase the dimension in order to accommodate high-order structures

\begin{figure}[t]
	\centering
	\includegraphics[width=\linewidth]{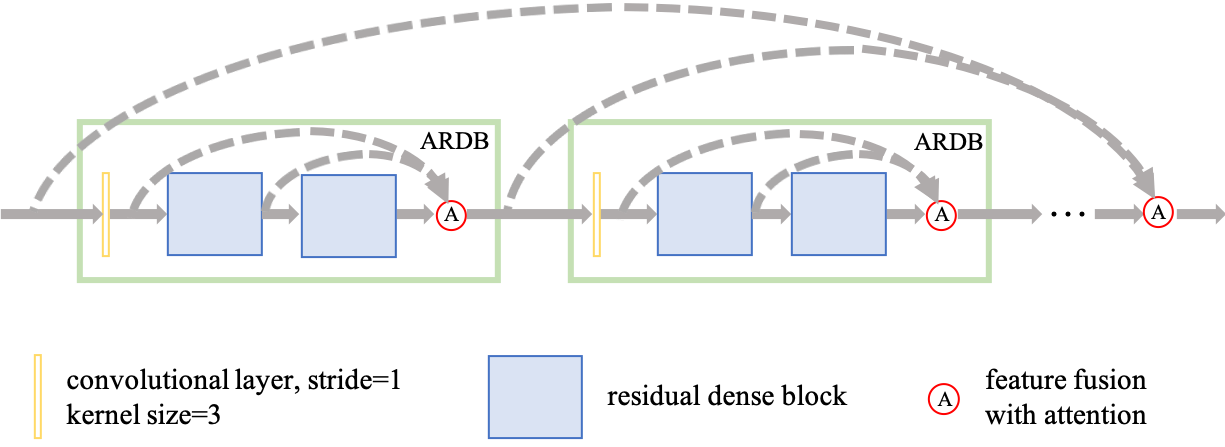}
	\caption{The architecture of the reconstruction module.}
	\label{fig:RM}
\end{figure}

\subsection{Reconstruction Module} \label{RM}

The reconstruction module is elaborately designed with two tasks in mind: 1) refining the fused feature associated with the input LR raw frames and 2) learning the ISP-specific information needed for color transformation and correction from the color reference frame.
It is built using
ARDBs  as shown in Fig.~\ref{fig:RM}. Each ARDB  consists of one convolutional layer and two residual dense blocks. We follow the default RDB settings in~\cite{zhang2018residual} except that the growth rate is set to $64$. The generated intermediate feature maps are fused under the guidance of the channel-wise attention. To this end, we adopt the SENet in~\cite{hu2018squeeze} to produce  attention weights to enhance/suppress the contributions of the associated feature maps in accordance with their relevance.
In addition to the local fusion in each ARDB, we also employ a global fusion with channel-wise attention to further strengthen the learning ability of the reconstruction module. To strike a balance between system performance and computational complexity, 
totally four ARDBs are used in our design.

\begin{table*}[h]
	\begin{center}
		\caption{Quantitative comparisons on the RawVD test data for $2\times$ and $4\times$ VSR. \textbf{\color{red}Red} and \underline{\color{blue}Blue} indicate the best and the second best performance, respectively.}
		\label{tab:canon}
		
		\begin{adjustbox}{width=\textwidth}
			\begin{tabular}{|c|c|c|c|c|c|c|c|c|c|c|c|}
				\hline
				\multicolumn{11}{|c|}{RawVD $2\times$} \vline \\
				\hline  \hline                                                    
				Name & Bicubic & VSRNet~\cite{kappeler2016video} & VESPCN~\cite{caballero2017real} & SPMC~\cite{tao2017detail} &DUF~\cite{jo2018deep} & RBPN~\cite{haris2019recurrent}  &  TDAN \cite{tian2018tdan} & EDVR~\cite{wang2019edvr} & RawVSR{$^\dagger$} & RawVSR \\ % <--
				\hline \hline
				% Clipname
				Store & 
				% Bicubic
				27.84/0.8299 & 
				% VSRNet
				28.99/0.8457 & 
				% VESPCN
				29.02/0.8494 & 
				% SPMC 
				29.41/0.8503 & 
				% VSRDUF
				30.75/0.8710 & 
				% RBPN,
				- &
				% TDAN
				31.27/0.8863 &
				% EDVR
				31.98/0.9060 &
				% Ours(RGB)
				\underline{\color{blue}33.04/0.9129} &
				% Ours(Raw)
				\textbf{\color{red}34.30/0.9263} \\ % <--
				
				%Painting
				% Clipname
				Painting & 
				% Bicubic
				28.80/0.8086 & 
				% VSRNet
				29.38/0.8225 & 
				% VESPCN
				29.45/0.8218 & 
				% SPMC 
				28.82/0.7998 & 
				% VSRDUF
				31.22/0.8512 & 
				% RBPN
				- &
				% TDAN
				31.17/0.8556 &
				% EDVR
				32.55/\underline{\color{blue}0.8816} &
				% Ours(RGB)
				\underline{\color{blue}32.80}/0.8773 &
				% Ours(Raw)
				\textbf{\color{red}33.79/0.8943} \\
				
				%Train
				% Clipname
				Train & 
				% Bicubic
				28.11/0.7738 &  
				% VSRNet
				29.18/0.8081 & 
				% VESPCN
				29.32/0.8121 & 
				% SPMC 
				30.01/0.8298 & 
				% VSRDUF
				31.25/0.8464 & 
				% RBPN
				- &
				% TDAN
				30.94/0.8396 &
				% EDVR
				31.42/\underline{\color{blue}0.8617} &
				% Ours(RGB)
				\underline{\color{blue}31.66}/0.8522 &
				% Ours(Raw)
				\textbf{\color{red}32.79/0.8729} \\
				
				%City
				% Clipname
				City & 
				% Bicubic
				27.76/0.7527 & 
				% VSRNet
				28.99/0.8012 & 
				% VESPCN
				29.22/0.8047 & 
				% SPMC 
				29.44/0.8210 & 
				% VSRDUF
				30.47/0.8190 & 
				% RBPN
				- &
				% TDAN
				30.26/0.8227 &
				% EDVR
				31.11/0.8517 &
				% Ours(RGB)
				\underline{\color{blue}31.41/0.8588} &
				% Ours(Raw)
				\textbf{\color{red}32.11/0.8637} \\

				%Tree
				% Clipname
				Walk & 
				% Bicubic
				26.43/0.7553 & 
				% VSRNet
				28.92/0.7871 & 
				% VESPCN
				28.78/0.7820 & 
				% SPMC 
				30.25/0.8439 & 
				% VSRDUF
				30.13/0.8117 & 
				% RBPN
				- &
				% TDAN
				29.26/0.8207 & 
				% EDVR
				\underline{\color{blue}30.58/0.8415} &
				% Ours(RGB)
				29.63/0.8282 &
				% Ours(Raw)
				\textbf{\color{red}31.20/0.8516}\\
				
				\hline\hline
				
				%Average
				% Clipname
				Average & 
				% Bicubic
				27.79/0.7841 & 
				% VSRNet
				29.09/0.8129 & 
				% VESPCN
				29.16/0.8140 & 
				% SPMC 
				29.59/0.8290 & 
				% VSRDUF
				30.76/0.8399 & 
				% RBPN
				- & 
				% TDAN
				30.58/0.8450 &
				% EDVR
				31.53/\underline{\color{blue}0.8685}&
				% Ours(RGB)
				\underline{\color{blue}31.71}/0.8659&
				% Ours(Raw)
				\textbf{\color{red}32.84/0.8818} \\
				\hline
			\end{tabular}
		\end{adjustbox}
		\vspace{1.5mm}
		
		\begin{adjustbox}{width=\textwidth}
			\begin{tabular}{|c|c|c|c|c|c|c|c|c|c|c|c|}
				\hline
				\multicolumn{11}{|c}{RawVD $4\times$} \vline \\
				\hline  \hline                                                    
				Name & Bicubic & VSRNet~\cite{kappeler2016video} & VESPCN~\cite{caballero2017real} & SPMC~\cite{tao2017detail} & DUF~\cite{jo2018deep} & RBPN~\cite{haris2019recurrent}  &  TDAN \cite{tian2018tdan} & EDVR~\cite{wang2019edvr} & RawVSR{$^\dagger$} & RawVSR \\ % <--
				\hline \hline
				% Clipname
				Store & 
				% Bicubic
				22.59/0.6507 & 
				% VSRNet
				25.24/0.7135 & 
				% VESPCN
				25.46/0.7250 & 
				% SPMC 
				26.34/0.7351 & 
				% VSRDUF
				27.84/0.8105 & 
				% RBPN
				27.53/0.7647 &
				% tdan
				26.67/0.7802 &
				% EDVR
				28.23/\underline{\color{blue}0.8202} &
				% Our RGB
				\underline{\color{blue}28.29}/0.8191 &
				% Ours(Raw)
				\textbf{\color{red}29.04/0.8400} \\ % <--
				
				%Painting
				% Clipname
				Painting & 
				% Bicubic
				24.78/0.6712 & 
				% VSRNet
				26.17/0.7119 & 
				% VESPCN
				26.02/0.7120 & 
				% SPMC 
				26.13/0.7199 & 
				% VSRDUF
				27.71/0.7833 & 
				% RBPN
				26.40/0.7209 &
				% tdan
				26.75/0.7539 &
				% EDVR
				27.74/0.7908 &
				% Our RGB
				\underline{\color{blue}28.51/0.7940} &
				% Ours(Raw)
				\textbf{\color{red}29.02/0.8104} \\
				
				%Train
				% Clipname
				Train & 
				% Bicubic
				24.12/0.6166 &  
				% VSRNet
				25.80/0.6671 & 
				% VESPCN
				25.96/0.6724 & 
				% SPMC 
				26.70/0.6984 & 
				% VSRDUF
				27.53/0.7377 & 
				% RBPN
				27.64/0.7358 &
				% tdan
				26.94/0.7205 &
				% EDVR
				27.81/\underline{\color{blue}0.7470} &
				% Our RGB
				\underline{\color{blue}28.02}/0.7443 &
				% Ours(Raw)
				\textbf{\color{red}28.59/0.7625} \\
				
				%City
				% Clipname
				City & 
				% Bicubic
				24.94/0.6330 & 
				% VSRNet
				26.59/0.6907 & 
				% VESPCN
				26.63/0.6927 & 
				% SPMC 
				27.03/0.6684 & 
				% VSRDUF
				27.93/0.7508 & 
				% RBPN
				28.39/0.7332 &
				% tdan
				27.55/0.7336 &
				% EDVR
				28.24/0.7558 &
				% Our RGB
				\underline{\color{blue}28.43/0.7640} &
				% Ours(Raw)
				\textbf{\color{red}29.08/0.7843} \\
				%Tree
				% Clipname
				Walk & 
				% Bicubic
				22.46/0.6229 & 
				% VSRNet
				25.52/0.6766 & 
				% VESPCN
				25.62/0.6726 & 
				% SPMC 
				25.30/0.6797 & 
				% VSRDUF
				27.30/0.7512 & 
				% RBPN
				26.46/0.7337 & 
				% tdan
				26.36/0.7229 &
				% EDVR
				27.02/0.7533 &
				% Our RGB
				\underline{\color{blue}27.90/0.7607} &
				% Ours(Raw)
				\textbf{\color{red}28.06/0.7724}\\
				
				\hline\hline
				
				%Average
				% Clipname
				Average & 
				% Bicubic
				23.78/0.6389 & 
				% VSRNet
				25.87/0.6919 & 
				% VESPCN
				25.94/0.6950 & 
				% SPMC 
				26.30/0.7002 & 
				% VSRDUF
				27.66/0.7667 & 
				% RBPN
				27.29/0.7377 & 
				% tdan
				26.85/0.7422 &
				% EDVR
				27.81/0.7734 &
				% Our RGB
				\underline{\color{blue}28.23/0.7765} &
				% Ours(Raw)
				\textbf{\color{red}28.76/0.7939} \\
				\hline
				
			\end{tabular}
		\end{adjustbox}
	\end{center}
\end{table*}

\begin{table}[h]
	\begin{center}
		\caption{Quantitative comparisons on the iPhone data for $4\times$ VSR. \textbf{\color{red}Red} and \underline{\color{blue}Blue} indicate the best and the second best performance, respectively.}
		\label{tab:iphone}
		\begin{adjustbox}{width=\linewidth}
			\begin{tabular}{|c|c|c|c|c|}
				\hline
				\multicolumn{5}{|c}{iPhone $4\times$} \vline \\
				\hline  \hline                                                    
				Name & DUF~\cite{jo2018deep} & EDVR~\cite{wang2019edvr} & RawVSR{$^\dagger$} & RawVSR \\ % <--
				\hline \hline
				% Clipname
				Tree & 
				% DUF
				27.15/0.6630 & 
				% EDVR
				27.26/0.6504 &
				% Our RGB
				\underline{\color{blue}28.66/0.7073} &
				% Ours(Raw)
				\textbf{\color{red}29.54/0.7381} \\ % <--
				
				% Clipname
				Door & 
				% DUF
				26.11/0.6334 & 
				% EDVR
				26.05/0.6231 &
				% Our RGB
				\underline{\color{blue}26.28/0.6694} &
				% Ours(Raw)
				\textbf{\color{red}28.62/0.6957} \\
				
				% Clipname
				Flower & 
				% DUF
				25.90/0.6225 & 
				% EDVR
				25.74/0.6062 &
				% Our RGB
				\underline{\color{blue}27.07/0.6771} &
				% Ours(Raw)
				\textbf{\color{red}28.17/0.7307} \\
				
				\hline\hline
				
				%Average
				% Clipname
				Average & 
				% DUF
				26.38/0.6396 & 
				% EDVR
				26.35/0.6266 &
				% Our RGB
				\underline{\color{blue}27.34/0.6846} &
				% Ours(Raw)
				\textbf{\color{red}28.78/0.7215} \\
				\hline
			\end{tabular}
		\end{adjustbox}
	\end{center}
\end{table}

\subsection{On the Practicability of raw VSR } \label{usefuleness}
Although the present work mainly focuses on the technical feasibility of raw VSR, here we would like to briefly comment on its practical prospect. In particular, we shall address the following two questions: 1) the availability of camera raw data, and 2) the application scenarios of raw VSR. As raw data record untouched radiance information that can be fruitfully exploited for many different purposes, they receive considerable attention from professional photographers. To facilitate their use, the major camera manufacturers (\textit{e.g.}, Canon and Nikon) have provided the official interface to access raw data. This removes a major hurdle for the wide adoption of raw VSR. Moreover, in most scenarios where conventional VSR is currently being considered, raw VSR can potentially be  a more favorable choice to its superior performance and consistency with the real imaging pipeline. It is worth mentioning that some recent works \cite{zhou2018deep, brooks2019unprocessing} attempt to generate raw-like data based on processed data. However, such raw-like data  inevitably suffer from the information loss issue due to the non-invertibility of ISP operations, and great caution should be exercised when using them in lieu of  authentic raw data such as those in our RawVD. More discussions can be found in Section \ref{sec:fakeraw}.

\section{Experimental Results}\label{sec:experiment}
Extensive experiments are conducted to demonstrate the competitive performance of the proposed RawVSR. They also provide 
solid justifications for our design (especially, the SDI module and the reconstruction module) and convincing evidence regarding the value of raw data.

\begin{figure*}[htbp]
	\centering
	\begin{minipage}[h]{0.161\linewidth} %0.193
		\centering
		\includegraphics[width=\linewidth]{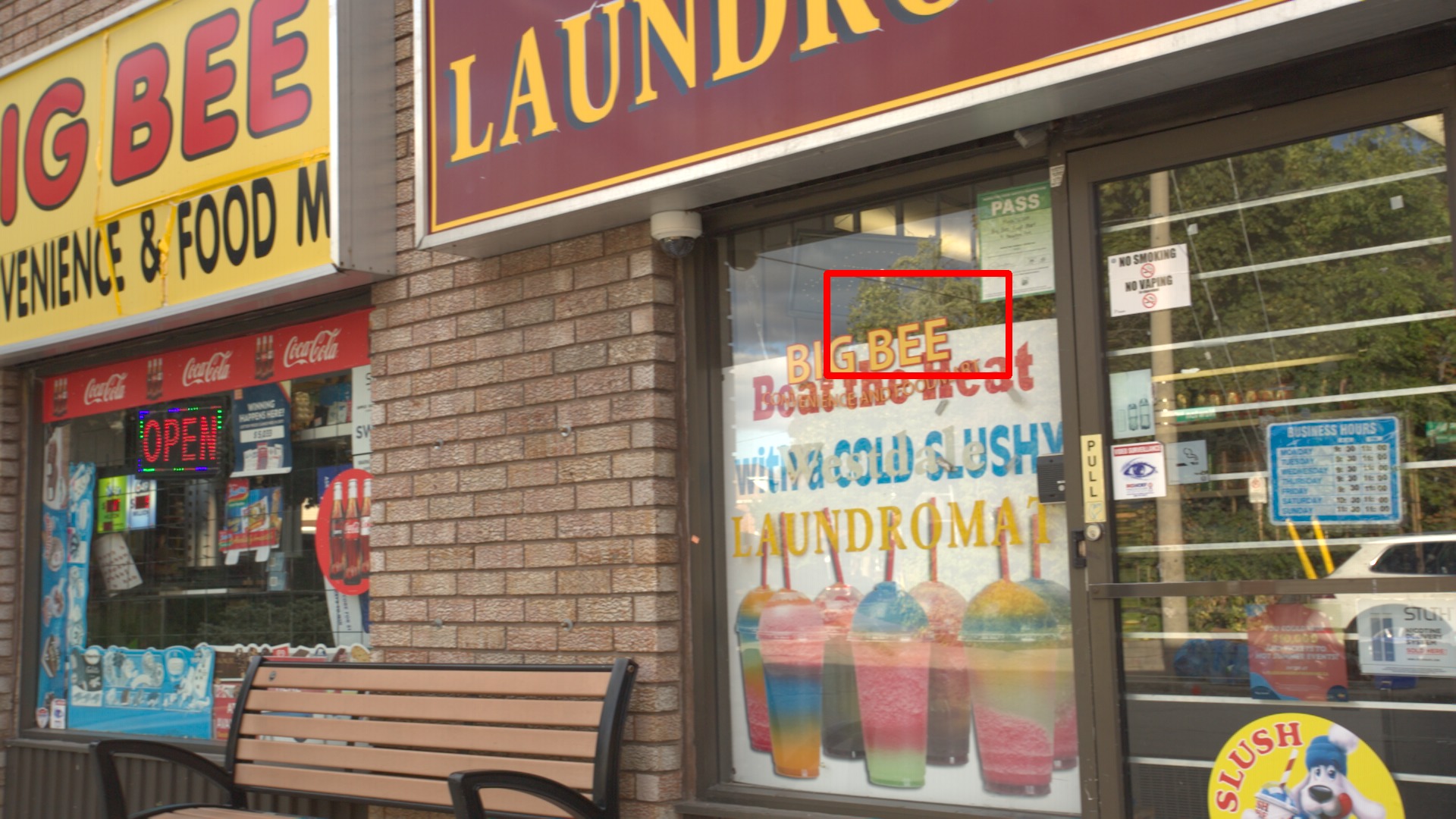}
		\scriptsize{(a) Store}
	\end{minipage}
	\begin{minipage}[h]{0.161\linewidth}
		\centering
		\includegraphics[width=\linewidth]{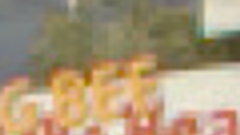}
		\scriptsize{(b) Bicubic}
	\end{minipage}
	\begin{minipage}[h]{0.161\linewidth}
		\centering
		\includegraphics[width=\linewidth]{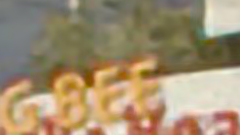}
		\scriptsize{(c) VSRNet~\cite{kappeler2016video}}
	\end{minipage}
	\begin{minipage}[h]{0.161\linewidth}
		\centering
		\includegraphics[width=\linewidth]{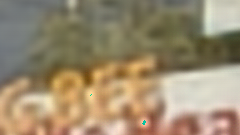}
		\scriptsize{(d) VESPCN~\cite{caballero2017real}}
	\end{minipage}
	\begin{minipage}[h]{0.161\linewidth}
		\centering
		\includegraphics[width=\linewidth]{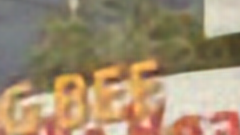}
		\scriptsize{(e) SPMC~\cite{tao2017detail}}
	\end{minipage}
	\begin{minipage}[h]{0.161\linewidth}
		\centering
		\includegraphics[width=\linewidth]{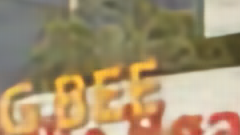}
		\scriptsize{(f) DUF~\cite{jo2018deep}}
	\end{minipage}
	\begin{minipage}[h]{0.161\linewidth}
		\centering
		\includegraphics[width=\linewidth]{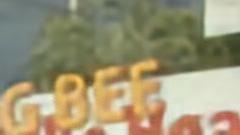}
		\scriptsize{(g) RBPN~\cite{haris2019recurrent}}
	\end{minipage}
	\begin{minipage}[h]{0.161\linewidth}
		\centering
		\includegraphics[width=\linewidth]{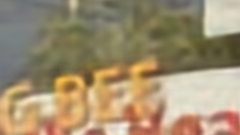}
		\scriptsize{(h) TDAN}
	\end{minipage}
	\begin{minipage}[h]{0.161\linewidth}
		\centering
		\includegraphics[width=\linewidth]{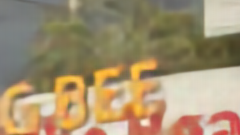}
		\scriptsize{(i) EDVR~\cite{wang2019edvr}}
	\end{minipage}
	\begin{minipage}[h]{0.161\linewidth}
		\centering
		\includegraphics[width=\linewidth]{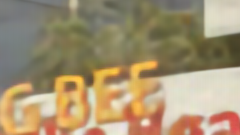}
		\scriptsize{(j) RawVSR$^\dagger$}
	\end{minipage}
	\begin{minipage}[h]{0.161\linewidth}
		\centering
		\includegraphics[width=\linewidth]{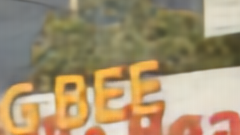}
		\scriptsize{(k) RawVSR}
	\end{minipage}
	\begin{minipage}[h]{0.161\linewidth}
		\centering
		\includegraphics[width=\linewidth]{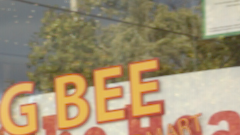}
		\scriptsize{(l) ground truth}
	\end{minipage}

	\centering
	\begin{minipage}[h]{0.161\linewidth}
		\centering
		\includegraphics[width=\linewidth]{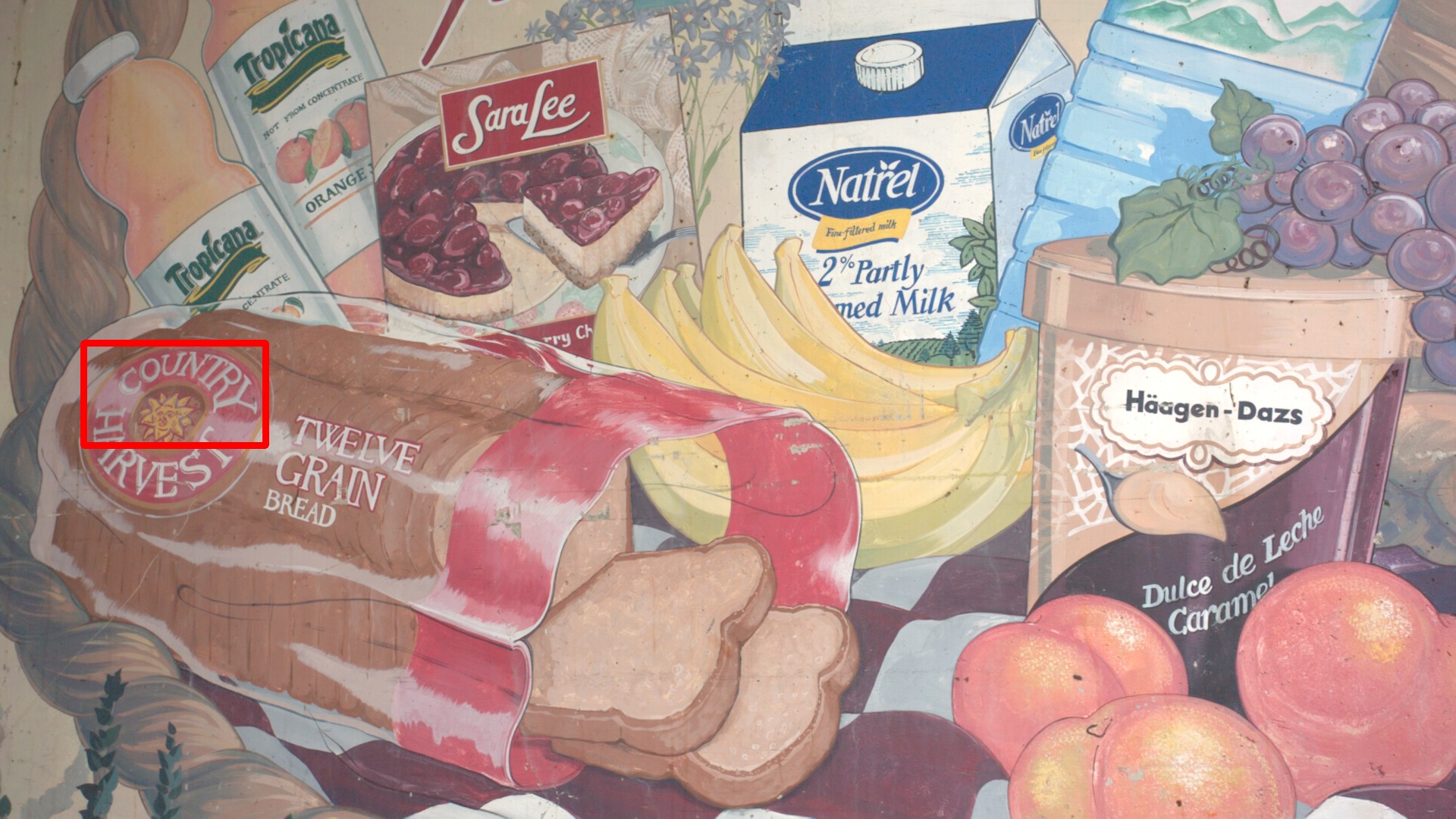}
		\scriptsize{(a) Painting}
	\end{minipage}
	\begin{minipage}[h]{0.161\linewidth}
		\centering
		\includegraphics[width=\linewidth]{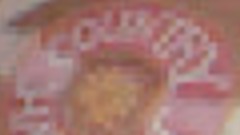}
		\scriptsize{(b) Bicubic}
	\end{minipage}
	\begin{minipage}[h]{0.161\linewidth}
		\centering
		\includegraphics[width=\linewidth]{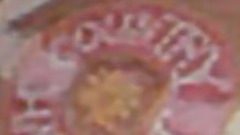}
		\scriptsize{(c) VSRNet~\cite{kappeler2016video}}
	\end{minipage}
	\begin{minipage}[h]{0.161\linewidth}
		\centering
		\includegraphics[width=\linewidth]{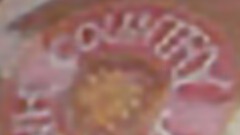}
		\scriptsize{(d) VESPCN~\cite{caballero2017real}}
	\end{minipage}
	\begin{minipage}[h]{0.161\linewidth}
		\centering
		\includegraphics[width=\linewidth]{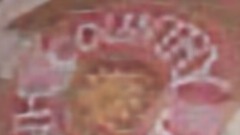}
		\scriptsize{(e) SPMC~\cite{tao2017detail}}
	\end{minipage}
	\begin{minipage}[h]{0.161\linewidth}
		\centering
		\includegraphics[width=\linewidth]{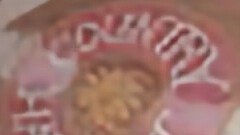}
		\scriptsize{(f) DUF~\cite{jo2018deep}}
	\end{minipage}
	\begin{minipage}[h]{0.161\linewidth}
		\centering
		\includegraphics[width=\linewidth]{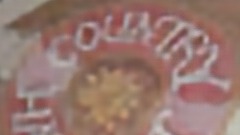}
		\scriptsize{(g) RBPN~\cite{haris2019recurrent}}
	\end{minipage}
	\begin{minipage}[h]{0.161\linewidth}
		\centering
		\includegraphics[width=\linewidth]{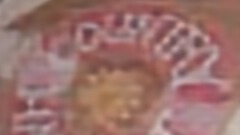}
		\scriptsize{(h) TDAN}
	\end{minipage}
	\begin{minipage}[h]{0.161\linewidth}
		\centering
		\includegraphics[width=\linewidth]{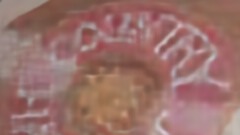}
		\scriptsize{(i) EDVR~\cite{wang2019edvr}}
	\end{minipage}
	\begin{minipage}[h]{0.161\linewidth}
		\centering
		\includegraphics[width=\linewidth]{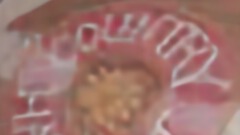}
		\scriptsize{(j) RawVSR$^\dagger$}
	\end{minipage}
	\begin{minipage}[h]{0.161\linewidth}
		\centering
		\includegraphics[width=\linewidth]{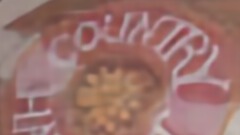}
		\scriptsize{(k) RawVSR}
	\end{minipage}
	\begin{minipage}[h]{0.161\linewidth}
		\centering
		\includegraphics[width=\linewidth]{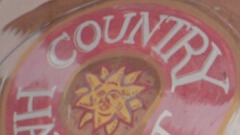}
		\scriptsize{(l) ground truth}
	\end{minipage}
	
	\centering
	\begin{minipage}[h]{0.161\linewidth}
		\centering
		\includegraphics[width=\linewidth]{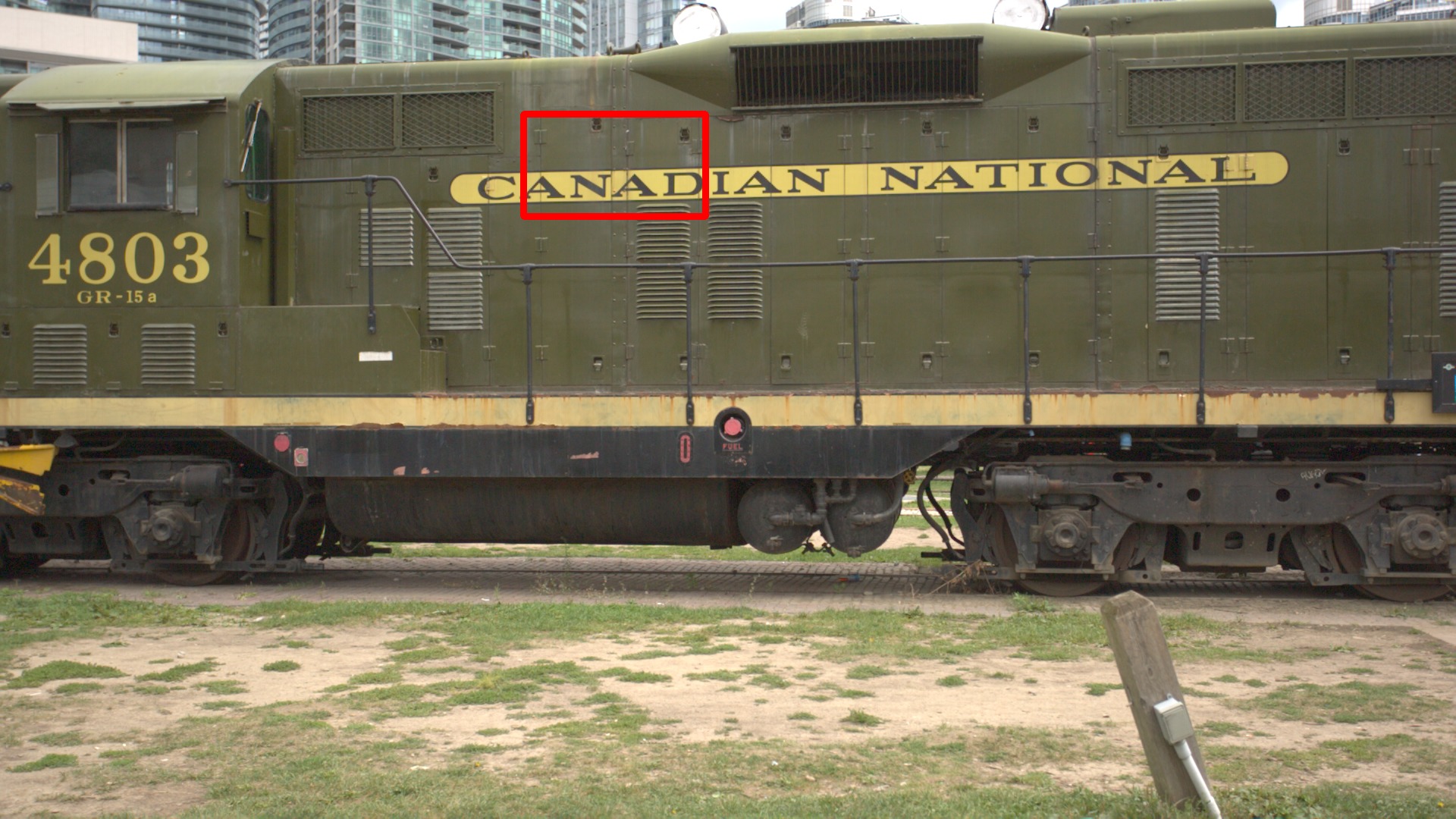}
		\scriptsize{(a) Train}
	\end{minipage}
	\begin{minipage}[h]{0.161\linewidth}
		\centering
		\includegraphics[width=\linewidth]{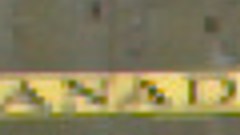}
		\scriptsize{(b) Bicubic}
	\end{minipage}
	\begin{minipage}[h]{0.161\linewidth}
		\centering
		\includegraphics[width=\linewidth]{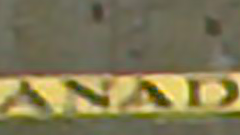}
		\scriptsize{(c) VSRNet~\cite{kappeler2016video}}
	\end{minipage}
	\begin{minipage}[h]{0.161\linewidth}
		\centering
		\includegraphics[width=\linewidth]{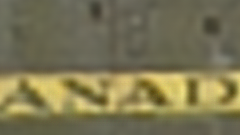}
		\scriptsize{(d) VESPCN~\cite{caballero2017real}}
	\end{minipage}
	\begin{minipage}[h]{0.161\linewidth}
		\centering
		\includegraphics[width=\linewidth]{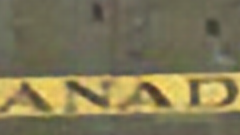}
		\scriptsize{(e) SPMC~\cite{tao2017detail}}
	\end{minipage}
	\begin{minipage}[h]{0.161\linewidth}
		\centering
		\includegraphics[width=\linewidth]{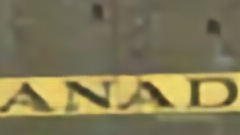}
		\scriptsize{(f) DUF~\cite{jo2018deep}}
	\end{minipage}
	\begin{minipage}[h]{0.161\linewidth}
		\centering
		\includegraphics[width=\linewidth]{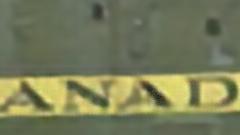}
		\scriptsize{(g) RBPN~\cite{haris2019recurrent}}
	\end{minipage}	
	\begin{minipage}[h]{0.161\linewidth}
		\centering
		\includegraphics[width=\linewidth]{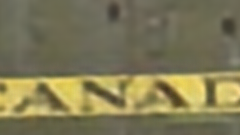}
		\scriptsize{(h) TDAN}
	\end{minipage}
	\begin{minipage}[h]{0.161\linewidth}
		\centering
		\includegraphics[width=\linewidth]{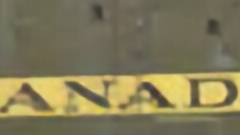}
		\scriptsize{(i) EDVR~\cite{wang2019edvr}}
	\end{minipage}
	\begin{minipage}[h]{0.161\linewidth}
		\centering
		\includegraphics[width=\linewidth]{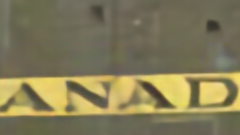}
		\scriptsize{(j) RawVSR$^\dagger$}
	\end{minipage}
	\begin{minipage}[h]{0.161\linewidth}
		\centering
		\includegraphics[width=\linewidth]{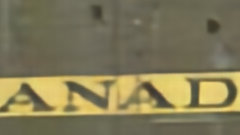}
		\scriptsize{(k) RawVSR}
	\end{minipage}
	\begin{minipage}[h]{0.161\linewidth}
		\centering
		\includegraphics[width=\linewidth]{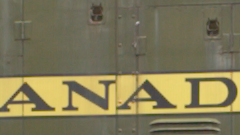}
		\scriptsize{(l) ground truth}
	\end{minipage}

	\centering
	\begin{minipage}[h]{0.161\linewidth}
		\centering
		\includegraphics[width=\linewidth]{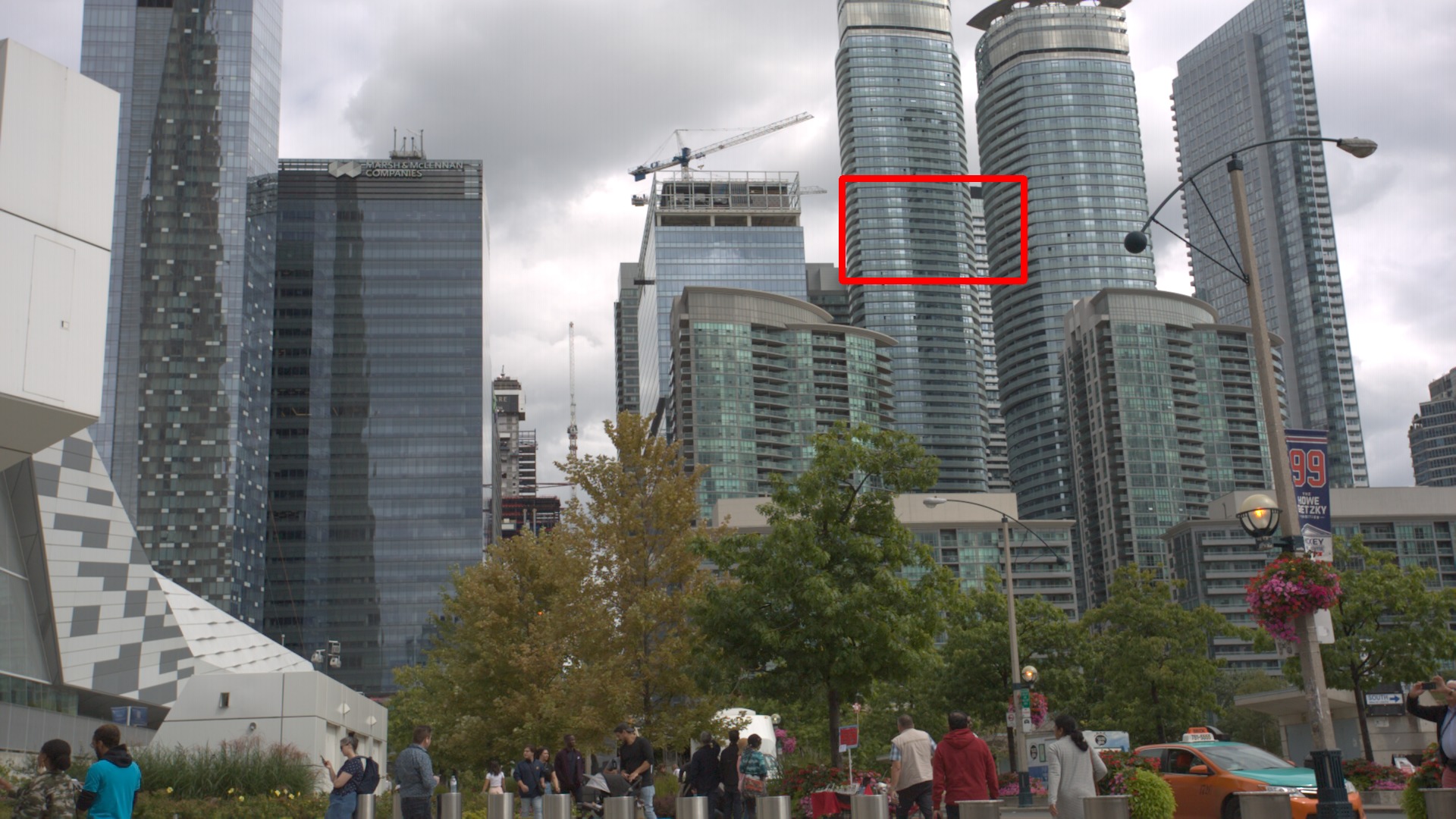}
		\scriptsize{(a) City}
	\end{minipage}
	\begin{minipage}[h]{0.161\linewidth}
		\centering
		\includegraphics[width=\linewidth]{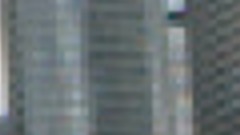}
		\scriptsize{(b) Bicubic}
	\end{minipage}
	\begin{minipage}[h]{0.161\linewidth}
		\centering
		\includegraphics[width=\linewidth]{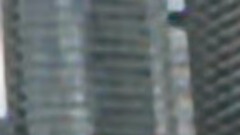}
		\scriptsize{(c) VSRNet~\cite{kappeler2016video}}
	\end{minipage}
	\begin{minipage}[h]{0.161\linewidth}
		\centering
		\includegraphics[width=\linewidth]{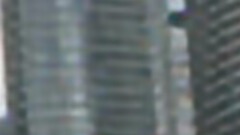}
		\scriptsize{(d) VESPCN~\cite{caballero2017real}}
	\end{minipage}
	\begin{minipage}[h]{0.161\linewidth}
		\centering
		\includegraphics[width=\linewidth]{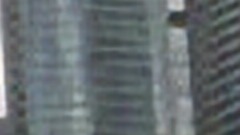}
		\scriptsize{(e) SPMC~\cite{tao2017detail}}
	\end{minipage}
	\begin{minipage}[h]{0.161\linewidth}
		\centering
		\includegraphics[width=\linewidth]{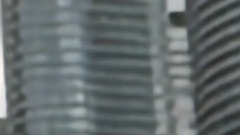}
		\scriptsize{(f) DUF~\cite{jo2018deep}}
	\end{minipage}
	\begin{minipage}[h]{0.161\linewidth}
		\centering
		\includegraphics[width=\linewidth]{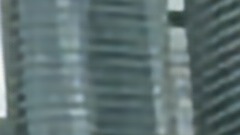}
		\scriptsize{(g) RBPN~\cite{haris2019recurrent}}
	\end{minipage}
	\begin{minipage}[h]{0.161\linewidth}
		\centering
		\includegraphics[width=\linewidth]{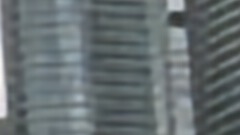}
		\scriptsize{(h) TDAN}
	\end{minipage}
	\begin{minipage}[h]{0.161\linewidth}
		\centering
		\includegraphics[width=\linewidth]{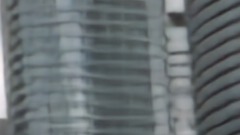}
		\scriptsize{(i) EDVR~\cite{wang2019edvr}}
	\end{minipage}
	\begin{minipage}[h]{0.161\linewidth}
		\centering
		\includegraphics[width=\linewidth]{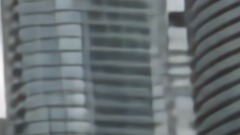}
		\scriptsize{(j) RawVSR$^\dagger$}
	\end{minipage}
	\begin{minipage}[h]{0.161\linewidth}
		\centering
		\includegraphics[width=\linewidth]{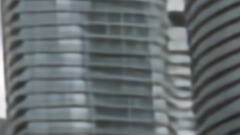}
		\scriptsize{(k) RawVSR}
	\end{minipage}
	\begin{minipage}[h]{0.161\linewidth}
		\centering
		\includegraphics[width=\linewidth]{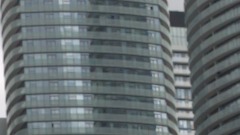}
		\scriptsize{(l) ground truth}
	\end{minipage}
	
	\centering
	\begin{minipage}[h]{0.161\linewidth}
		\centering
		\includegraphics[width=\linewidth]{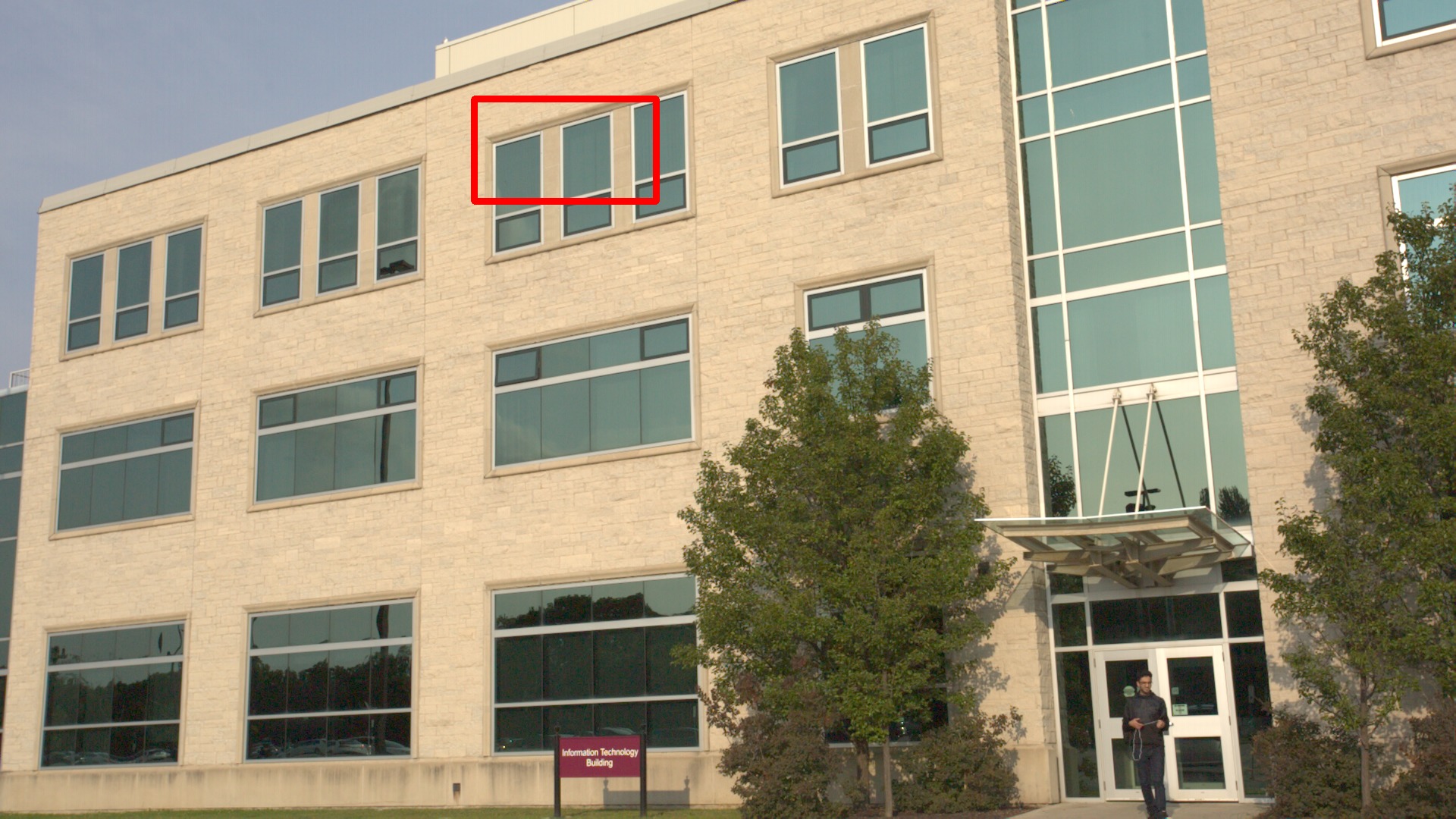}
		\scriptsize{(a) Walk}
	\end{minipage}
	\begin{minipage}[h]{0.161\linewidth}
		\centering
		\includegraphics[width=\linewidth]{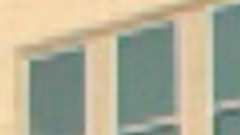}
		\scriptsize{(b) Bicubic}
	\end{minipage}
	\begin{minipage}[h]{0.161\linewidth}
		\centering
		\includegraphics[width=\linewidth]{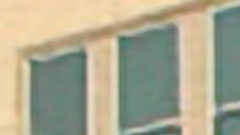}
		\scriptsize{(c) VSRNet~\cite{kappeler2016video}}
	\end{minipage}
	\begin{minipage}[h]{0.161\linewidth}
		\centering
		\includegraphics[width=\linewidth]{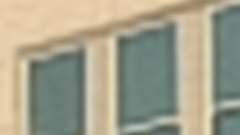}
		\scriptsize{(d) VESPCN~\cite{caballero2017real}}
	\end{minipage}
	\begin{minipage}[h]{0.161\linewidth}
		\centering
		\includegraphics[width=\linewidth]{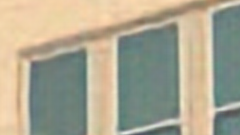}
		\scriptsize{(e) SPMC~\cite{tao2017detail}}
	\end{minipage}
	\begin{minipage}[h]{0.161\linewidth}
		\centering
		\includegraphics[width=\linewidth]{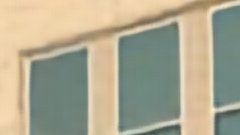}
		\scriptsize{(f) DUF~\cite{jo2018deep}}
	\end{minipage}
	\begin{minipage}[h]{0.161\linewidth}
		\centering
		\includegraphics[width=\linewidth]{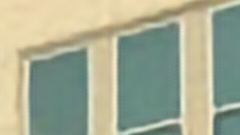}
		\scriptsize{(g) RBPN~\cite{haris2019recurrent}}
	\end{minipage}
	\begin{minipage}[h]{0.161\linewidth}
		\centering
		\includegraphics[width=\linewidth]{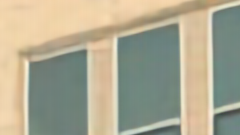}
		\scriptsize{(h) TDAN}
	\end{minipage}
	\begin{minipage}[h]{0.161\linewidth}
		\centering
		\includegraphics[width=\linewidth]{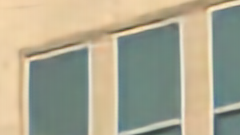}
		\scriptsize{(i) EDVR~\cite{wang2019edvr}}
	\end{minipage}
	\begin{minipage}[h]{0.161\linewidth}
		\centering
		\includegraphics[width=\linewidth]{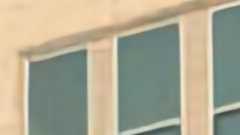}
		\scriptsize{(j) RawVSR$^\dagger$}
	\end{minipage}
	\begin{minipage}[h]{0.161\linewidth}
		\centering
		\includegraphics[width=\linewidth]{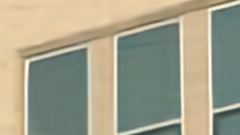}
		\scriptsize{(k) RawVSR}
	\end{minipage}
	\begin{minipage}[h]{0.161\linewidth}
		\centering
		\includegraphics[width=\linewidth]{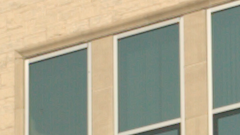}
		\scriptsize{(l) ground truth}
	\end{minipage}
	
	\caption{Qualitative comparisons on the RawVD test data for $4\times$ VSR. Zoom in for better visualization.}
	\label{fig:canon}
\end{figure*}

\begin{figure}[htbp]
	\centering
	\begin{minipage}[h]{0.189\linewidth}
		\centering
		\includegraphics[width=\linewidth]{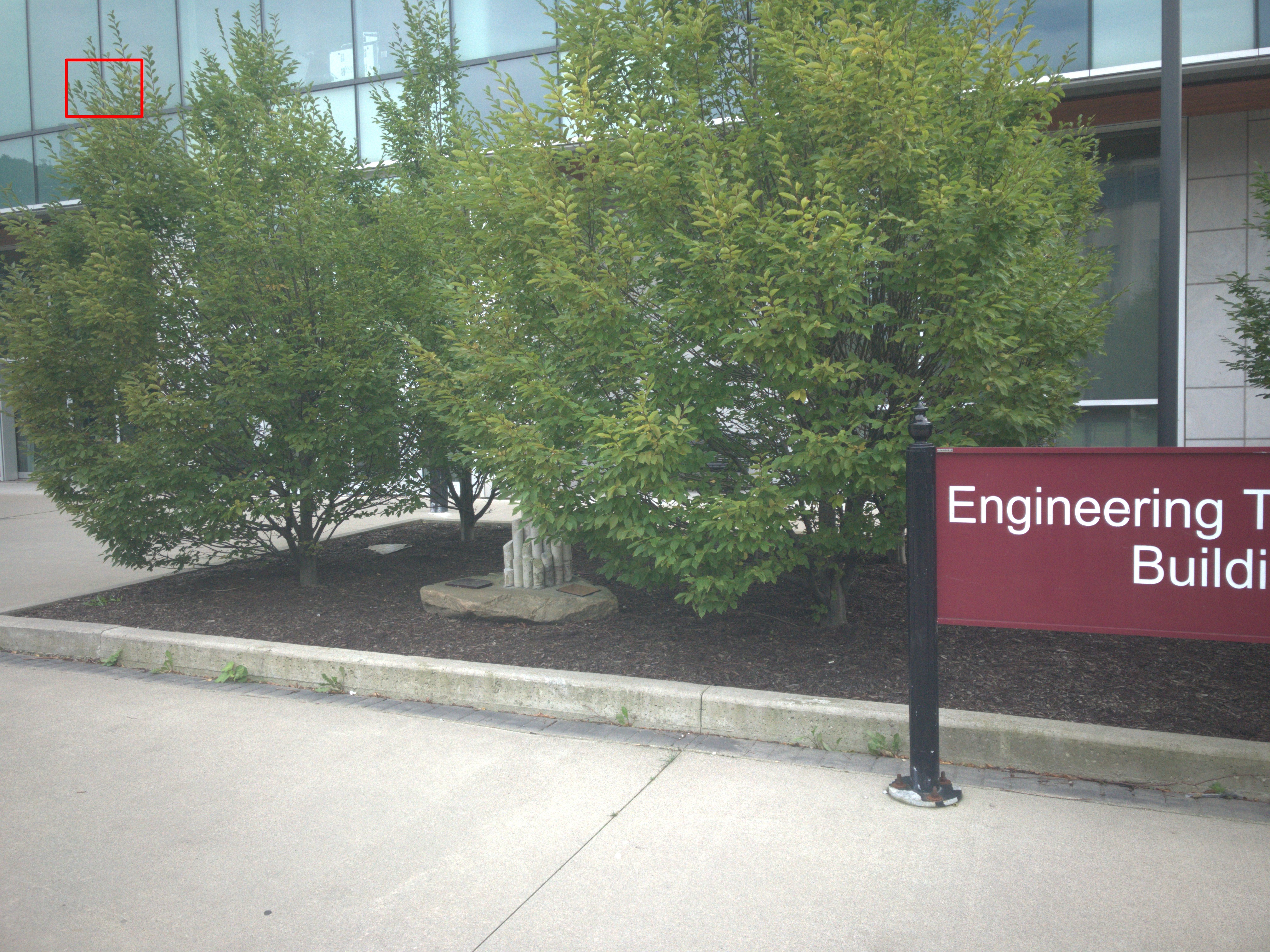}
	\end{minipage}
	\vspace{0.5mm}
	\begin{minipage}[h]{0.189\linewidth}
		\centering
		\includegraphics[width=\linewidth]{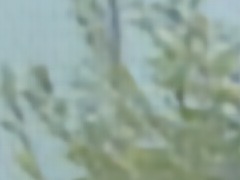}
	\end{minipage}
	\begin{minipage}[h]{0.189\linewidth}
		\centering
		\includegraphics[width=\linewidth]{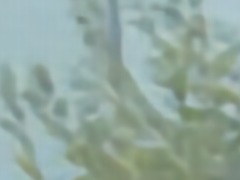}
	\end{minipage}
	\begin{minipage}[h]{0.189\linewidth}
		\centering
		\includegraphics[width=\linewidth]{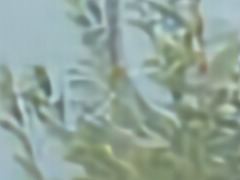}
	\end{minipage}
	\begin{minipage}[h]{0.189\linewidth}
		\centering
		\includegraphics[width=\linewidth]{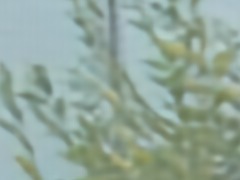}
	\end{minipage}
	\vspace{0.5mm}
	\begin{minipage}[h]{0.189\linewidth}
		\centering
		\includegraphics[width=\linewidth]{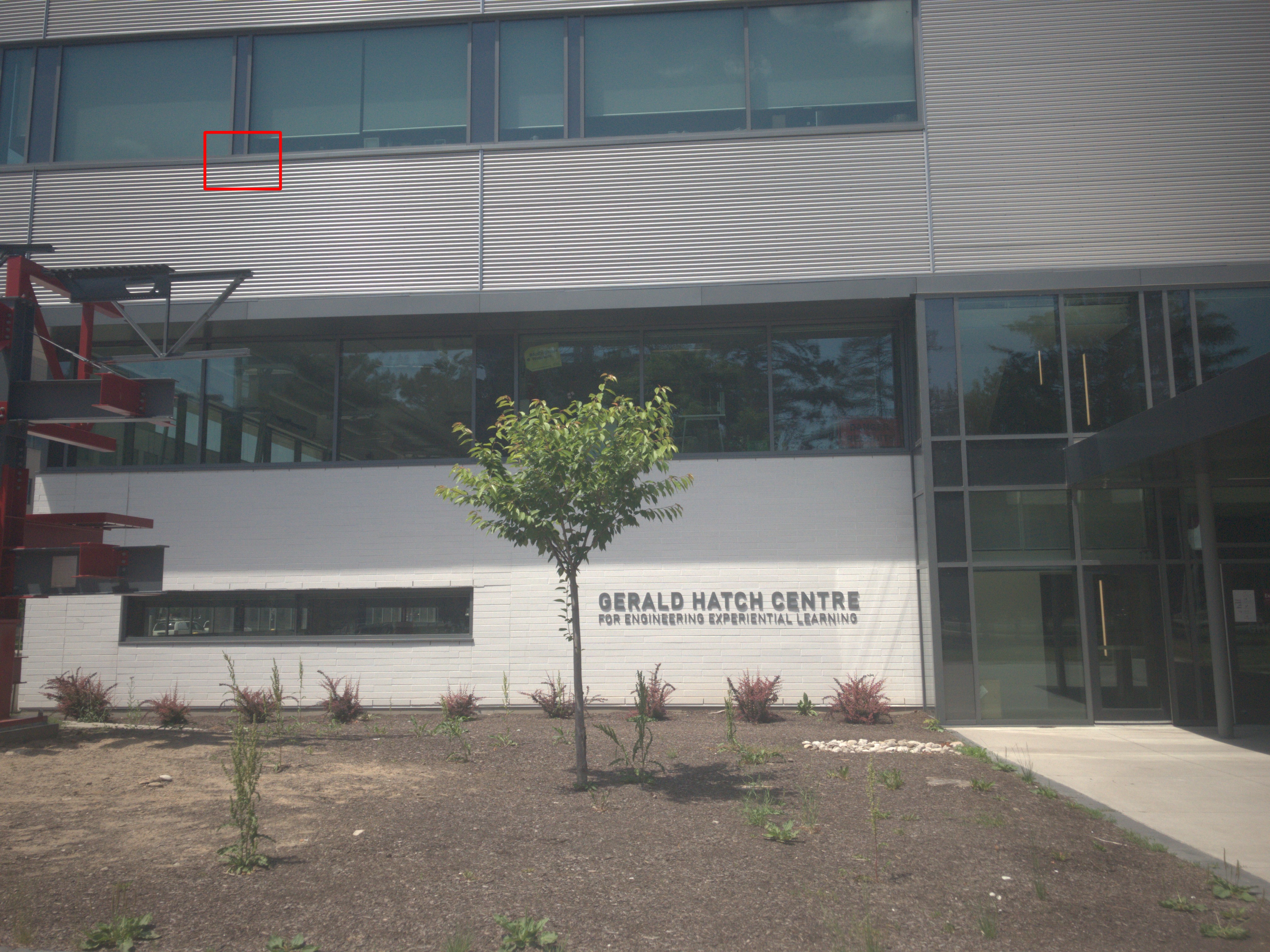}
	\end{minipage}
	\begin{minipage}[h]{0.189\linewidth}
		\centering
		\includegraphics[width=\linewidth]{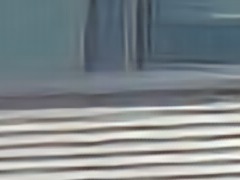}
	\end{minipage}	
	\begin{minipage}[h]{0.189\linewidth}
		\centering
		\includegraphics[width=\linewidth]{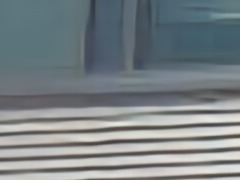}
	\end{minipage}
	\begin{minipage}[h]{0.189\linewidth}
		\centering
		\includegraphics[width=\linewidth]{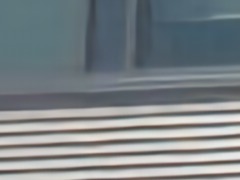}
	\end{minipage}
	\begin{minipage}[h]{0.189\linewidth}
		\centering
		\includegraphics[width=\linewidth]{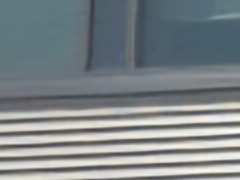}
	\end{minipage}
	\vspace{0.5mm}
	\begin{minipage}[h]{0.189\linewidth}
		\centering
		\includegraphics[width=\linewidth]{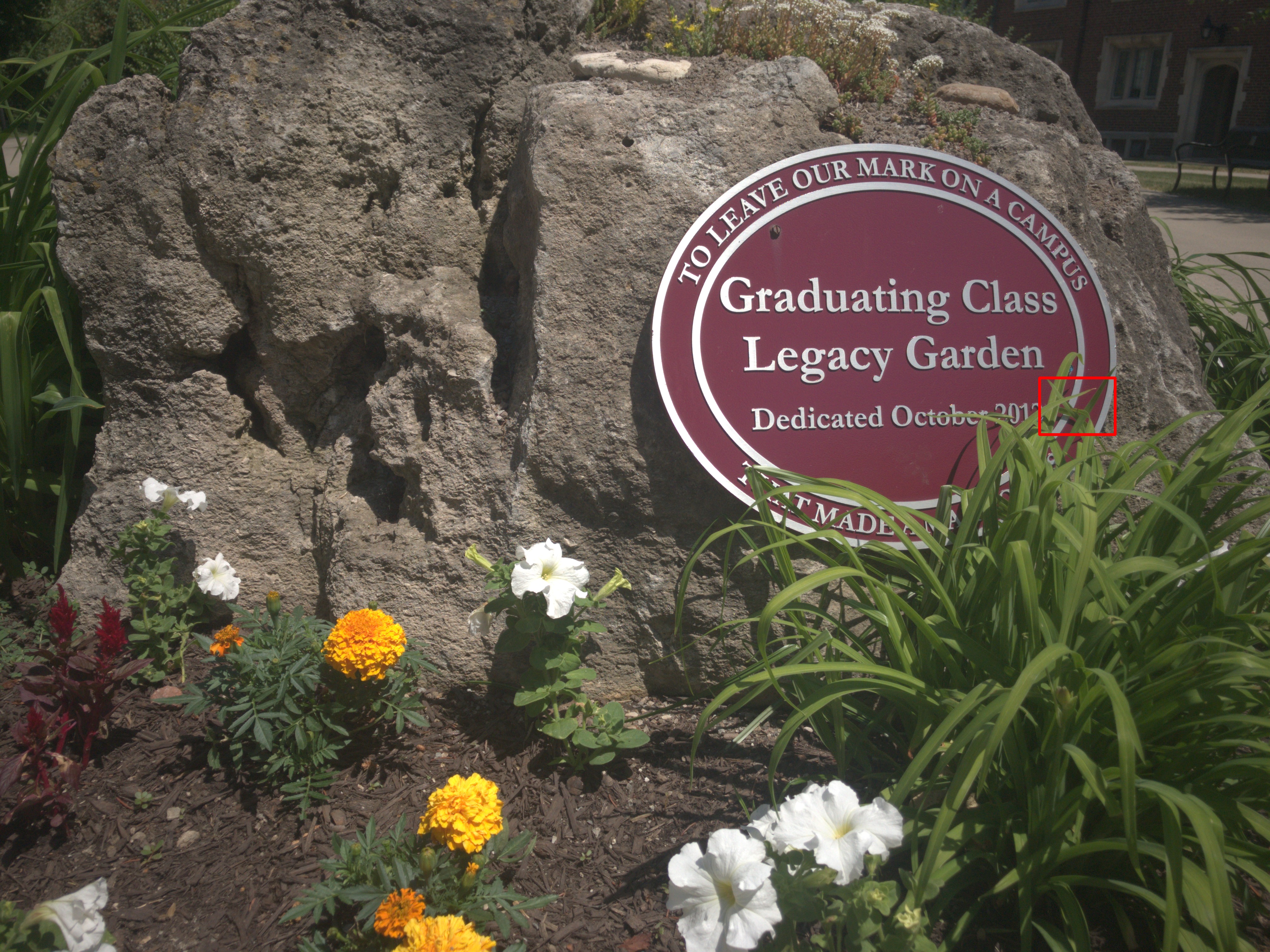}
		\scriptsize{(a) ground truth}
	\end{minipage}
	\begin{minipage}[h]{0.189\linewidth}
		\centering
		\includegraphics[width=\linewidth]{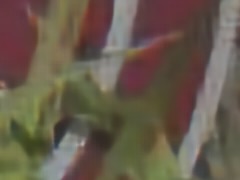}
		\scriptsize{(b) DUF~\cite{jo2018deep}}
		\end{minipage}
	\begin{minipage}[h]{0.189\linewidth}
		\centering
		\includegraphics[width=\linewidth]{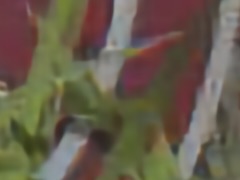}
		\scriptsize{(c) EDVR~\cite{wang2019edvr}}
		\end{minipage}
	\begin{minipage}[h]{0.189\linewidth}
		\centering
		\includegraphics[width=\linewidth]{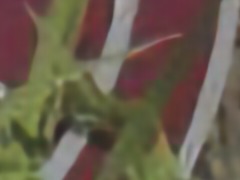}
		\scriptsize{(d) RawVSR$^\dagger$}
		\end{minipage}
	\begin{minipage}[h]{0.189\linewidth}
		\centering
		\includegraphics[width=\linewidth]{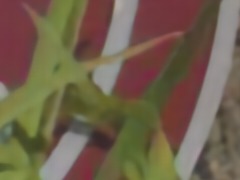}
		\scriptsize{(e) RawVSR}
		\end{minipage}
	\caption{Qualitative comparisons on the iPhone data for $4\times$ VSR. The names of video sequences from the first row to the last row are ``Tree'', ``Door'', and ``Flower'', respectively. Zoom in for better visualization.}
	\label{fig:iphone}
\end{figure}

\subsection{Implementation} \label{Implementation}

\begin{figure*}[t]
	\centering
	\includegraphics[width=\linewidth]{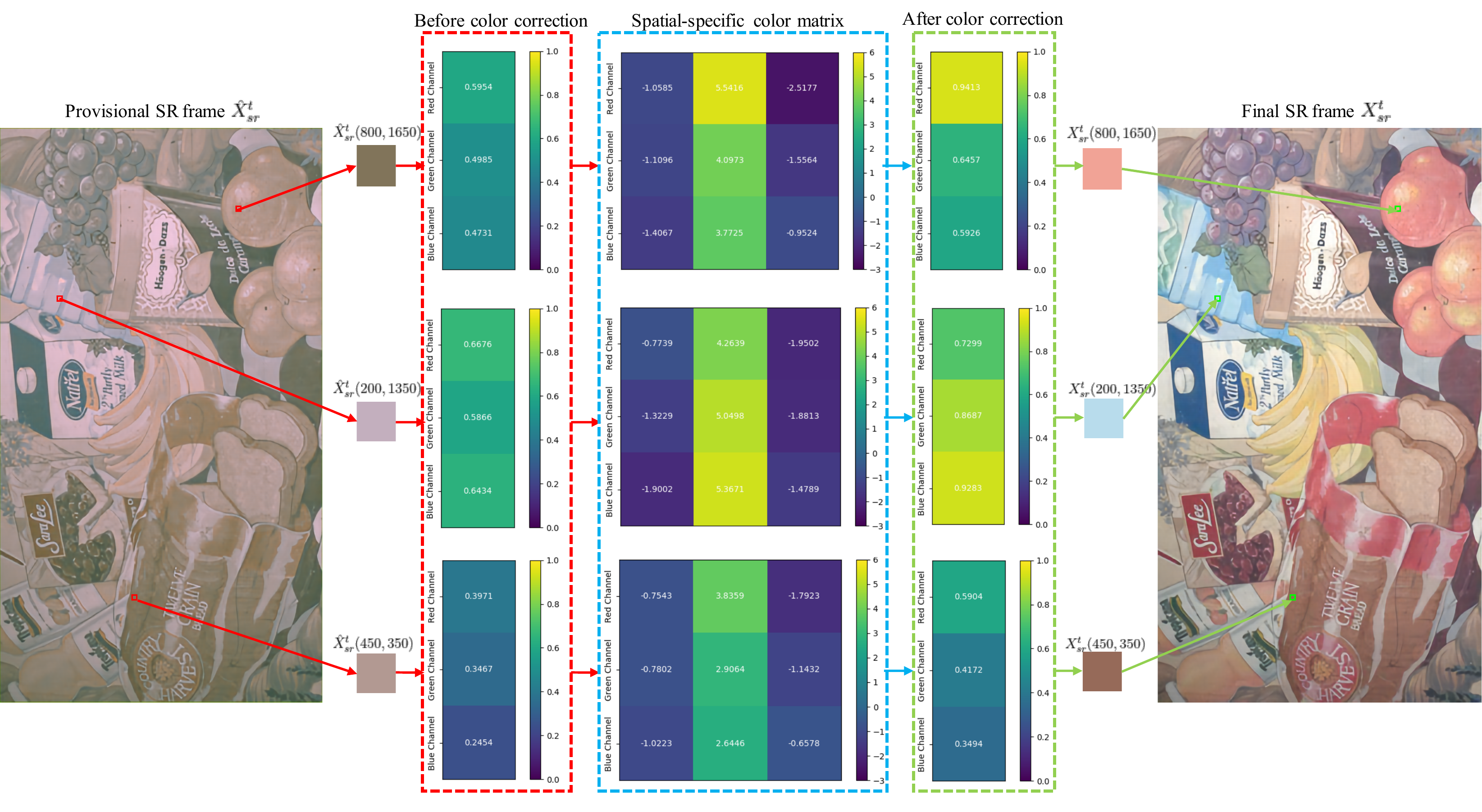}
	\caption{Visualization of the color correction process (zoom-in for details).}
	\label{fig:visualization}
\end{figure*}

The proposed RawVSR is end-to-end trainable without any additional supervision or pre-training for sub-modules. We adopt RawVD for training and testing as it is the only available video dataset for raw-data-based VSR methods. The scale ratio $S$ is set to $4$; the size of the defocus blur is randomly chosen from $\{3, 5, 7, 9, 11\}$; the two parameters $\sigma^2_1$ and $\sigma^2_2$ of heteroscedastic Gaussian noise are randomly sampled from $[0, 0.1]$ and $[0, 0.02]$, respectively. We use patches of size $64\times64$ cropped from LR raw videos as the training inputs, and augment the training data through inverting the frame order,  horizontal flips and rotations~\cite{timofte2016seven}. Following~\cite{lim2017enhanced}, batch normalization is not used in RawVSR. Without bells and whistles, we adopt a weighted version of  Mean Square Error (MSE) and Structural SIMilarity index (SSIM)~\cite{wang2004image} as our loss function to measure the difference between the final reconstructed SR frame and the ground truth. Furthermore, we provide an additional supervision to the provisional SR frame to enforce the learning of texture detail using an extra SSIM loss. The total loss function $L$ can be formulated as:
\begin{equation} \label{loss function}
L = L^{MSE}_F + \lambda_FL^{SSIM}_F + \lambda_PL^{SSIM}_P,  
\end{equation}
where $L^{MSE}_F$, $L^{SSIM}_F$, and $L^{SSIM}_P$ are respectively the MSE loss, the SSIM loss from the final SR frame, and the SSIM loss from the provisional SR frame. The trade-off parameters $\lambda_F$ and $\lambda_P$ are both set to $0.005$ to balance the value ranges of these sub-level losses. 
%Note that the SSIM loss is designed to measure the structural similarity between two images and  is relatively insensitive to color distortions (\textit{e.g.}, two images photographed in the same scene under different light conditions might have a high SSIM value but a low PSNR value), rendering it ideal for assessing the quality of the provisional SR frame.
Since the SSIM loss is designed to measure the structural similarity between two images and is relatively insensitive to color distortions, two images photographed in the same scene under different light conditions might have a high SSIM value but a low PSNR value. This fact inspires us to employ SSIM for assessing the quality of the provisional SR frame.
The textural consistency between the provisional SR frame and the ground truth then enables the spatial-specific transformation to focus on color correction.	In this way, the texture restoration branch and the color correction branch are effectively functionally disentangled. It will be seen in Section \ref{sec:visualization} that the additional supervision on the provisional SR frame improves the interpretability of our network and makes it more compatible with the real imaging pipeline, where the super-resolution process is carried out before color correction. We would point out that the linear measurement {$X_{lin}$} provided in RawVD is not employed as the supervision reference of the provisional SR frame. 
This is because the proposed RawVSR is intended to be device-independent, whereas the use of $X_{lin}$ might jeopardize its generalization ability due to the fact that $X_{lin}$ is produced by a specific ISP (which is Rawpy in the current setting).

 To accelerate network training, the Adam optimizer~\cite{kingma2014adam} is used with a batch size of $30$ and the default parameter values $\beta_1=0.9$  and $\beta_2=0.999$. We set the initial learning rate to $2e$-$4$, which is reduced by half every $20$ epochs until a total of $100$ epochs is reached. The training process is carried out on a PC with two NVIDIA GTX $1080$Ti. All PSNR/SSIM values are evaluated using the test data in RawVD.

\subsection{Quantitative and Qualitative Comparisons} 

	We compare the proposed RawVSR with several existing processed-data-based VSR methods, including  VSRNet~\cite{kappeler2016video}, VESPCN~\cite{caballero2017real}, SPMC~\cite{tao2017detail}, DUF~\cite{jo2018deep}, RBPN~\cite{haris2019recurrent}, TDAN~\cite{tian2018tdan}, and EDVR~\cite{wang2019edvr}. In particular, EDVR is the champion of the NTIRE 2019 challenge on video deblurring and super-resolution~\cite{nah2019ntire} and can be considered as the current state-of-the-art. For fair comparisons, we laboriously retrain all the aforementioned methods using the same strategy as described in Section~\ref{Implementation} until convergence.
	These methods are trained with ($Y_{rgb}$, $X_{rgb}$) pairs instead of ($Y_{raw}$, $X_{rgb}$) pairs (see Fig.~\ref{fig:RawVD}) since they are incapable of handling raw data. In contrast, the proposed RawVSR can leverage both raw data and processed data for training.  To distinguish it from the original version of RawVSR for which the training is based on ($Y_{raw}$, $X_{rgb}$) pairs, 	
	the version trained with
	($Y_{rgb}$, $X_{rgb}$) pairs is denoted as RawVSR$^\dagger$, where the color correction branch is removed since the color reference is not needed in this case. Quantitative comparisons on the RawVD test data for $2\times$ and $4\times$ VSR are shown in Table~\ref{tab:canon} (with the only exception of RBPN for which we are unable to implement the same training strategy on our PC for $2\times$ VSR  due to its large model size).
	For qualitative comparisons, we only demonstrate in Fig.~\ref{fig:canon} $4\times$ VSR since it is more visually distinguishable than $2\times$ VSR. It can be seen that RawVSR$^\dagger$ outperforms the other processed-data-based methods in terms of PSNR and SSIM metrics, and visually restores sharper edges and finer details in most cases (see, \textit{e.g.}, the word \textit{Country} in  video ``Painting'', and the balcony part of the building in video ``City''), which provides solid justifications for our overall network design.
		  As compared to RawVSR$^\dagger$, the results of RawVSR are even more appealing, both quantitatively and qualitatively. This improvement provides convincing evidence regarding the benefits of raw data since it is clearly attributed to their richer information content. 		  
		  In addition, to validate that the proposed method is device-independent, three raw $4$K videos (named ``Tree'', ``Door'', and ``Flower'') filmed by Adobe Lightroom software\footnote {https://www.adobe.com/ca/products/photoshop-lightroom.html} on an iPhone 8p are used to complement our RawVD test data for $4\times$ VSR. The quantitative and qualitative comparisons are illustrated in Table~\ref{tab:iphone} and Fig.~\ref{fig:iphone}, respectively. The superior VSR results indicate that the proposed RawVSR,  trained only on RawVD, continues to perform  well on the data collected by other devices without fine-tuning, and thus possesses good generalization abilities. More evidence can be found in Section \ref{sec:independence}.

\subsection{Visualization of the Color Correction Process} \label{sec:visualization}

To show that the texture restoration branch and the color correction branch indeed fulfill their designated roles,  Fig.~\ref{fig:visualization} visualizes how  a provisional SR frame $\hat{X}_{sr}^t$ undergoes the color correction process to produce the corresponding final SR frame $X_{sr}^t$. In particular, we highlight three representative pixels (\textit{i.e.}, $\hat{X}_{sr}^t(800, 1650)$, $\hat{X}_{sr}^t(200, 1350)$, and $\hat{X}_{sr}^t(450, 350)$) extracted from a provisional SR frame in video ``Painting" and their 
color-corrected counterparts (\textit{i.e.}, $X_{sr}^t(800, 1650)$, $X_{sr}^t(200, 1350)$, and $X_{sr}^t(450, 350)$) together with the associated 
spatial-specific color transformation matrices.  Clearly, although the provisional SR frame $\hat{X}_{sr}^t$ suffers severe color distortion, it exhibits correct details and is texturally consistent with the final SR frame $X_{sr}^t$. This validates the effectiveness of additional SSIM-based supervision in guiding the texture restoration branch to accomplish its desired purpose.
The color-corrected pixels are produced by applying the spatial-specific transformation matrices on their respective  provisional pixels according to Equ.~(\ref{color_matrix}). It is worth noting that even though the pixels extracted from $\hat{X}_{sr}^t$ are color-wise quite similar, their associated color transformation matrices are distinctively different, resulting in visually more distinguishable color-corrected pixels. This provides strong supporting evidence for the usefulness of the color correction branch.

\begin{figure}[htbp]
	\centering
	
	%     \vspace{0.5cm}
	%	\begin{minipage}[h]{0.24\linewidth}
	%		\centering
	%		\includegraphics[width=\linewidth]{ablation/store/store_full_gt.jpg}
	%		\scriptsize{  }
	%	\end{minipage}	
	\begin{minipage}[h]{0.3\linewidth}
		\centering
		\includegraphics[width=\linewidth]{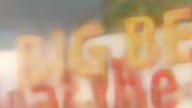}
	\end{minipage}
	\begin{minipage}[h]{0.3\linewidth}
		\centering
		\includegraphics[width=\linewidth]{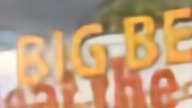}
	\end{minipage}
	\vspace{1mm}
	\begin{minipage}[h]{0.3\linewidth}
		\centering
		\includegraphics[width=\linewidth]{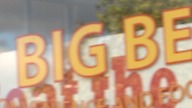}
	\end{minipage}
	%	\begin{minipage}[h]{0.24\linewidth}
	%		\centering
	%		\includegraphics[width=\linewidth]{Supplementary/raw_rgb_x4/00023_000030_gt.jpg}
	%		\scriptsize{ }
	%	\end{minipage}
	\begin{minipage}[h]{0.3\linewidth}
		\centering
		\includegraphics[width=\linewidth]{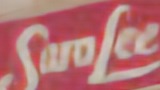}
	\end{minipage}
	\begin{minipage}[h]{0.3\linewidth}
		\centering
		\includegraphics[width=\linewidth]{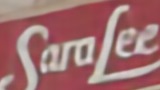}
	\end{minipage}
	\vspace{1mm}
	\begin{minipage}[h]{0.3\linewidth}
		\centering
		\includegraphics[width=\linewidth]{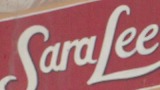}
	\end{minipage}
	%     \vspace{0.5cm}
	%	\begin{minipage}[h]{0.24\linewidth}
	%		\centering
	%		\includegraphics[width=\linewidth]{Supplementary/raw_rgb_x4/00042_000060_gt.jpg}
	%		\scriptsize{ }
	%	\end{minipage}
	\begin{minipage}[h]{0.3\linewidth}
		\centering
		\includegraphics[width=\linewidth]{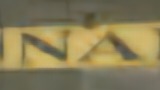}
		\scriptsize{(a) w/ ISP-inverted data}
	\end{minipage}
	\begin{minipage}[h]{0.3\linewidth}
		\centering
		\includegraphics[width=\linewidth]{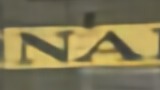}
		\scriptsize{(b) w/ authentic raw data}
	\end{minipage}
	\begin{minipage}[h]{0.3\linewidth}
		\centering
		\includegraphics[width=\linewidth]{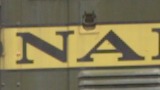}
		\scriptsize{(c) ground truth}
	\end{minipage}
	\caption{The proposed RawVSR with ISP-inverted data and authentic raw data respectively on the RawVD test dataset for $4\times$ VSR.}
	\label{fig:rawvsfake}
\end{figure}

\subsection{Authentic Raw Data v.s. ISP-Inverted Data} \label{sec:fakeraw}

Several recent works \cite{brooks2019unprocessing, zhou2018deep} have attempted to
 invert the camera ISP to produce raw-like data (which will be referred to as ISP-inverted data) from processed data and use them in lieu of authentic raw data for training. However, the resemblance between ISP-inverted data and raw data 
should not be overestimated. Indeed, such inversion  can never be perfect due to the presence of non-invertible ISP components (\textit{e.g.}, quantization); actually it may even cause additional information loss. To gain a concrete understanding, we follow the same procedure carried out in \cite{zhou2018deep} to produce ISP-inverted data from processed data $Y_{rgb}$ in RawVD, and input them to our network for testing. The results are then compared with those based on authentic raw data as shown in Fig.~\ref{fig:rawvsfake}. It can be seen that ISP-inverted data incur two evident degradations: 1) significant color distortion (which is likely caused by the distribution mismatch between ISP-inverted data and raw data), 2) blurred texture detail (which is a consequence of information loss). Therefore, authentic raw data as those collected in RawVD remain indispensable for raw VSR.

\begin{table*}[t]
	\begin{center}
		\caption{Ablation studies for different variants for $4\times$ VSR. \textbf{Bold} indicates the best performance.}
		\label{tab:ablation}
		\begin{adjustbox}{width=\textwidth}
			\begin{tabular}{|c||c|c|c||c|c||c|c||c|}
				
				\hline
				Ablation& \multicolumn{3}{c||}{SDI} & \multicolumn{2}{c||}{Reconstruction module} &\multicolumn{2}{c||}{Input frame number} & Baseline \\
				
				\hline
				Variant & w/ EF & w/ improved EF &w/ SF & w/o attention & w/ RDB~\cite{zhang2018residual} &w/ 3 frames &w/ 5 frames & RawVSR \\ % <--
				\hline \hline
				% Name
				Store & 
				% w/o align
				28.00/0.8226 & 
				% w/ direct align
				28.61/0.8304 & 
				% w/ intermediate align
				28.82/0.8351 &
				% w/o attention 29.17/0.8439
				28.52/0.8303 & 
				% w/ resnet 28.78/0.8354
				28.48/0.8277 &
				% w/ 3 frames
				28.30/0.8176 &
				% w/ 5 frames
				28.75/0.8324 &
				% proposed
				\textbf{29.04/0.8400} \\
				
				%Painting
				% Name
				Painting & 
				% w/o align
				28.05/0.7910 & 
				% w/ direct align
				28.71/0.8007 & 
				% w/ intermediate align
				28.91/0.8034 &
				% w/o attention 28.96/0.8165
				28.16/0.8006 & 
				% w/ resnet 28.52/0.8055
				28.00/0.7981 &
				% w/ 3 frames
				28.36/0.7927 &
				% w/ 5 frames
				28.59/0.8057 &
				% proposed
				\textbf{29.02/0.8104} \\
				
				%Train
				% Name
				Train & 
				% w/o align
				27.44/0.7481 & 
				% w/ direct align
				28.08/0.7565 &
				% w/ intermediate align
				28.29/0.7556 & 
				% w/o attention 28.54/0.7682
				28.07/0.7568 & 
				% w/ resnet 28.26/0.7614
				28.08/0.7555 &
				% w/ 3 frames
				28.01/0.7481 &
				% w/ 5 frames
				28.27/0.7554 &
				% proposed
				\textbf{28.59/0.7625} \\
				
				%City
				% Name
				City & 
				% w/o align
				28.44/0.7699 & 
				% w/ direct align
				28.84/0.7769 &
				% w/ intermediate align
				28.94/0.7775 & 
				% w/o attention 29.10/0.7872
				28.73/0.7750 & 
				% w/ resnet 28.96/0.7813
				28.74/0.7743 &
				% w/ 3 frames
				28.64/0.7684 &
				% w/ 5 frames
				28.96/0.7776 &
				% proposed
				\textbf{29.08/0.7843} \\

				%Tree
				% Name
				Walk & 
				% w/o align
				26.26/0.7525 & 
				% w/ direct align
				27.78/0.7649 &
				% w/ intermediate align
				27.87/0.7628 & 
				% w/o attention 27.27/0.7685
				26.69/0.7559 & 
				% w/ resnet 27.15/0.7628
				27.10/0.7567 &
				% w/ 3 frames
				27.95/0.7558 &
				% w/ 5 frames
				28.04/0.7695 &
				% proposed
				\textbf{28.06/0.7724}\\
				
				\hline\hline
				
				%Average
				% Name
				Average & 
				% w/o align
				27.64/0.7768 & 
				% w/ direct align
				28.40/0.7859 &
				% w/ intermediate align
				28.57/0.7869 & 
				% w/o attention 28.61/0.7968
				28.03/0.7837 & 
				% w/ resnet 28.33/0.7893
				28.08/0.7825 &
				% w/ 3 frames
				28.25/0.7765 &
				% w/ 5 frames
				28.52/0.7881 &
				% proposed
				\textbf{28.76/0.7939} \\
				\hline
			\end{tabular}
		\end{adjustbox}
	\end{center}
\end{table*}

\subsection{Validation of Device Independence} \label{sec:independence}
Due to the separation of the super-resolution process and the color correction process, the proposed RawVSR can adapt easily to different camera-ISPs. We have demonstrated this desirable characteristic in Table~\ref{tab:canon} and Fig.~\ref{fig:iphone} by blindly testing RawVSR, trained with data from simulated Canon-ISP (\textit{i.e.}, Rawpy), on a new iPhone dataset. Here we shall provide more evidence. Due to the lack of 
 Digital Single-Lens Reflex (DSLR) cameras (other than Canon 5D3) at our disposal, we choose to generate LR color references by simulating ISPs of three major camera manufacturers (Nikon, Sony, and Fujifilm) based on the analysis of their color styles and preferences. According to our observation, the video frames produced by Nikon-ISP exhibit the lowest color temperature as compared to the other two, which leads to an overall cold tone; in contrast, those by Fujifilm-ISP show the highest color temperature, yielding a warm tone; Sony-ISP is somewhere in between in terms of color temperature, yet produces the highest luminance. Based on these insights, we can closely approximate the effects of these ISPs by suitably adjusting the exposure degree, contrast, and color temperature of LR color references in RawVD. 
 
 \begin{figure}[t]
 	\centering
 	\begin{minipage}[h]{0.325\linewidth}
 		\centering
 		\includegraphics[width=\linewidth]{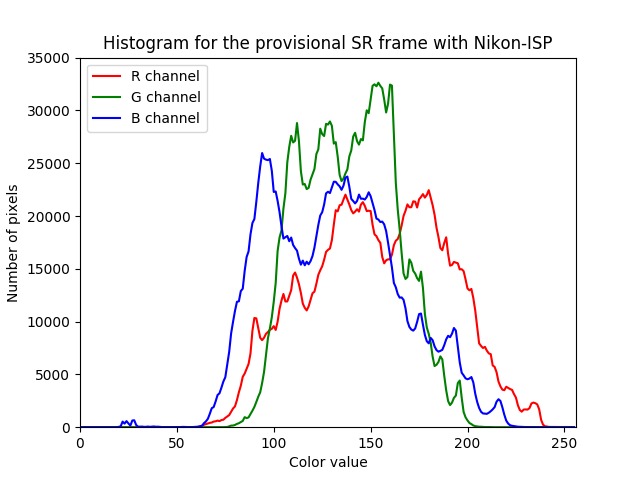}
 	\end{minipage}
 	\begin{minipage}[h]{0.325\linewidth}
 		\centering
 		\includegraphics[width=\linewidth]{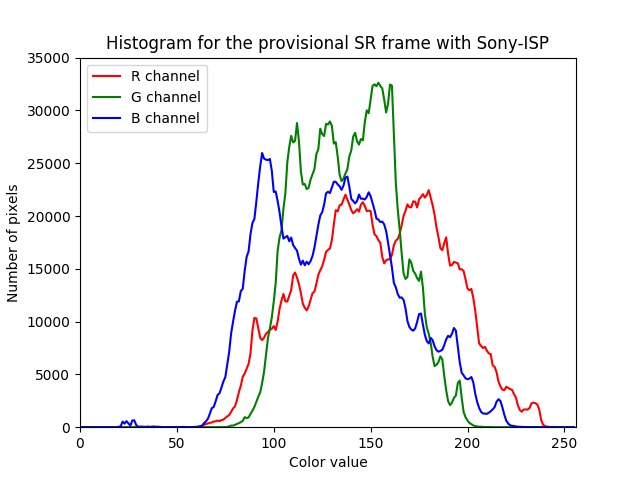}
 	\end{minipage}
 	\begin{minipage}[h]{0.325\linewidth}
 		\centering
 		\includegraphics[width=\linewidth]{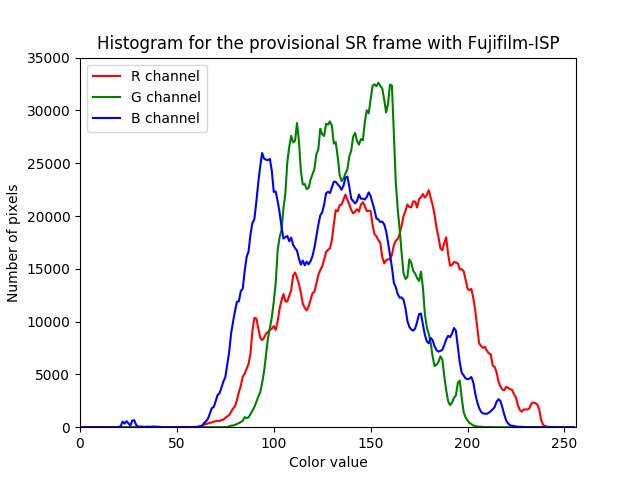}
 	\end{minipage}
 	\vspace{1mm}
 	\begin{minipage}[h]{0.325\linewidth}
 		\centering
 		\includegraphics[width=\linewidth]{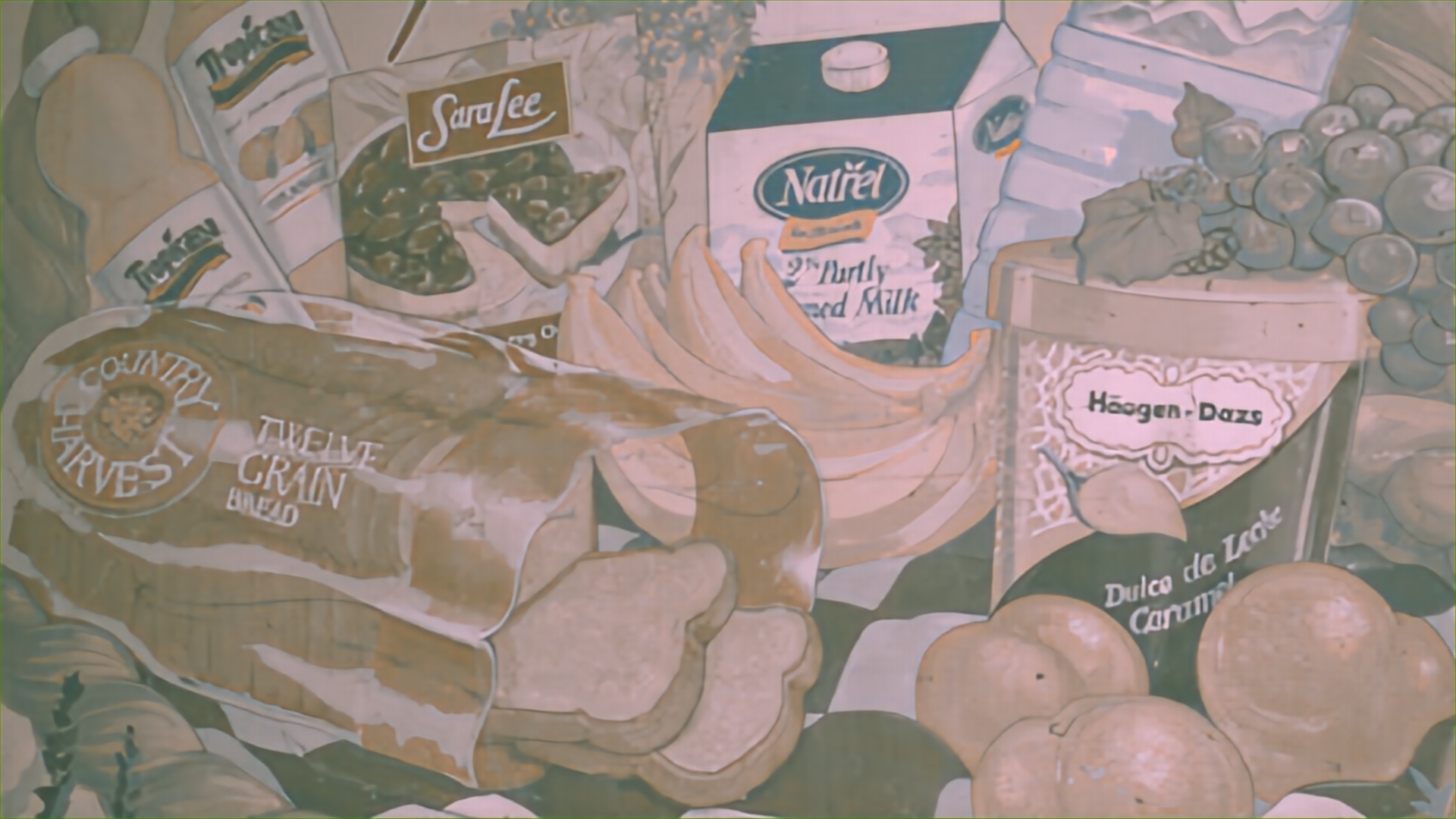}
 	\end{minipage}
 	\begin{minipage}[h]{0.325\linewidth}
 		\centering
 		\includegraphics[width=\linewidth]{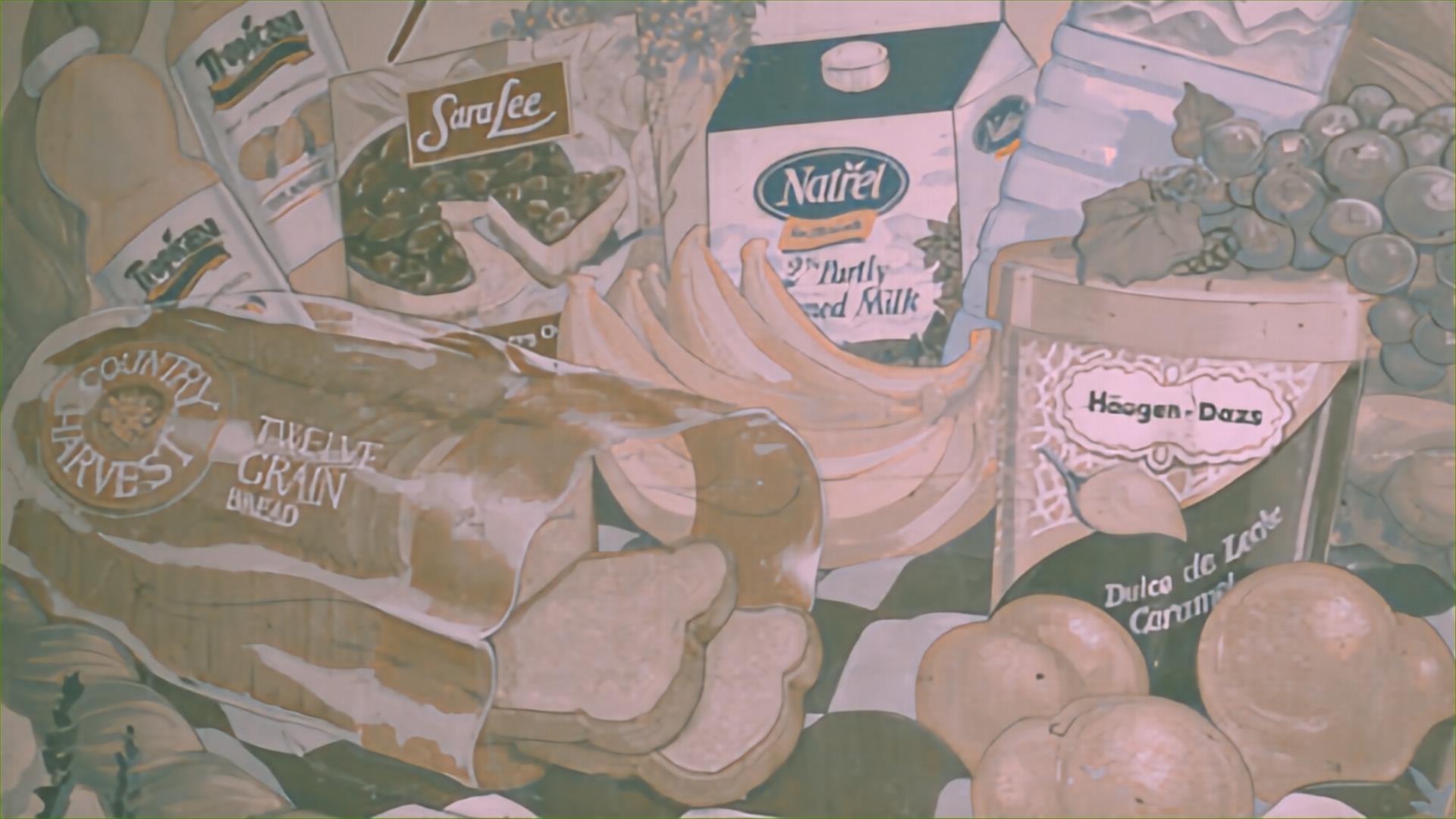}
 	\end{minipage}
 	\begin{minipage}[h]{0.325\linewidth}
 		\centering
 		\includegraphics[width=\linewidth]{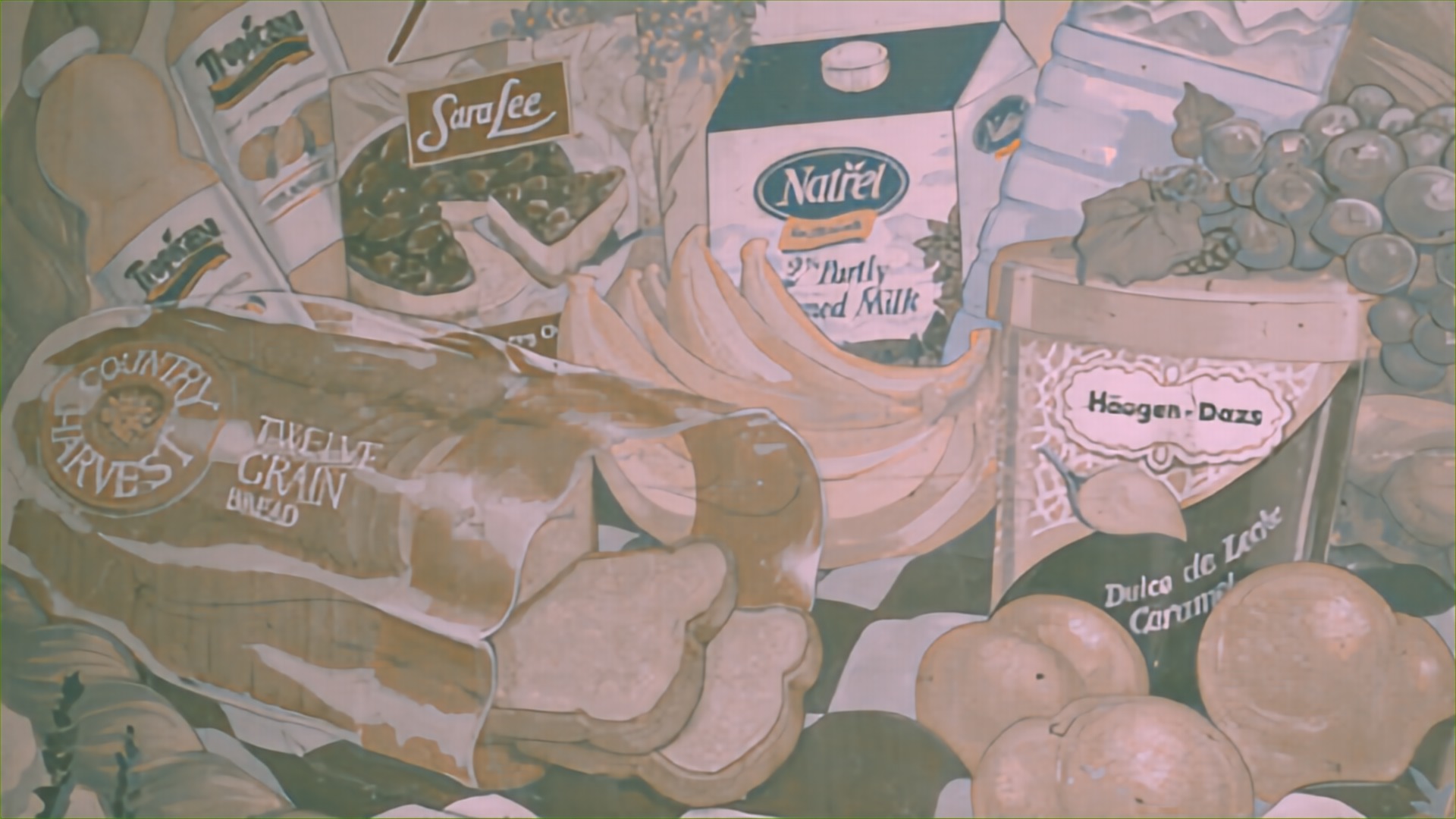}
 	\end{minipage}
 	\begin{minipage}[h]{0.325\linewidth}
 		\centering
 		\includegraphics[width=\linewidth]{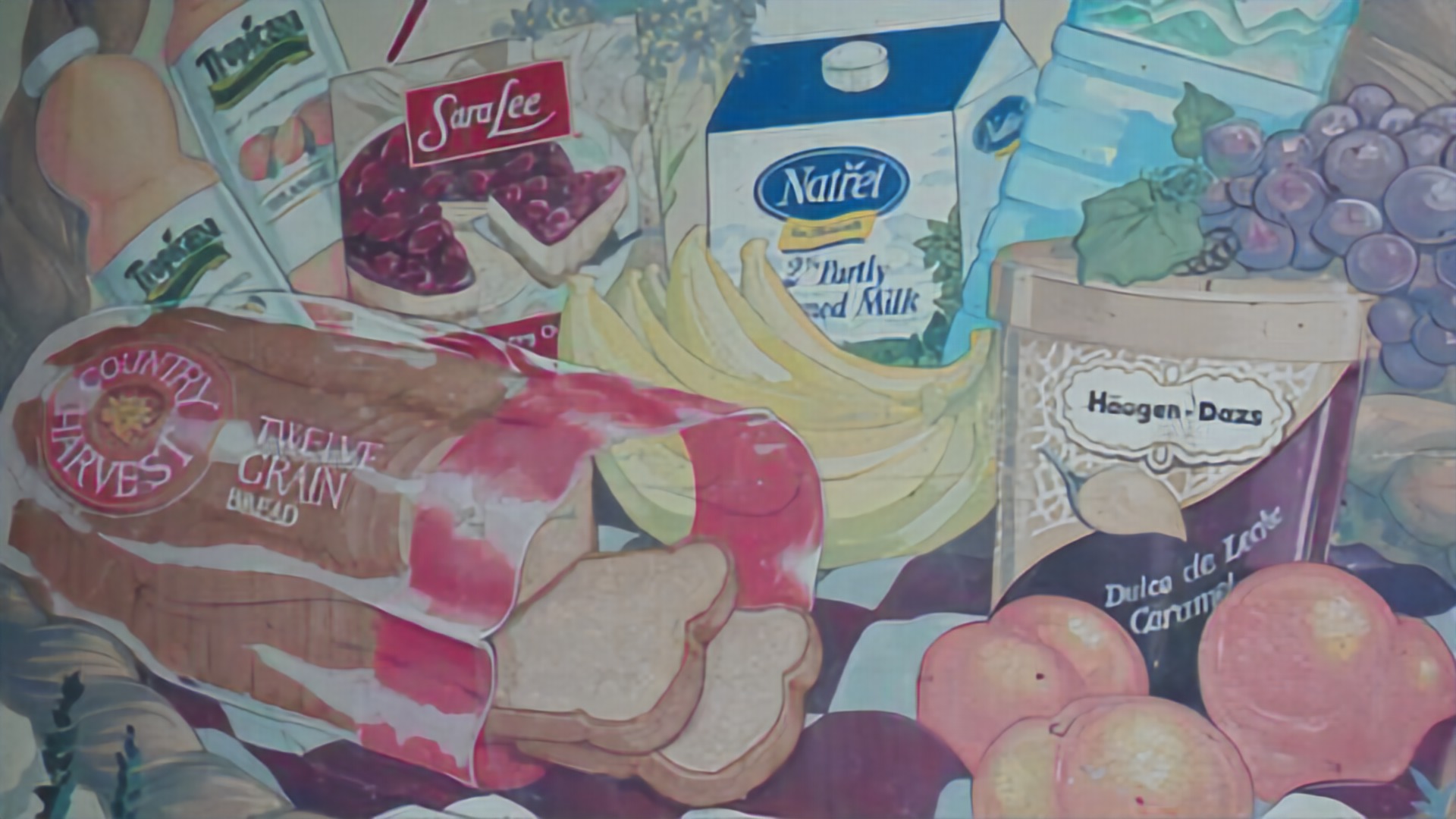}
 	\end{minipage}
 	\begin{minipage}[h]{0.325\linewidth}
 		\centering
 		\includegraphics[width=\linewidth]{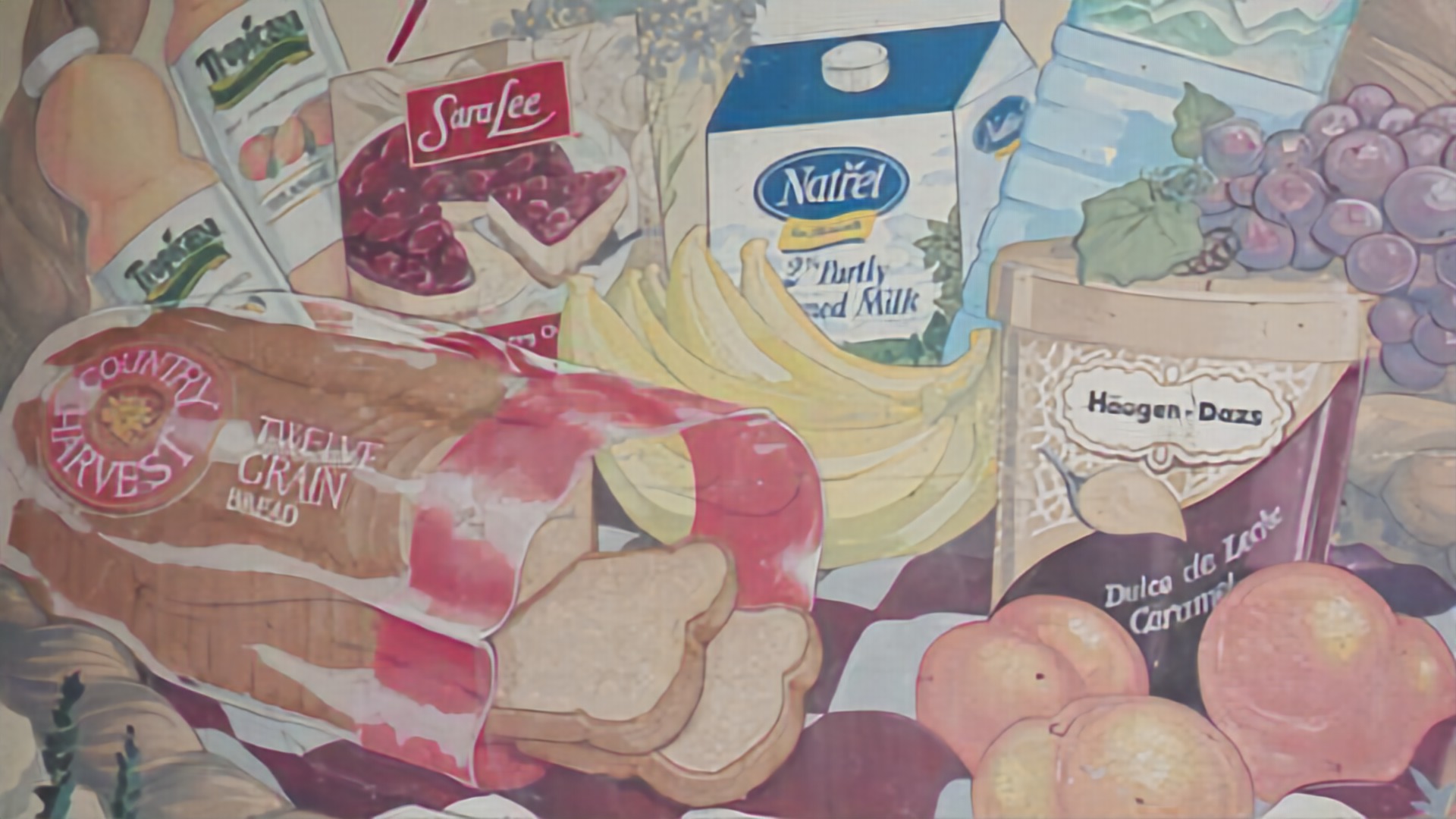}
 	\end{minipage}
 	\begin{minipage}[h]{0.325\linewidth}
 		\centering
 		\includegraphics[width=\linewidth]{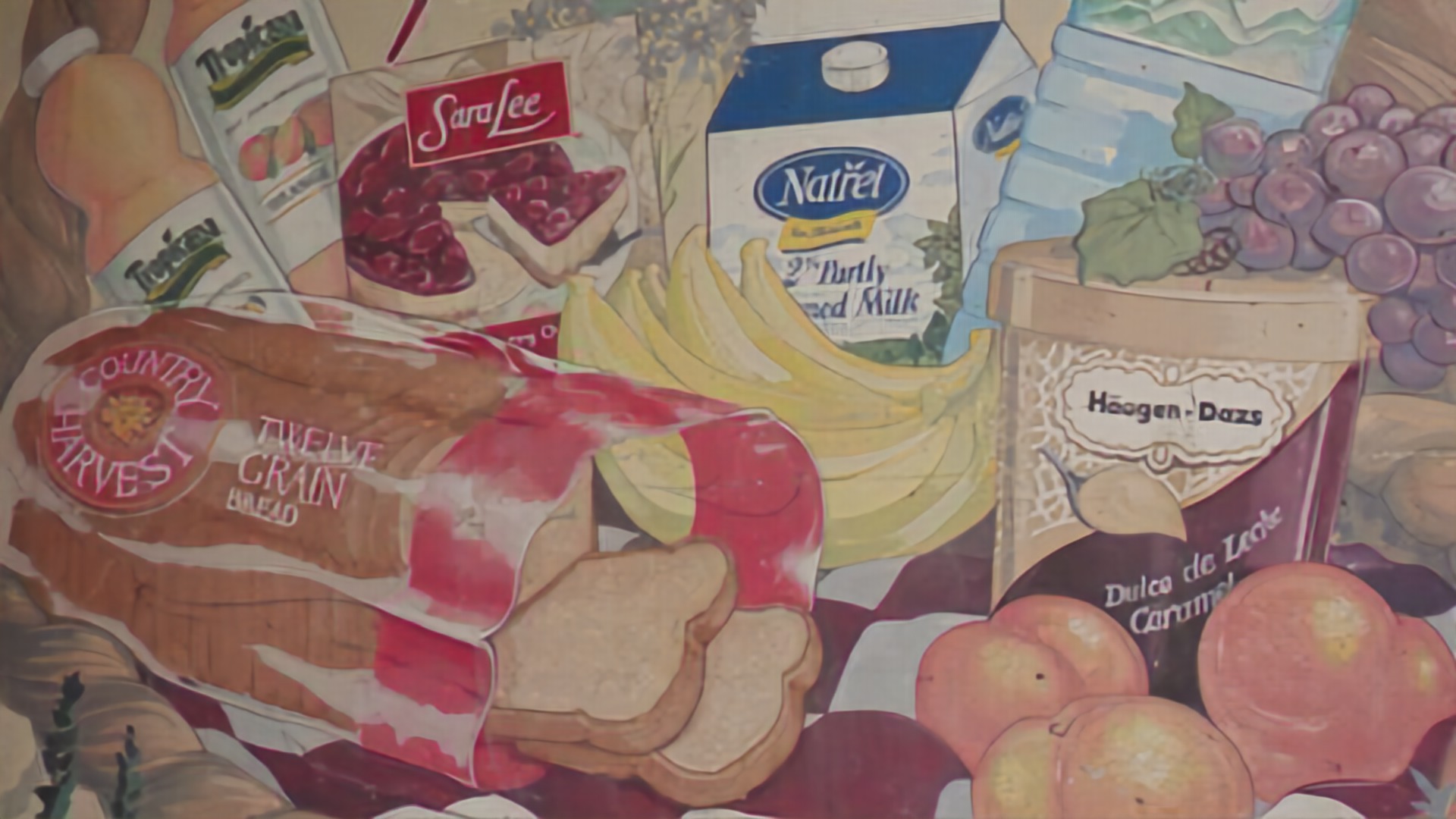}
 	\end{minipage}
 	\begin{minipage}[h]{0.325\linewidth}
 		\centering
 		\includegraphics[width=\linewidth]{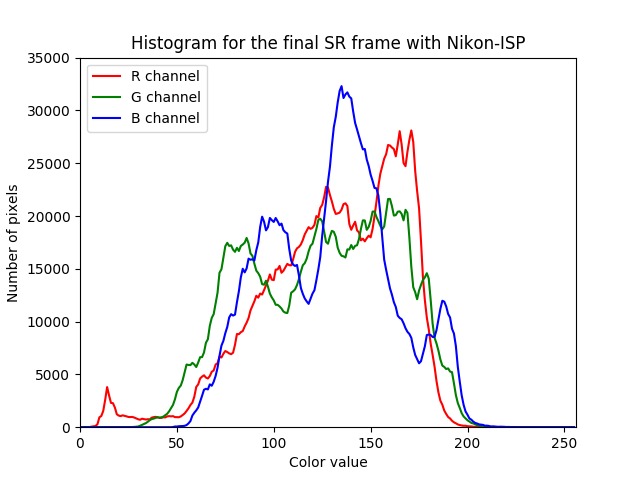}
 		\scriptsize{(a) w/ Nikon-ISP}
 	\end{minipage}
 	\begin{minipage}[h]{0.325\linewidth}
 		\centering
 		\includegraphics[width=\linewidth]{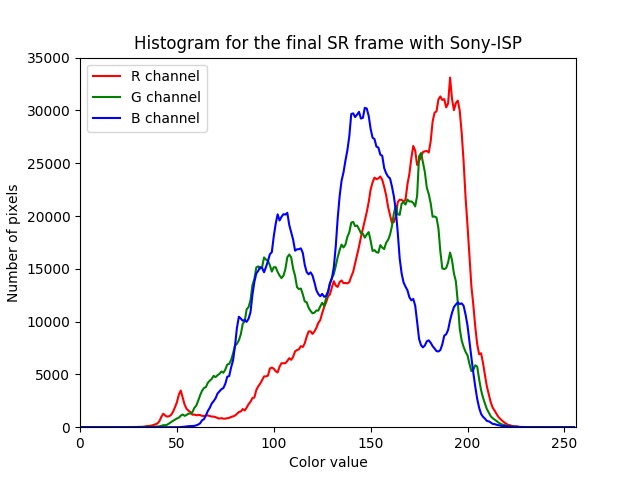}
 		\scriptsize{(b) w/ Sony-ISP}
 	\end{minipage}
 	\begin{minipage}[h]{0.325\linewidth}
 		\centering
 		\includegraphics[width=\linewidth]{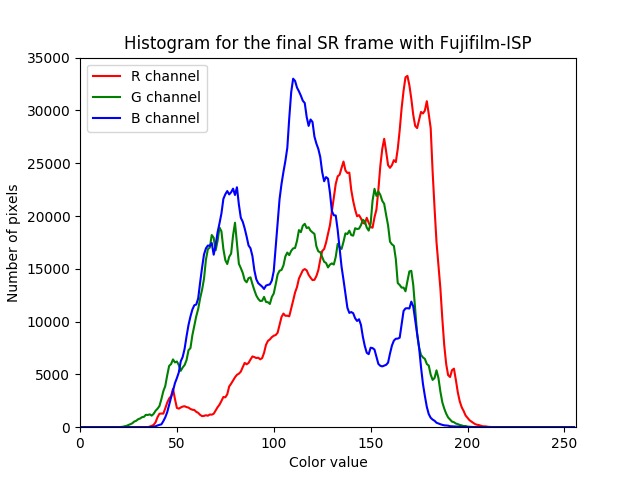}
 		\scriptsize{(c) w/ Fujifilm-ISP}
 	\end{minipage}
 	\caption{Visualization of the provisional frames (in the 2nd row), the final SR frames (in the 3rd row), and their associated color histograms (in the 1st and 4th rows) with simulated Nikon, Sony and Fujifilm-ISPs, to validate the device-independence characteristic of RawVSR.}
 	\label{fig:differentcolor}
 \end{figure}
 
 Specifically, we increase the color temperature from $6000$K to $8000$K to approximate the warm tone of Fujifilm-ISP, and decrease it to $5200$K to approximate the cold tone of Nikon-ISP. Since the Sony-ISP has a unique preference, simply adjusting one color setting does not yield good approximation. Instead, we increase the exposure degree from the default value $0$ to $0.7$, decrease the contrast from $0$ to $-8$, and change the color temperature to $6500$K in order to achieve the effect similar to that of Sony-ISP.
  
 Fig.~\ref{fig:differentcolor} shows the final SR frames  with simulated Nikon, Sony, and Fujifilm-ISPs, respectively, and the corresponding provisional SR frames as well as their associated color histograms for a representative frame in video ``Painting''.  
 Here, the input to the texture restoration branch is fixed, and different color references (based on simulated ISPs) are fed into the color correction branch without additional training. 
 Note that the choice of simulated ISP has no impact on the provisional frame (see the 2nd row in Fig.~\ref{fig:differentcolor}) and its associated color histogram (see the 1st row in Fig.~\ref{fig:differentcolor}). As such, the texture restoration branch and the color correction branch are indeed functionally disentangled, which is beneficial to the interpretability of overall network design.
 On the other hand, the color of the final SR frame changes in accordance with the provided color reference (see the 3rd row in Fig.~\ref{fig:differentcolor}), which is also reflected in the color histograms  (see the 4th row in Fig.~\ref{fig:differentcolor}). This shows convincingly that the proposed RawVSR can automatically adapt to the selected ISP, thus is device-independent.

\subsection{Validation of Temporal Consistency}
\begin{figure}[htbp]
	\centering
	\includegraphics[width=0.75\linewidth]{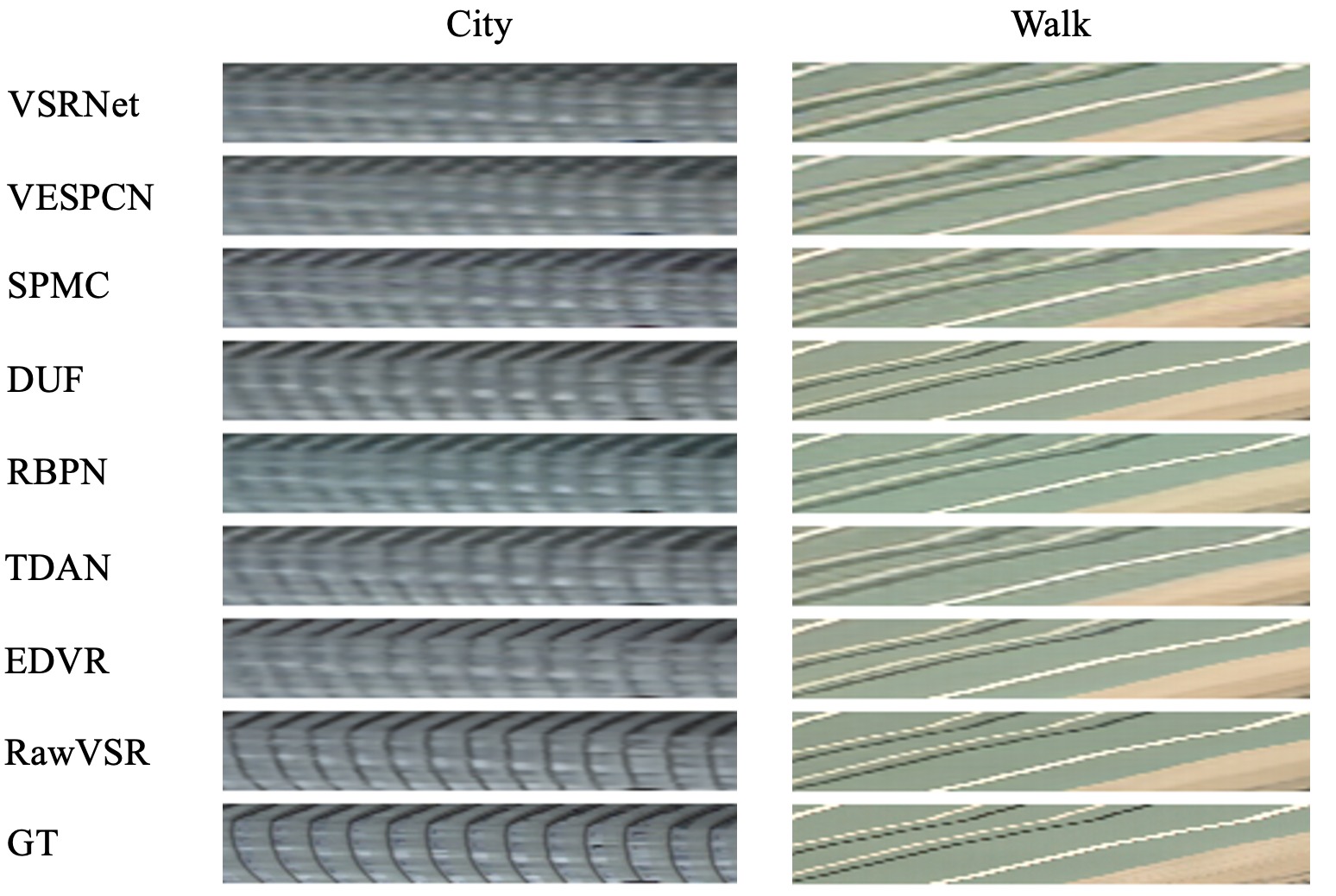}
	\caption{Visualization of temporal profiles produced by the proposed RawVSR and some existing methods.}
	\label{fig:temporal_consistency}
\end{figure}

A lack of temporal consistency may lead to flickering artifacts (\textit{e.g.}, in the form of jagged edges) that can degrade the quality of reconstructed SR frames~\cite{liu2020end}. To compare the proposed RawVSR with the existing methods from this perspective, we extract some co-located columns from the reconstructed ``City'' sequence, rotate and then stack them vertically   to produce a temporal profile \cite{sajjadi2018frame}; we also generate another temporal profile based on co-located rows from the reconstructed  ``Walk'' sequence. It can be seen from Fig.~\ref{fig:temporal_consistency} that RawVSR yields the most consistent temporal profiles in terms of texture clarity, edge sharpness, and detail accuracy. For example, in the temporal profile of the ``Walk'' sequence, the proposed RawVSR successfully recovers most line patterns whereas the other methods fail to distinguish dashed lines from solid ones.

%as well as some co-located rows from the reconstructed ``Walk'' sequence 

%the co-located column and row of the reconstructed ``City'' and ``walk'' sequences are extracted from the RawVSR and compared methods respectively, and then stacked vertically to depict a temporal profile \cite{sajjadi2018frame}.  It can be seen from Fig.~\ref{fig:temporal_consistency} that RawVSR yields the most consistent temporal profiles. More importantly, as demonstrated in temporal profiles, the RawVSR restores the favorable texture, sharpest edges, and finest details. For example, in ``walk'' frames, all of the compared methods is not capable of distinguishing dash lines from solid ones, while the proposed RawVSR successfully recovers most of dash patterns.

\subsection{Ablation Studies}

To gain a better understanding of key constituents  of the proposed network as well as the influence of the input sequence length, we conduct elaborately designed ablation studies by examining various alternative implementations. All such implementations are trained from scratch with the same training strategy described in  Section~\ref{Implementation}. The proposed RawVSR (trained on raw data) will be referred to as the baseline.

%Here, we implement several variants to conduct ablation studies, including the validation of SDI and reconstruction modules, and the influence of input frame number. All alternative implementations are trained from scratch with the same training strategy described in  Section~\ref{Implementation}. The proposed RawVSR (trained on raw data) will be referred to as the baseline.

% and the RawVSR (the one trained on raw data) is determined as the baseline.

\subsubsection{Justification of the SDI Module} Most existing VSR methods adopt early fusion, slow fusion, or 3D convolutional fusion to merge input frames  \cite{caballero2017real}. For early fusion, all input frames are concatenated at the very beginning and fed jointly into the network.  As an example, VSRNet \cite{kappeler2016video} follows this fusion strategy. In contrast, the slow fusion strategy merges input frames progressively according to a prescribed rule.
The scheme of the proposed SDI module falls into this general category; one of our main contributions is, in a certain sense, an optimal fusion rule deduced via HMM analysis. 3D convolutional fusion performs convolution on input frames in both temporal and spatial domains, and tends to be computationally expensive. To justify the design of the SDI module, three variants are considered, each associated with a different fusion scheme. The first variant, named w/ EF, simply follows the early fusion strategy. The second variant, named w/ improved EF, is similar to the first one except that all other frames are first aligned to the middle frame. The last one, named w/ SF, adopts a slow fusion scheme, 
 where all other frames are first aligned and fused with the middle frame in a pairwise manner before joint fusion is performed. For fair comparisons, these alternative fusion modules are designed to have roughly the same size as that of the SDI module (no structural change is made to the rest of the network).  Note that 
 3D convolutional fusion can not be implemented under this size constraint as it requires significantly more parameters. The comparison results are shown in Table~\ref{tab:ablation}. It can be seen that the baseline performs favorably in terms of PSNR and SSIM metrics, which provides strong evidence in support of our design of the SDI module.

\subsubsection{Justification of the Reconstruction Module} To justify the design of the reconstruction module,  two variants are considered for comparison. The first one (w/o attention) has  all channel-wise feature fusion removed in order to show the effectiveness of the attention mechanism, and the second one (w/ RDB) replaces the reconstruction module with a module of similar size --- a cascade of $8$ RDBs to verify the strength of the reconstruction module in terms of dual-task performance. Quantitative comparisons in Table \ref{tab:ablation} provide strong evidence in favor of our original design.

\subsubsection{Influence of the Input Sequence Length} Here we investigate how the overall VSR performance depends on the number of input frames. 
To this end, we consider the cases with $3$ consecutive frames as inputs (w/~$3$ frames),  $5$ consecutive frames as inputs (w/~$5$ frames), and $7$ consecutive frames as inputs (baseline). The corresponding quantitative results are shown in Table~\ref{tab:ablation}.  It is clear that the VSR performance improves as  the number of input frames increases, which is expected since one can more likely find the missing information when there are more frames available for exploitation. It is also worth mentioning that due to the effective design of the SDI module, the proposed RawVSR can flexibly handle any input length without adjusting the network size (although increasing the number of input frames typically leads to longer processing time).

\subsection{Runtime and Model Size Comparisons}

\begin{figure}[h]
	\centering
	\begin{minipage}[h]{\linewidth}
		\centering
		\includegraphics[width=0.9\linewidth]{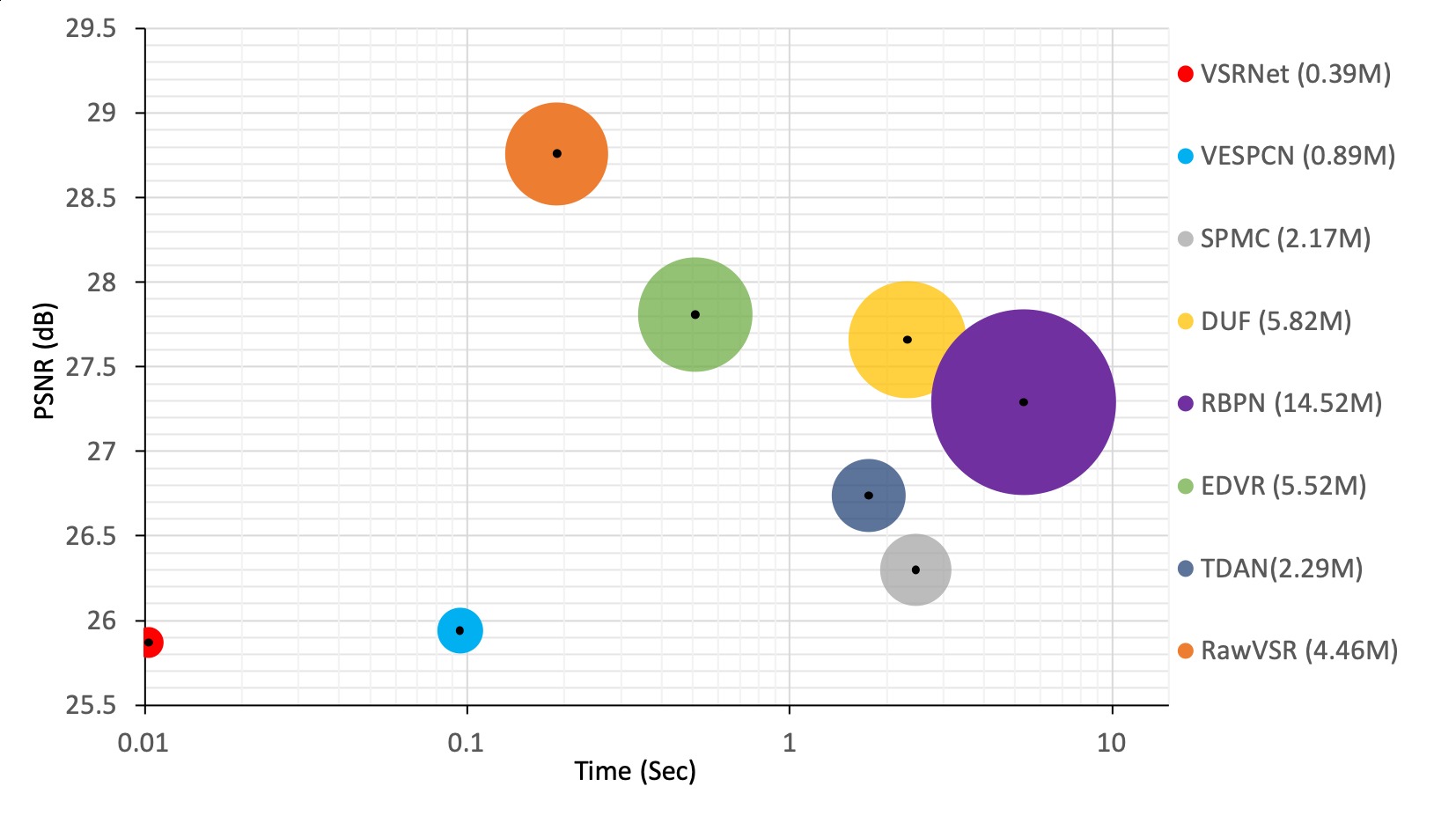}
	\end{minipage}
	\caption{Runtime and model size comparisons of different VSR methods for $4\times$ VSR.}
	\label{fig:time}
\end{figure}

Fig.~\ref{fig:time} compares different VSR methods in terms of runtime and model size as well as PSNR values.
Here runtime refers to the average time needed for producing one $1080$p frame (without taking into account the data loading phase) for $4\times$ VSR.
It can be seen that the proposed RawVSR is able to achieve the highest PSNR value with moderate model size and near-real-time processing.

%We compare the runtime and model size along with PSNR values of different VSR methods in . 

%The runtime is logged during the generation of one 1080p frame without counting in data loading time. It can be seen that the proposed method achieves the best VSR performance with relatively less network parameters. More importantly, our method is close to real-time VSR (0.13s per frame) that can be potentially embedded in camera ISP.  

\section{Conclusion}

We have proposed a new deep-learning-based method that can effectively exploit raw data for VSR.  The associated neural network has a dual-branch structure that disentangles the super-resolution process from the color correction process; as a consequence, it better fits the real imaging pipeline and has the desirable property of being device-independent. Its SDI module  is designed according to the architectural principle obtained via the analysis of a suitable probabilistic graphical model; this design philosophy is likely applicable to a wide range of problems, especially those involving data fusion. Its reconstruction module employs an attention mechanism to enhance the learning ability. Moreover, weight sharing is adopted to reduce the model size and to ensure the latent consistency of intermediate features in the two branches. Finally, it is also expected that the new raw video dataset collected in this paper can potentially benefit other video tasks including, among others, video interpolation and visual enhancement via exploiting raw data.

\ifCLASSOPTIONcaptionsoff
\newpage
\fi



%\bibliographystyle{IEEEtran}
%\begin{thebibliography}{1}
%\end{thebibliography}
\bibliographystyle{IEEEtran}
\bibliography{IEEEabrv,egbib}

\end{document}